\documentclass[english,a4paper]{aa}
\usepackage{charter}
\usepackage[T1]{fontenc}
\usepackage[latin1]{inputenc}
\usepackage{rotating}
\usepackage{graphicx}
\usepackage{amssymb}

\makeatletter

\newcommand{\noun}[1]{\textsc{#1}}

\providecommand{\tabularnewline}{\\}

\usepackage{charter}

\usepackage{rotating}

\usepackage{color}

\makeatletter

\makeatletter

\makeatletter

\makeatletter

\makeatletter

\titlerunning{L-Z relation of dIrr galaxies}
\authorrunning{I. Saviane et al.}

\makeatletter

\makeatletter

\usepackage{rotating}

\usepackage{txfonts}

\usepackage{color}

\def\RMS{0.11}   

\def\lz{L-Z}

\newcommand{\kms}{km s$^{\rm -1}$}

\newcommand{\micron}{\mbox{$\mu$m}}

\makeatother

\usepackage{babel}
\makeatother

\begin{document}

\title{Luminosity-Metallicity Relation for dIrr Galaxies in the Near-Infrared}

\author{Ivo Saviane\inst{1} Valentin D. Ivanov\inst{1} Enrico V. Held
\inst{2} Danielle Alloin\inst{3} R. Michael Rich\inst{4}
Fabio Bresolin \inst{5} \and Luca Rizzi \inst{6} \thanks{Based on data
collected at the European Southern Observatory, La Silla, Chile,
Proposal N. 70.B-0424(A,B); Based on observations made at Lick (UCO)
Observatory; Based on observations made with the William Herschel
Telescope operated on the island of La Palma by
the Isaac Newton Group in the Spanish Observatorio del Roque de los
Muchachos of the Instituto de Astrofísica de Canarias. }}

\institute{European Southern Observatory, Alonso de Cordova 3107, Santiago,
Chile\\
 \email{isaviane,vivanov@eso.org} \and OAPD, vicolo Osservatorio
5, I-35122 Padova, Italy\\
 \email{held@pd.astro.it} \and AIM, CEA/DSM/IRFU-Université Paris
7, Service d'Astrophysique, CEA/Saclay, 91191, Gif-sur-Yvette Cedex,
France \\
 \email{danielle.alloin@cea.fr} \and Department of Physics and
Astronomy, 430 Portola Plaza, UCLA, Los Angeles, CA 90095-1547, USA
\\
\email{rmr@astro.ucla.edu}\and Institute for Astronomy, University
of Hawaii at Manoa, 2680 Woodlawn Drive, Honolulu, HI 96822, USA\\
 \email{bresolin@ifa.hawaii.edu} \and Joint Astronomy Centre,
660 N. A\textquoteright{}ohoku Place, University Park, Hilo, HI 96720,
USA\\
\email{l.rizzi@jach.hawaii.edu}
}

\offprints{isaviane@eso.org}

\date{Received 05/04/2007; Accepted 22/05/2008}

\abstract{The luminosity-metallicity relation is one of the fundamental constraints
in the study of galaxy evolution; yet none of the relations available
today has been universally accepted by the community. } {The present
work is a first step to collect homogeneous abundances and near-infrared
(NIR) luminosities for a sample of dwarf irregular (dIrr) galaxies,
located in nearby groups. The use of NIR luminosities is intended to
provide a better proxy to mass than the blue luminosities commonly used
in the literature; in addition, selecting group members reduces the
impact of uncertain distances. Accurate abundances are derived to assess
the galaxy metallicity. } {Optical spectra are collected for \ion{H}{ii}
regions in the dIrrs, allowing the determination of oxygen abundances by
means of the temperature-sensitive method. For each dIrr galaxy $H$-band
imaging is performed and the total magnitudes are measured via surface
photometry.} {This high-quality database allows us to build a
well-defined luminosity-metallicity relation in the range $-13\geq
M_{H}\geq-20$. The scatter around its linear fit is reduced to
$\RMS$~dex, the lowest of all relations currently available. There might
exist a difference between the relation for dIrrs and the relation for
giant galaxies, although a firm conclusion should await direct abundance
determinations for a significant sample of massive galaxies.} {This new
dataset provides an improved luminosity-metallicity relation, based on a
standard NIR band, for dwarf star-forming galaxies.  The relation can
now be compared with some confidence to the predictions of models of
galaxy evolution. Exciting follow-ups of this work are (a) exploring
groups with higher densities, (b) exploring nearby galaxy clusters to
probe environmental effects on the luminosity-metallicity relation, and
(c) deriving direct oxygen abundances in the central regions of
star-forming giant galaxies, to settle the question of a possible
dichotomy between the chemical evolution of dwarfs and that of massive
galaxies.
\keywords{Galaxies: fundamental parameters; Infrared: galaxies; Galaxies:
dwarf; Galaxies: irregular; Galaxies: ISM} 
}


\maketitle

\section{Introduction}

According to standard scenarios of galactic chemical evolution, a
luminosity-metallicity (\lz) relation is established through galactic
winds -- induced by supernova explosions -- removing the interstellar
medium (ISM) before it has been totally converted into stars (see
the seminal work of Larson \cite{larson74}; and also Tinsley \& Larson
\cite{tinsley_larson79}, Dekel \& Silk \cite{dekel_silk86}, Lynden-Bell
\cite{lynden-bell92}, among others). As this process is more efficient
in low-mass (low escape velocity) galaxies, less metals should be
trapped in dwarf galaxies. Indeed, a well-defined \lz\ relation
is observed in spheroidal galaxies (e.g. Caldwell et al. \cite{caldwell_etal92}),
mostly field galaxies, in which old stellar populations dominate,
and for which the chemical evolution has stopped. Do we expect a similar
behaviour in the case of dwarf irregular galaxies (dIrr)? The situation
is more complex. First, dIrrs are still active (in the sense of star
formation), exhibiting substantial gas fractions and a broad range
of star-formation rates (SFR). So, one would expect to encounter in
Irrs, ISMs of different chemical maturities and rapidly evolving luminosities:
all factors which would loosen any established relation between mass
and metallicity. Second, the available sample of dIrrs is made of
galaxies belonging to groups rather than distributed in the field.
It is well known that the evolution of a galaxy gas content depends
on its environment, and that gas stripping is boosted through galaxy
interactions in dense groups and cluster sub-clumps. So, while one
can see reasons why the terminal chemical evolution in dIrrs might
depend on the galactic mass -- similarly to the sample of gas-free
spheroidal galaxies --, there are a number of extra processes blurring
the situation. Hence, it is unclear whether dIrr galaxies should exhibit
a luminosity-metallicity relation.

Under such a scenario dwarf galaxies lose substantial fractions of their
mass in the course of their evolution, and should be major contributors
to the chemical evolution of the inter-galactic medium (IGM). In
particular, mass loss through galactic winds should occur in dIrr
galaxies: their study contributes to probe the fraction of the IGM
enrichment which is due to dwarf galaxies in general {(e.g.,
Garnett \cite{garnett02})}.  Losses of enriched gas in dwarf galaxies
have indeed been observed: Martin et al. (\cite{martin_etal03}) detected
a galactic wind of $\sim6\times10^{6}\mathcal{M}{}_{\odot}$ from
NGC~1569, and for the first time, they could measure its
metallicity. The authors concluded at $3\times10^{4}\mathcal{M}_{\odot}$
of oxygen in the wind, almost as much as in the disk of the dwarf
itself. Hence, the cosmic chemical evolution and the existence of an
\lz\ relation appear to be closely inter-related: this was an additional
motivation to pursue the study presented here.

The case of another dIrr, SagDIG, further illustrates this point.
In one of our previous studies (Saviane et al. \cite{sagdig}) we
measured a very low oxygen content ($7.26\leq12+\log{\textrm{(O/H)}}\leq7.50$),
a result which is not compatible with a closed-box evolution. According
to the model, such a low abundance would imply a large gas mass fraction
$\mu=m_{{\rm gas}}/(m_{{\rm gas}}+m_{{\rm stars}})\approx0.97$, whereas
the observed gas mass fraction is only $\mu\approx0.86$. It is easy
to compute that $\approx1.5\times10^{6}\mathcal{M}_{\odot}$ of gas
are missing, and, since this mass is smaller than that of the NGC~1569
galactic wind, it is plausible to conclude that SagDIG lost some of
its gas into the IGM.

{On the other hand, a cautionary remark should be added, since
a general consensus on the role of galactic winds in the evolution
of dwarf galaxies and the IGM has not been reached yet. For example
by means of chemo-photometric models Calura \& Matteucci (\cite{calura_matteucci06})
conclude that dIrr galaxies play a negligible role in the enrichment
of the IGM, and the numerical models of Silich and Tenorio-Tagle (\cite{silich_tagle98})
show that that galactic winds never reach the escape velocity of these
dwarfs. And in the case of dwarf spheroidal (dSph) galaxies, the 3D
hydrodynamic simulations of Marcolini et al. (\cite{marcolini_etal06})
show that no galactic winds develop in these objects.}

In our attempt to place SagDIG on previously established \lz\ relations
for dIrr galaxies, we realized that the existence of such a relation
was controversial. Some studies had concluded at very well-defined
correlations (Skillman et al. \cite{skh89}; Richer \& McCall \cite{rm95};
Pilyugin \cite{pilyugin01b}), others have found only mild relations
with substantial scatter (e.g. Skillman et al. \cite{skillman_scl_hii}),
or no correlation at al (Hidalgo Gámez \& Olofsson \cite{anamaria};
Hunter \& Hoffman \cite{hunter_hoffman99}). Perhaps the major source
of confusion in these investigations comes from the ill-defined samples
used in the analysis. Abundance data are taken from different sources
(with spectra of variable quality and processed through different
reduction and analysis techniques); apparent luminosities and distances
are taken from catalogs, meaning that they are largely approximate.
Moreover, although the \lz\ relation is expected to depend on the
environment, this is rarely taken into account in the analysis. Another
main source of uncertainty is the common use of the blue absolute
magnitude as a tracer of the mass. Optical luminosities can in fact
be extremely misleading in Irr galaxies, because of their star-bursting
activity (e.g. Tosi et al. \cite{tosi_etal92} or Schmidt et al. \cite{schmidt_etal95}).
Already, Bruzual \& Charlot (\cite{bruzual_charlot83}) showed that
at $400\,{\rm nm}$ a $1$~Myr old burst is $\sim$ three orders
of magnitude brighter than an underlying old ($15$~Gyr) stellar
population of comparable mass. In other words, a dwarf galaxy hosting
a recent starburst could be as luminous, in the blue, as a galaxy
orders of magnitudes more massive but lacking a recent star formation
episode. In comparison, a recent starburst is at most $10$ times
brighter in the near-infrared (NIR) than its underlying old population
of same mass. Therefore, the NIR window being more stable with regard
to star formation episodes is more appropriate for probing the galaxy
basic properties, such as its mass.

In order to {address these issues}, we embarked on a medium-term
project aimed at gathering nebular oxygen abundances and NIR luminosities
for a sample of dIrrs belonging to the three nearest groups of galaxies.
In this way we can test the existence of an \lz\ relation in well-defined
environments (characterized by the group density), for which the scatter
in apparent distance modulus is low, and for which an homogeneous
set of abundances can be obtained.

Since we started the project, a few investigations have been published,
in relation with the context presented above. The first \lz\ relation
using NIR luminosities (aside the work by Saviane et al. \cite{sidney})
is that by Salzer et al. (\cite{salzer05}; hereafter S05). Their
sample is dominated by massive galaxies, and their relation was derived
using 2MASS data for the luminosities, and proprietary spectra from
the KPNO International Spectroscopic Survey (KISS) for the metallicities.
The 2MASS survey having a relatively shallow limiting magnitude, for
a few additional dwarf galaxies NIR photometry was supplemented by
the authors. At the other extreme, the sample assembled by Mendes
de Oliveira et al. (\cite{mendes-de-oliveira_etal06}) is entirely
made of dwarf galaxies. They gathered $K$-band luminosities and metallicities
-- obtained through the direct method -- for $29$ dIrrs. The NIR
data are from 2MASS or from Vaduvescu et al. (\cite{vaduvescu_etal05}),
and metallicities have been picked up from a variety of sources. Finally
Lee et al. (\cite{lee_etal06}) have used the \emph{Spitzer} Infrared
Array Camera to compute $4.5$~\micron\ luminosities for $\sim30$
nearby dIrrs, the distance of which have been derived using standard
candles. With oxygen abundances collected from the literature, they
{constructed a} 4.5~\micron\ \lz\ relation.

Although these studies have certainly improved the situation, they
do not represent the ideal case yet. S05 abundances are derived through
a new but indirect method, and the sample is very scarce in dwarf
galaxies. Yet, S05 is the only study which includes $H$-band photometry,
so a comparison with their \lz\ relation is carried out later in
this paper, in Sect.~\ref{sec:dichot}. As extensively discussed
in Sect.~\ref{subsec:Comparison-with-other}, the use of heterogeneous
data from the literature may produce a different \lz\ relation than
the one obtained with an homogeneous dataset, a likely consequence
of distance uncertainties. Therefore, we anticipate that the relation
by Mendes de Oliveira et al. (\cite{mendes-de-oliveira_etal06}) will
need to be revised once a better controlled sample, with $K$-band
photometry, becomes available. Finally, the use of a non-standard
passband does not allow an easy comparison of the \lz\ relation
by Lee et al. (\cite{lee_etal06}) to other \lz\ relations. For
example they need to make several successive assumptions in order
to convert their luminosities into masses, and then compare the mass-metallicity
relation to the SDSS one. For all these reasons, we believe that so
far our approach stands as the one with the smallest number of uncertainties
and the broadest applicability.

The paper is structured as described below. In Sect.~2 we explain
how the targets were selected, and give a brief account of the data
reduction techniques -- described in more detail in the appendices
--. The computation of chemical abundances is described in Sect.~3,
where we also compare our results with previous abundance determinations.
The resulting NIR luminosity-metallicity relation for dIrr galaxies
is presented in Sect.~4. It is discussed in Sect.~5, with a possible
explanation of its origin presented in Sect.~5.1 and its comparison
with a relation based on literature data in Sect.~5.2. In Sect.~5.3
we discuss the possible dwarf vs. giant galaxy dichotomy, through
a comparison of our results to those by Salzer et al. (\cite{salzer05}).
Finally a summary and the conclusions of this study are provided in
Sect.~6.

\section{The data set \label{sec:The-data-set}}

\subsection{Target selection}

\begin{table}
\caption{Target list \label{tabcap:Name-aliases-of}}

\begin{centering}
\begin{tabular}{lll}
\multicolumn{3}{c}{}\tabularnewline
\hline
\hline 
Main ID  & \multicolumn{2}{c}{Other IDs}\tabularnewline
\hline 
 & \multicolumn{2}{c}{}\tabularnewline
\multicolumn{3}{c}{M81 group}\tabularnewline
DDO~42  & NGC~2366  & UGC 03851\tabularnewline
KDG~52  & M81~dw~A  & PGC 023521\tabularnewline
DDO~53  & UGC~4459  & VII Zw 238\tabularnewline
UGC~4483  &  & \tabularnewline
KDG~54  & UGC~4945  & UGCA 158\tabularnewline
candidate 5  & {[}FM2000] 5  & \tabularnewline
BK~3N  & PGC~028529  & \tabularnewline
DDO~66  & UGC 05336  & Holmberg~IX\tabularnewline
DDO~82  & UGC~5692  & \tabularnewline
DDO~165  & UGC~8201  & VII Zw 499\tabularnewline
 &  & \tabularnewline
\multicolumn{3}{c}{Sculptor group}\tabularnewline
ESO 347-G017  &  & \tabularnewline
UGCA 442  & ESO 471- G 006  & AM 2341-321 \tabularnewline
ESO 348-G009  & AM 2346-380  & UKS 2346-380\tabularnewline
NGC 59  & ESO 539- G 004  & \tabularnewline
ESO 294-G010  & AM 0024-420  & \tabularnewline
ESO 473-G024  & AM 0028-23  & \tabularnewline
AM0106-382  & LEDA 166061  & {[}KK98] 011\tabularnewline
NGC 625  & ESO 297- G 005  & AM 0132-414\tabularnewline
ESO 245-G005  & AM 0142-435  & A143\tabularnewline
\hline
\end{tabular}
\par\end{centering}
\end{table}

The next two sections describe how the targets were selected in each
of the two groups. The most popular galaxy names are listed in Table~\ref{tabcap:Name-aliases-of},
together with other aliases.

\subsubsection{Sculptor group}

\begin{figure}
\begin{centering}
\begin{tabular}{|c|c|c|}
\hline 
DSS  & $H$  & H$\alpha$, R\tabularnewline
\hline
\hline 
\multicolumn{3}{|c|}{NGC 625}\tabularnewline
\hline 
\includegraphics[width=0.1\textheight]{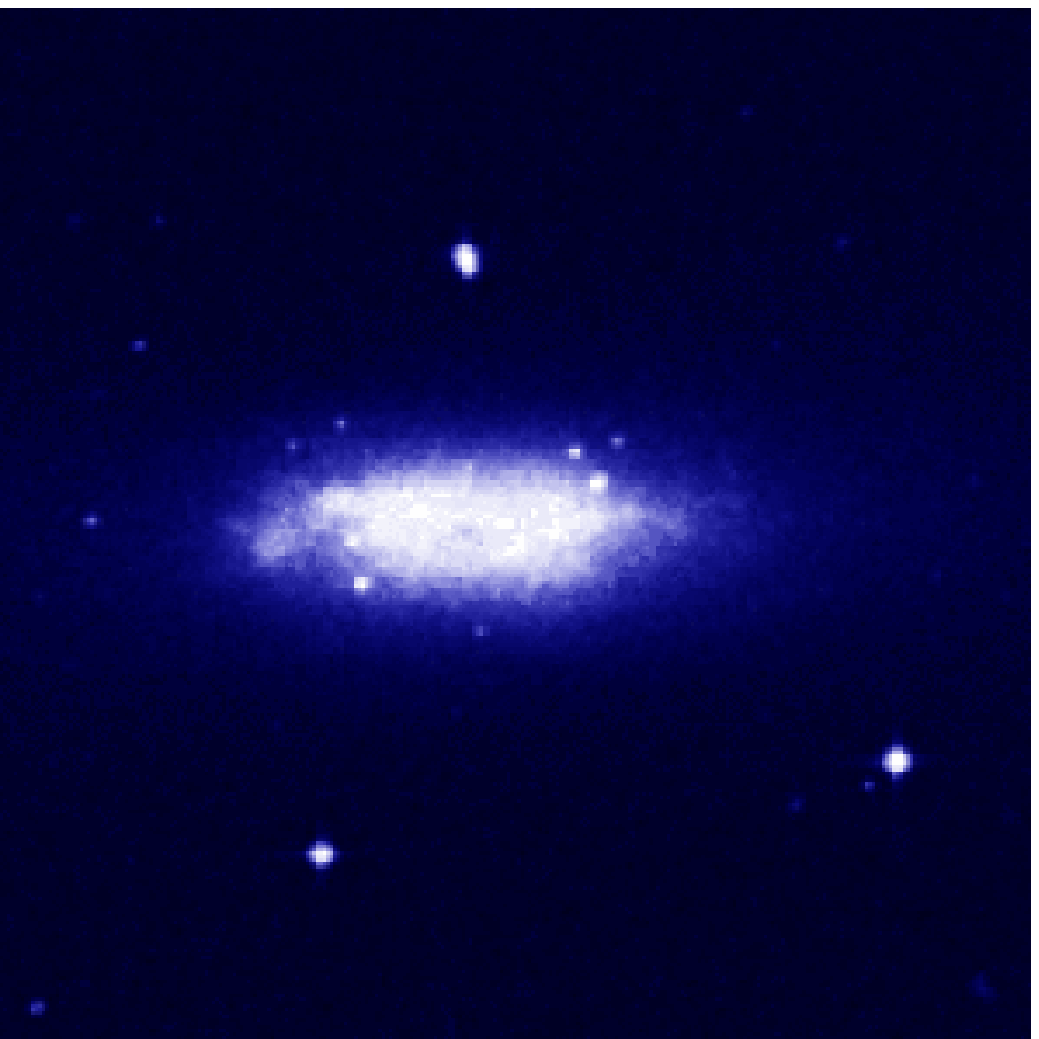}  & \includegraphics[width=0.1\textheight]{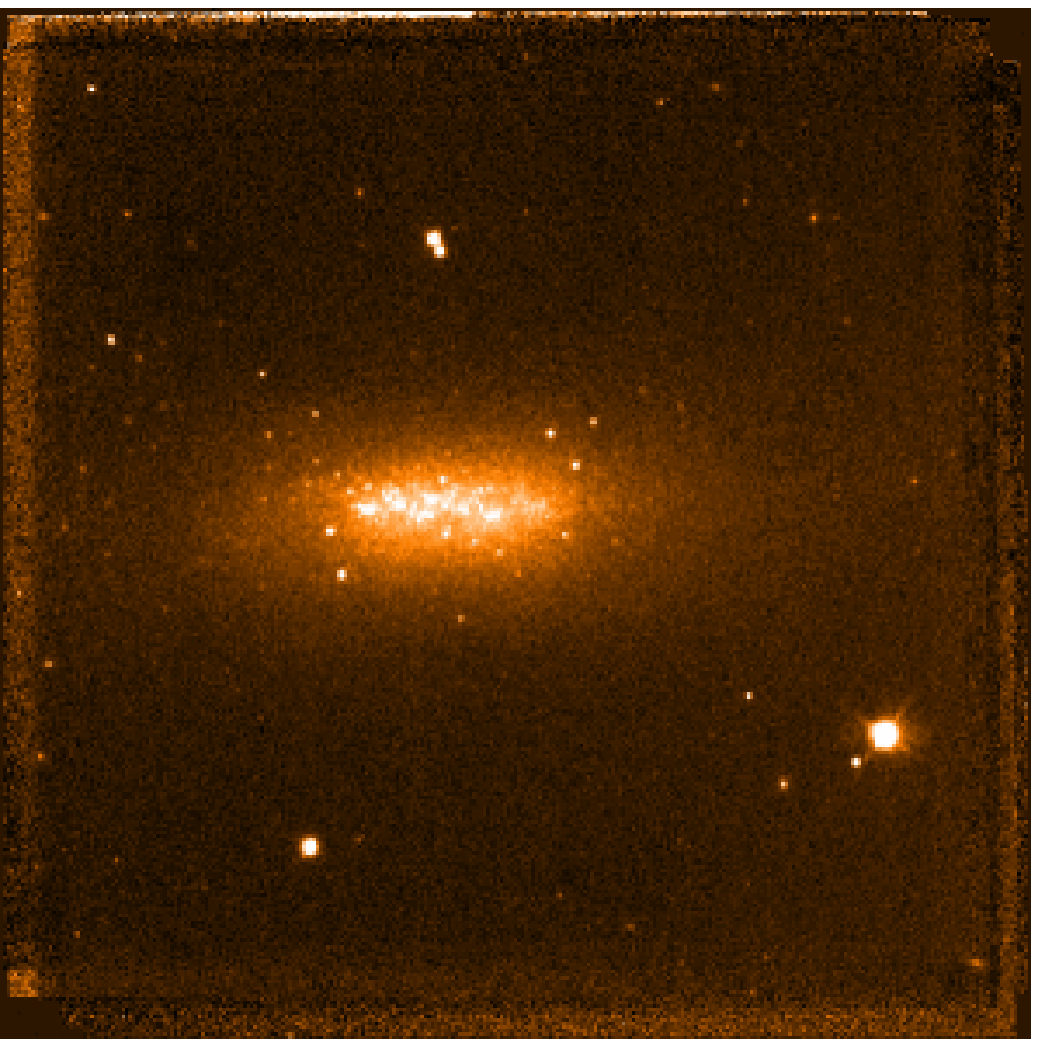}  & \includegraphics[width=0.1\textheight]{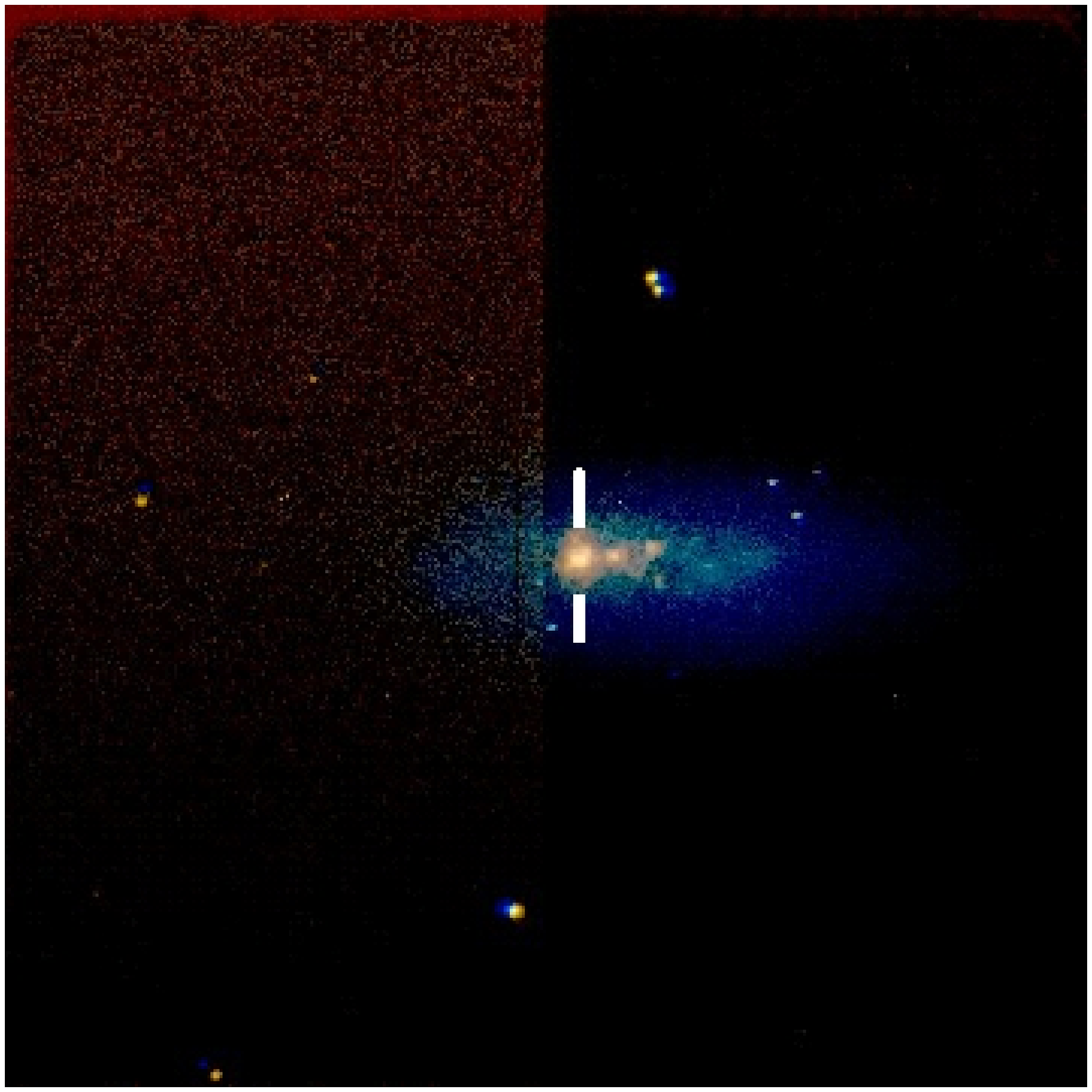}\tabularnewline
\hline 
\multicolumn{3}{|c|}{NGC 59}\tabularnewline
\hline 
\includegraphics[width=0.1\textheight]{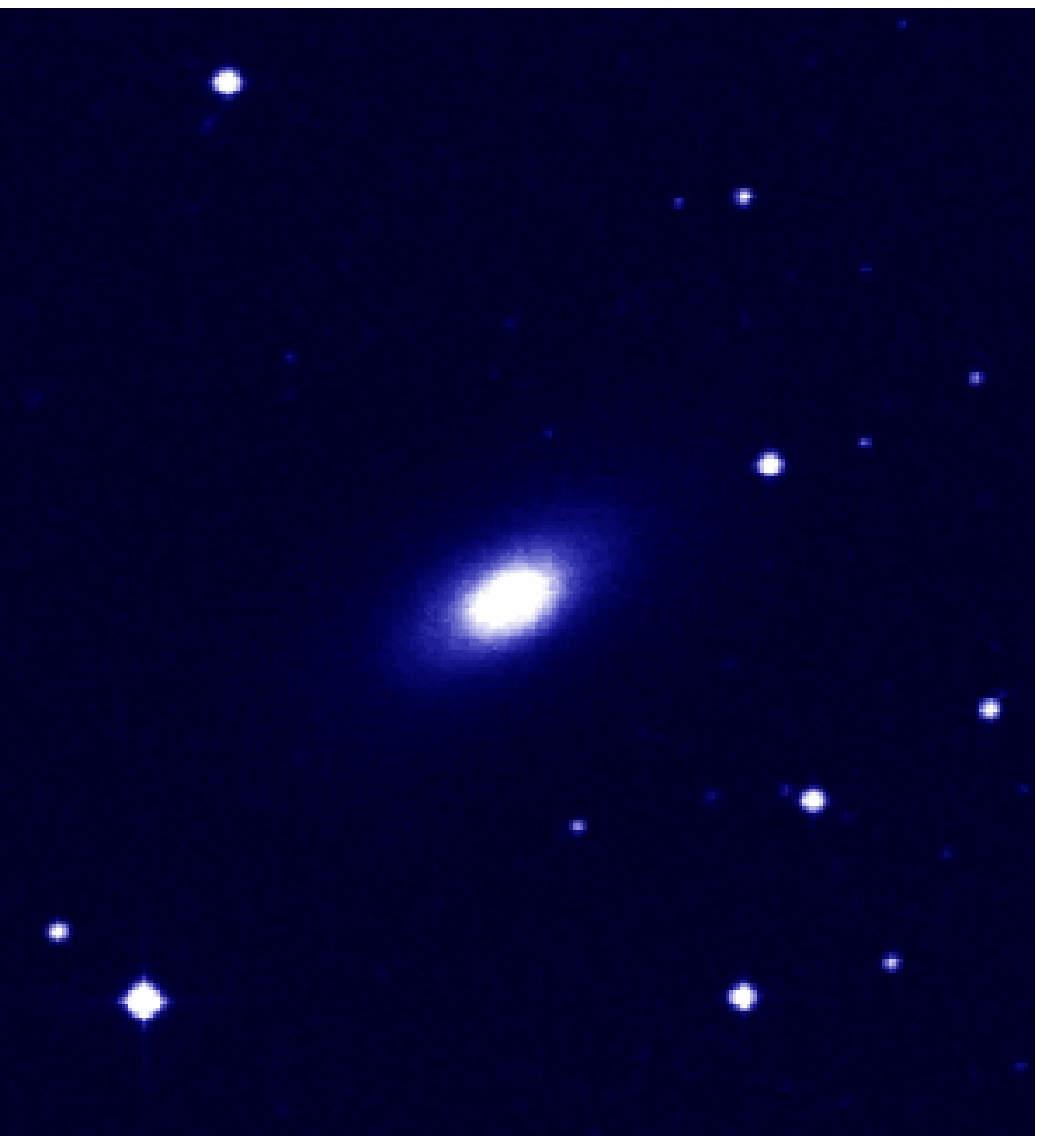}  & \includegraphics[width=0.11\textheight]{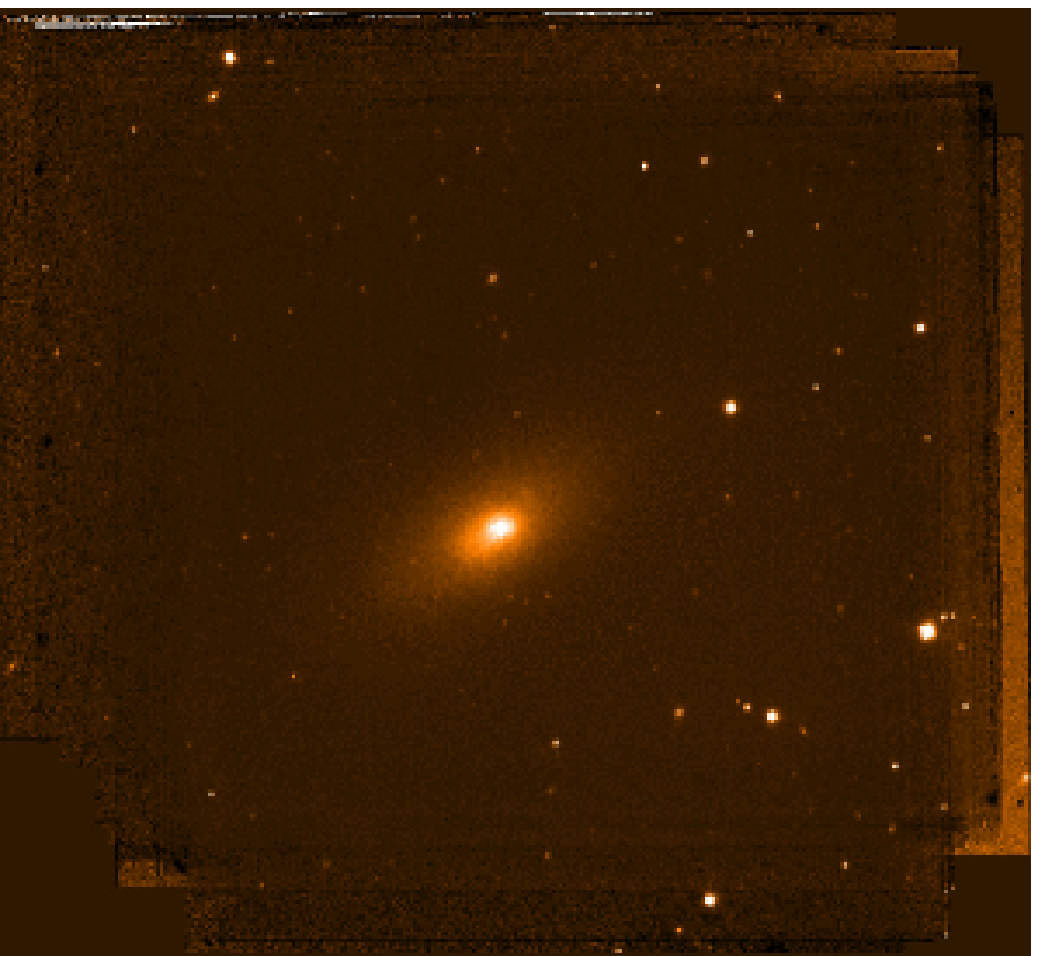}  & \includegraphics[width=0.1\textheight]{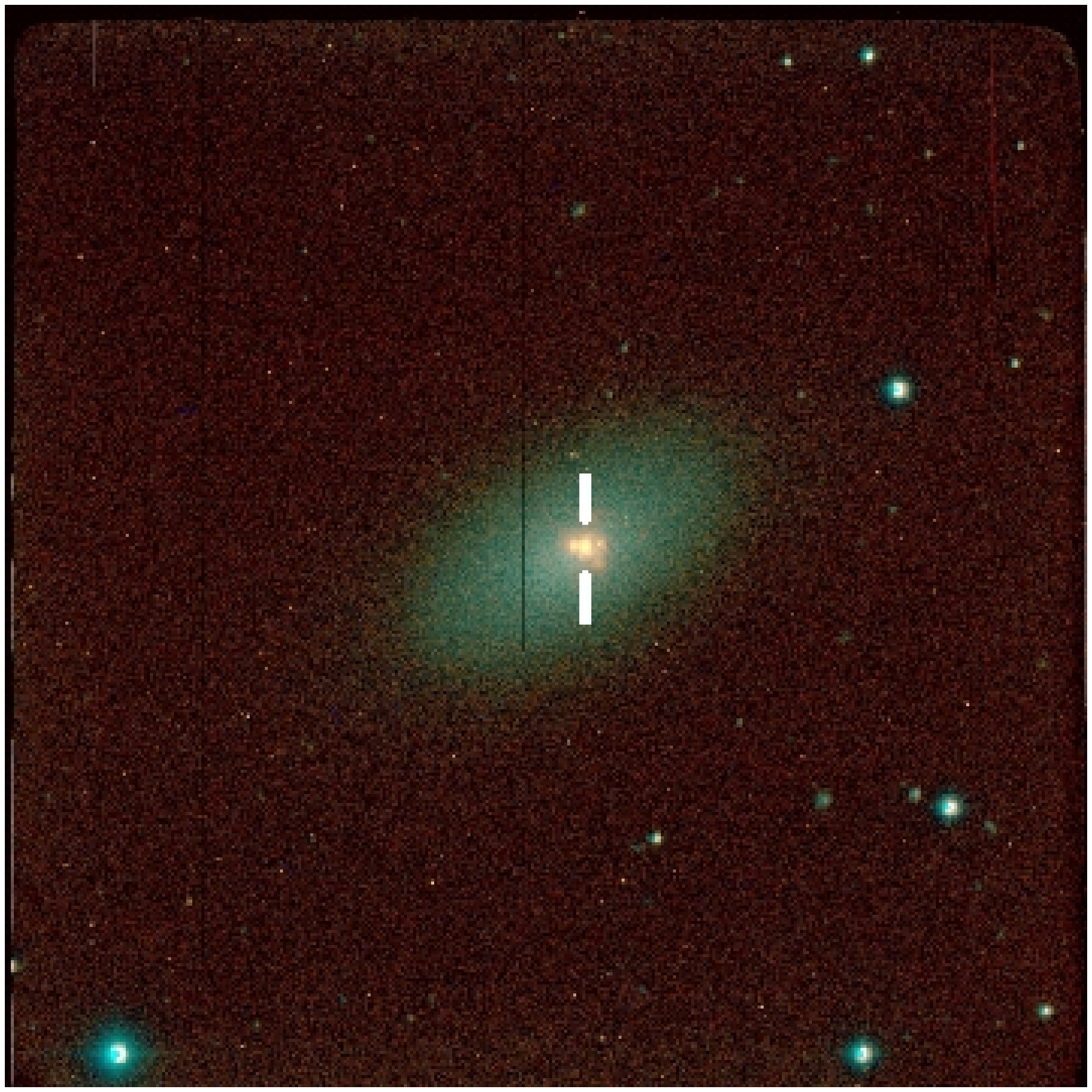}\tabularnewline
\hline 
\multicolumn{3}{|c|}{ESO 245-G005}\tabularnewline
\hline 
\includegraphics[width=0.1\textheight]{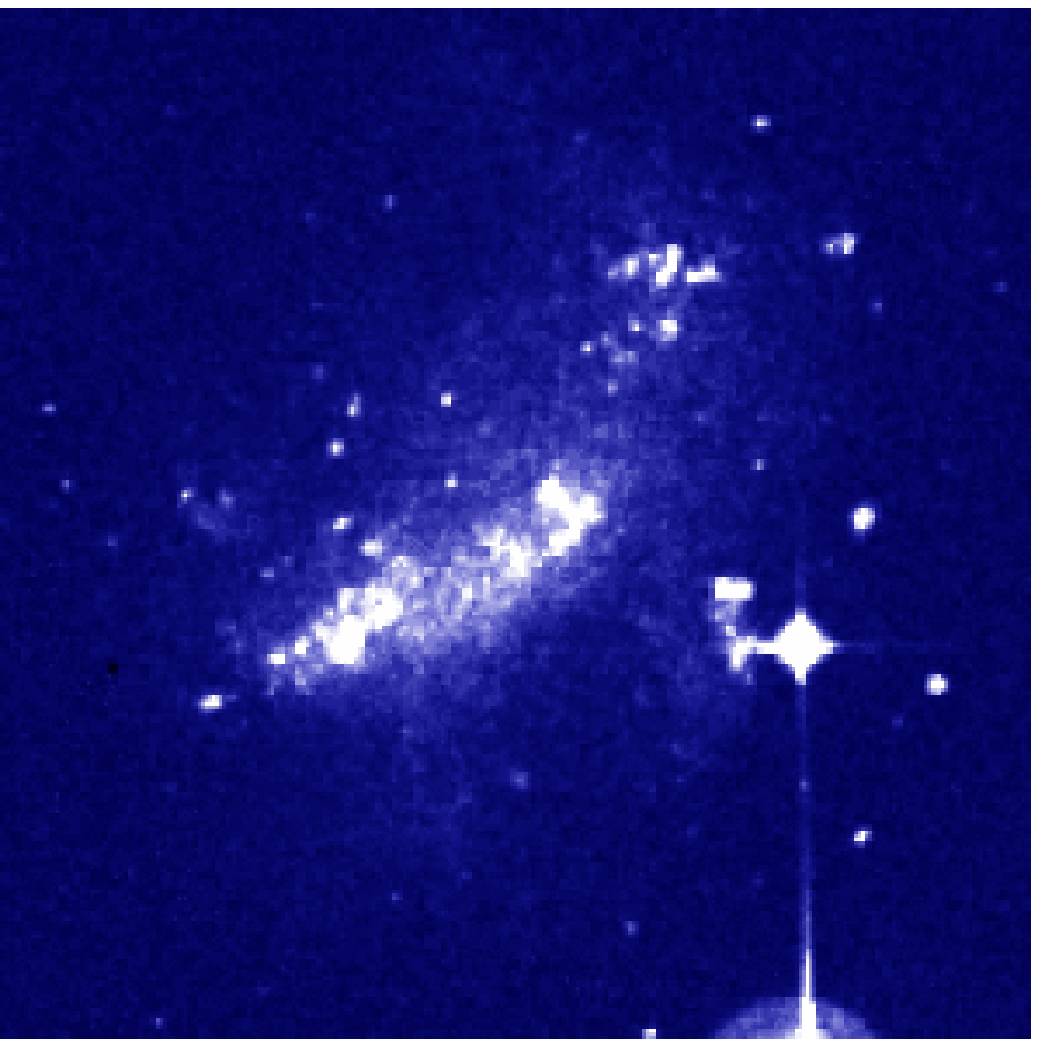}  & \includegraphics[width=0.1\textheight]{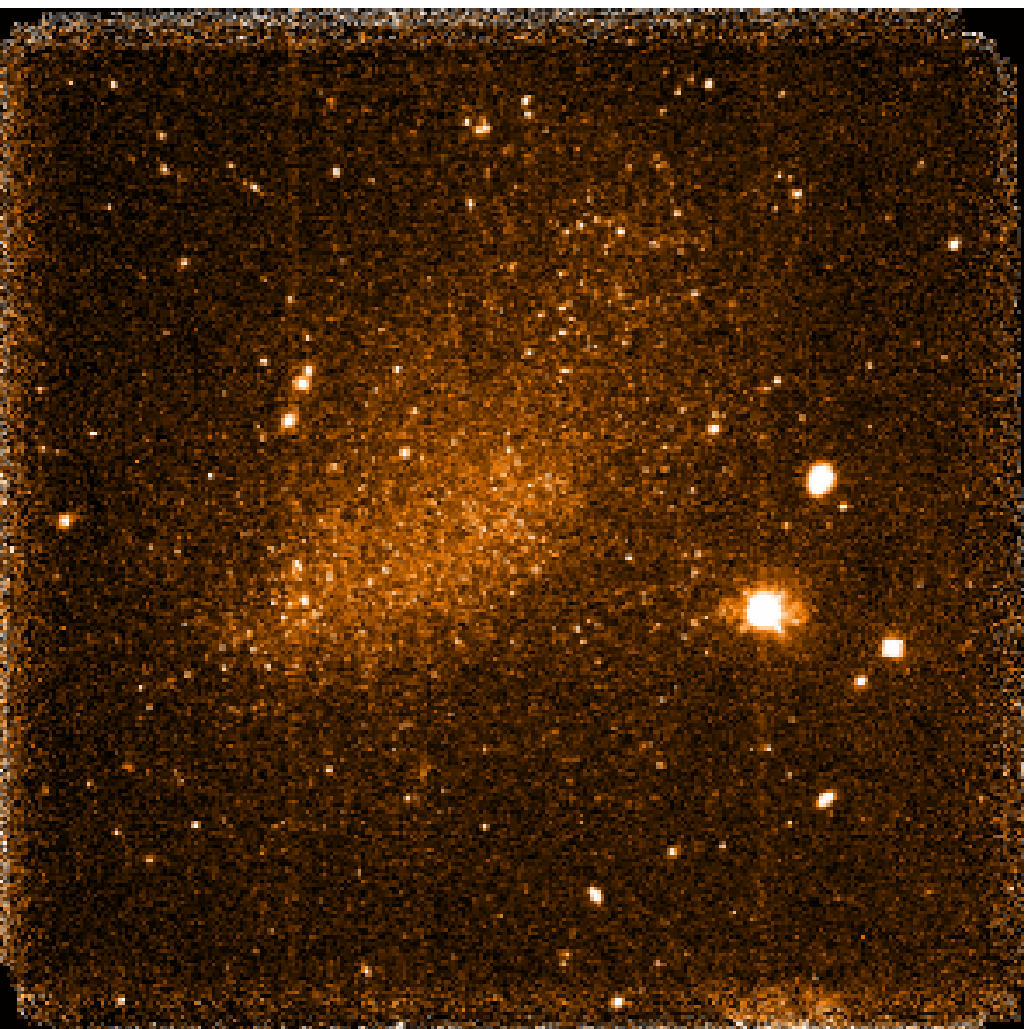}  & \includegraphics[width=0.1\textheight]{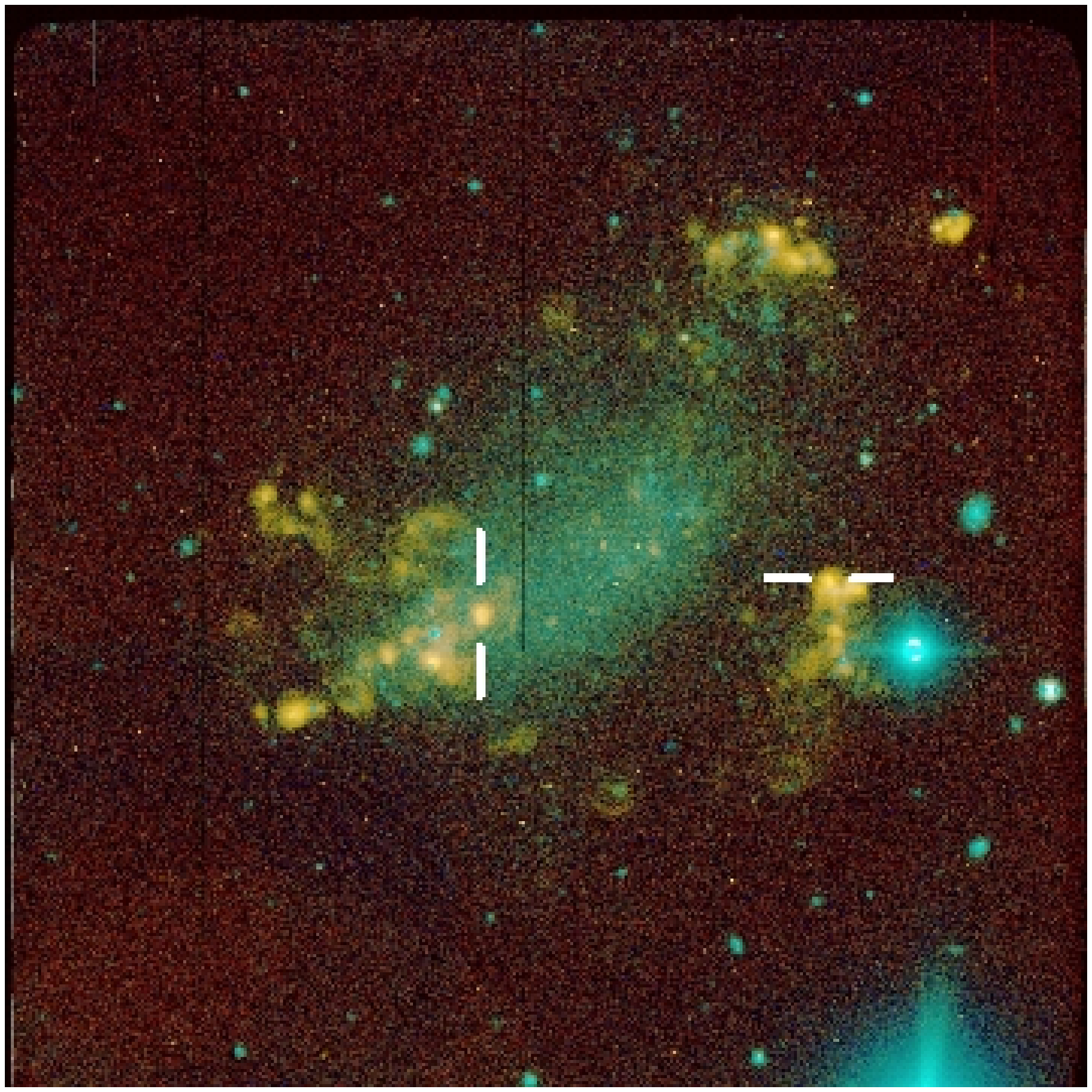}\tabularnewline
\hline 
\multicolumn{3}{|c|}{UGCA 442}\tabularnewline
\hline 
\includegraphics[width=0.1\textheight]{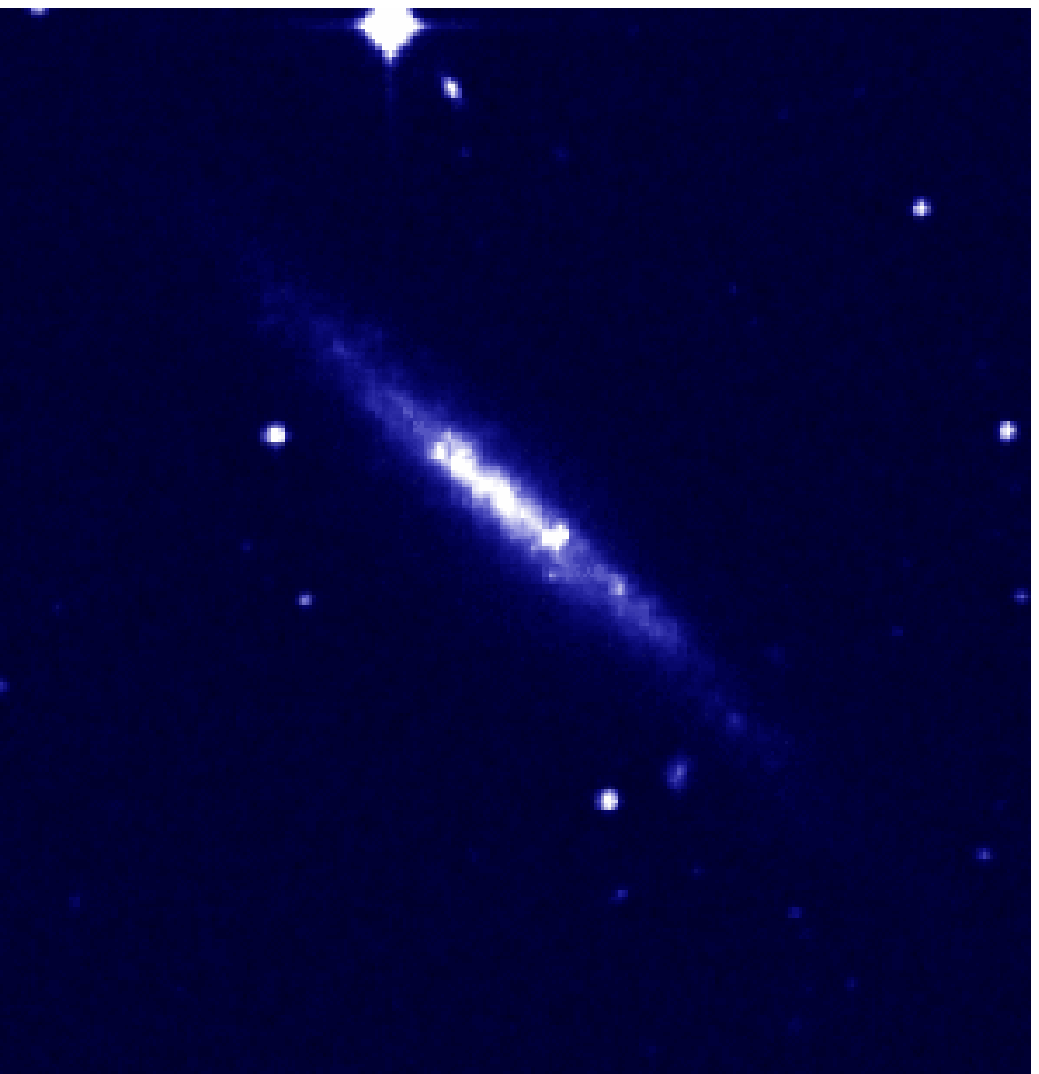}  & \includegraphics[width=0.1\textheight]{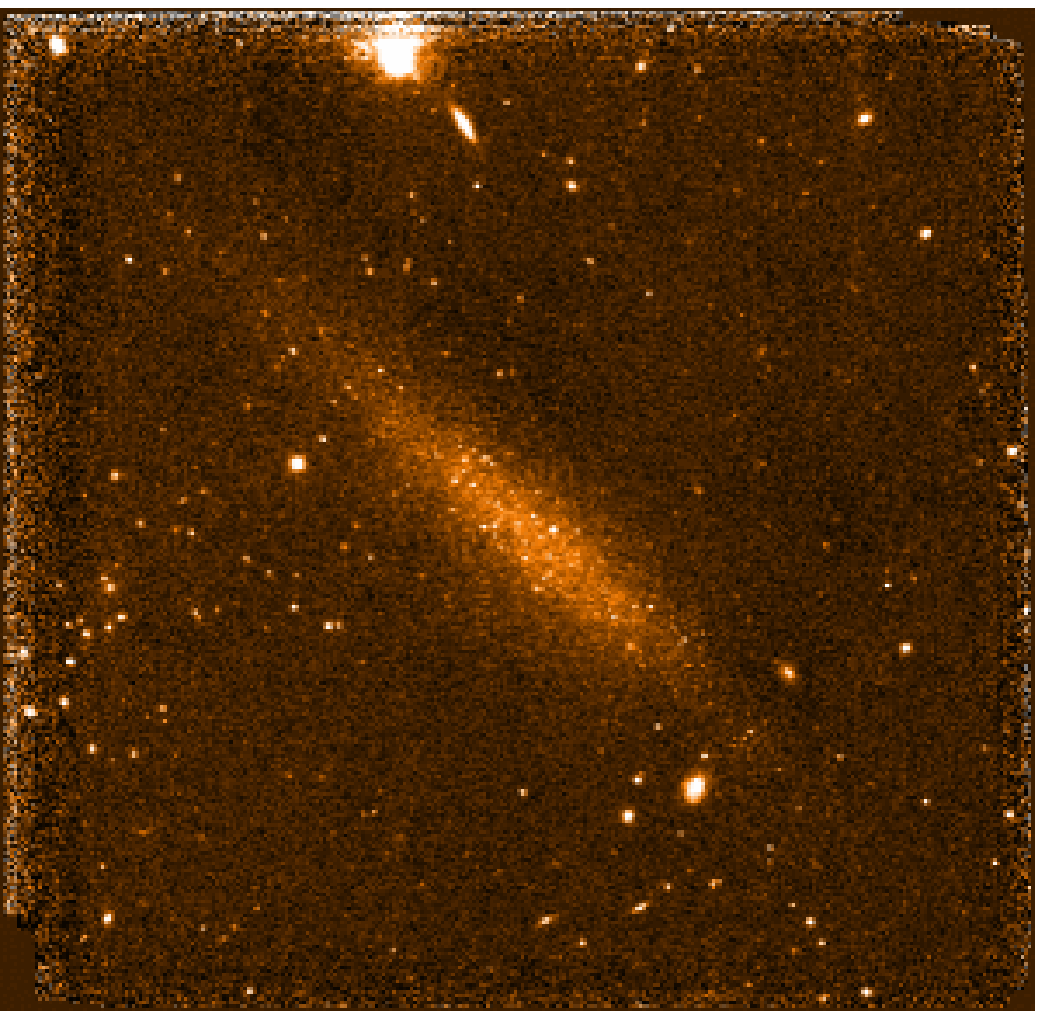}  & \tabularnewline
\hline 
\multicolumn{3}{|c|}{ESO 347-G017}\tabularnewline
\hline 
\includegraphics[width=0.1\textheight]{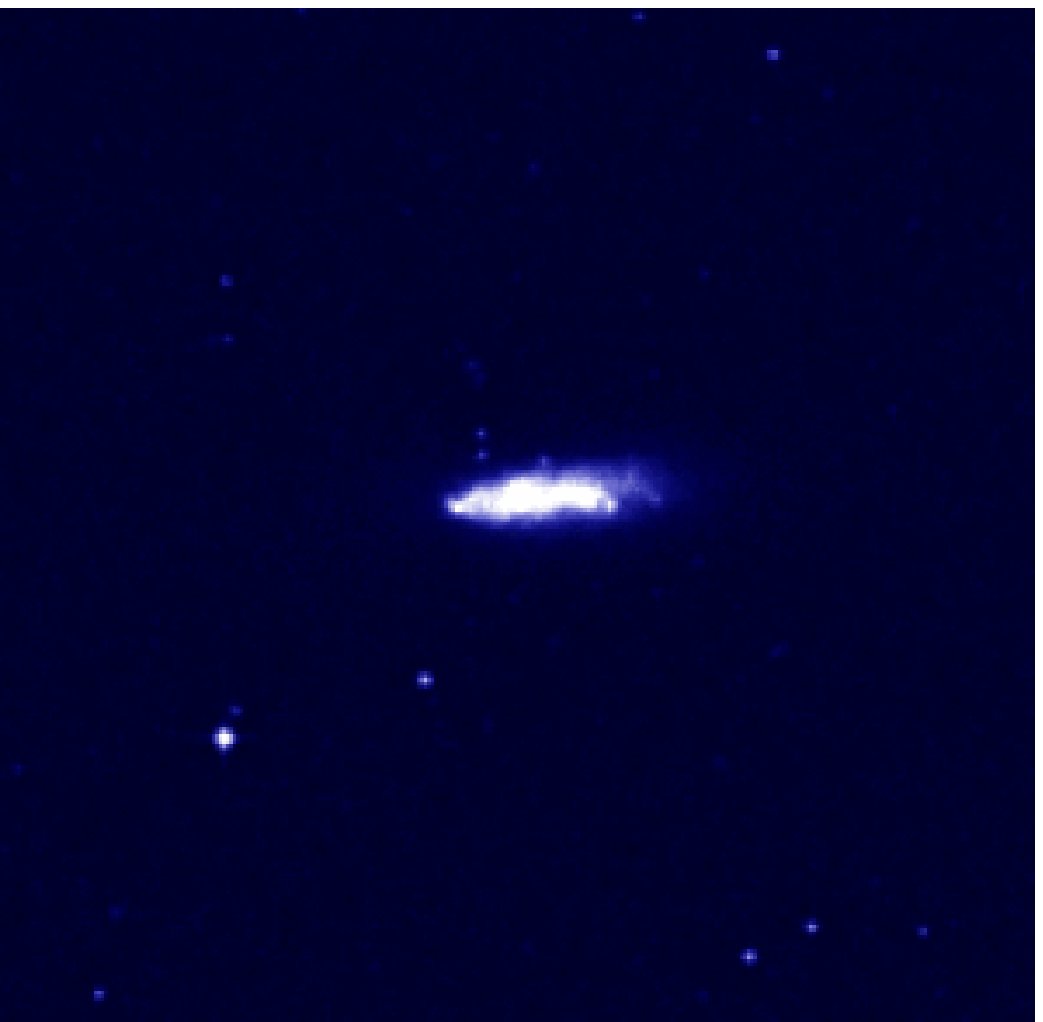}  & \includegraphics[width=0.1\textheight]{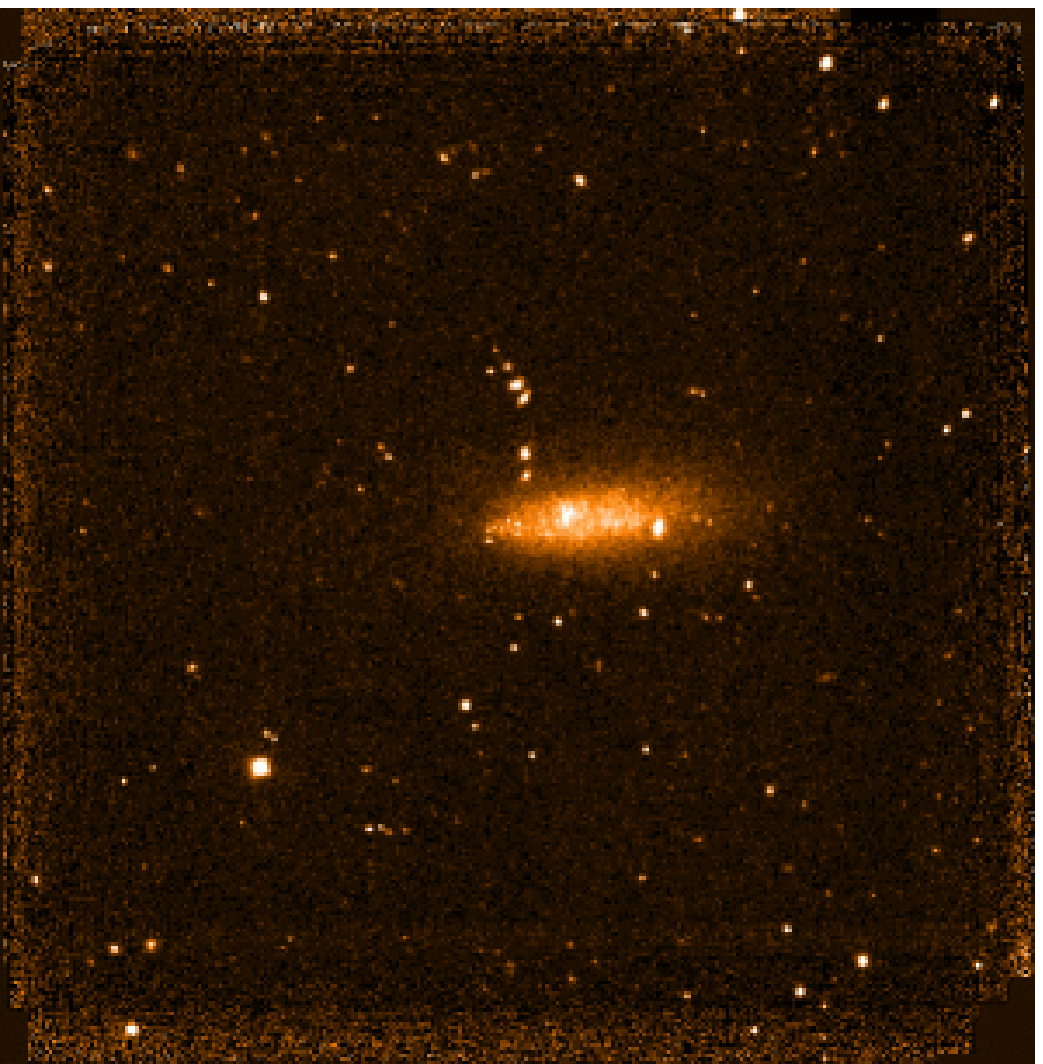}  & \includegraphics[width=0.1\textheight]{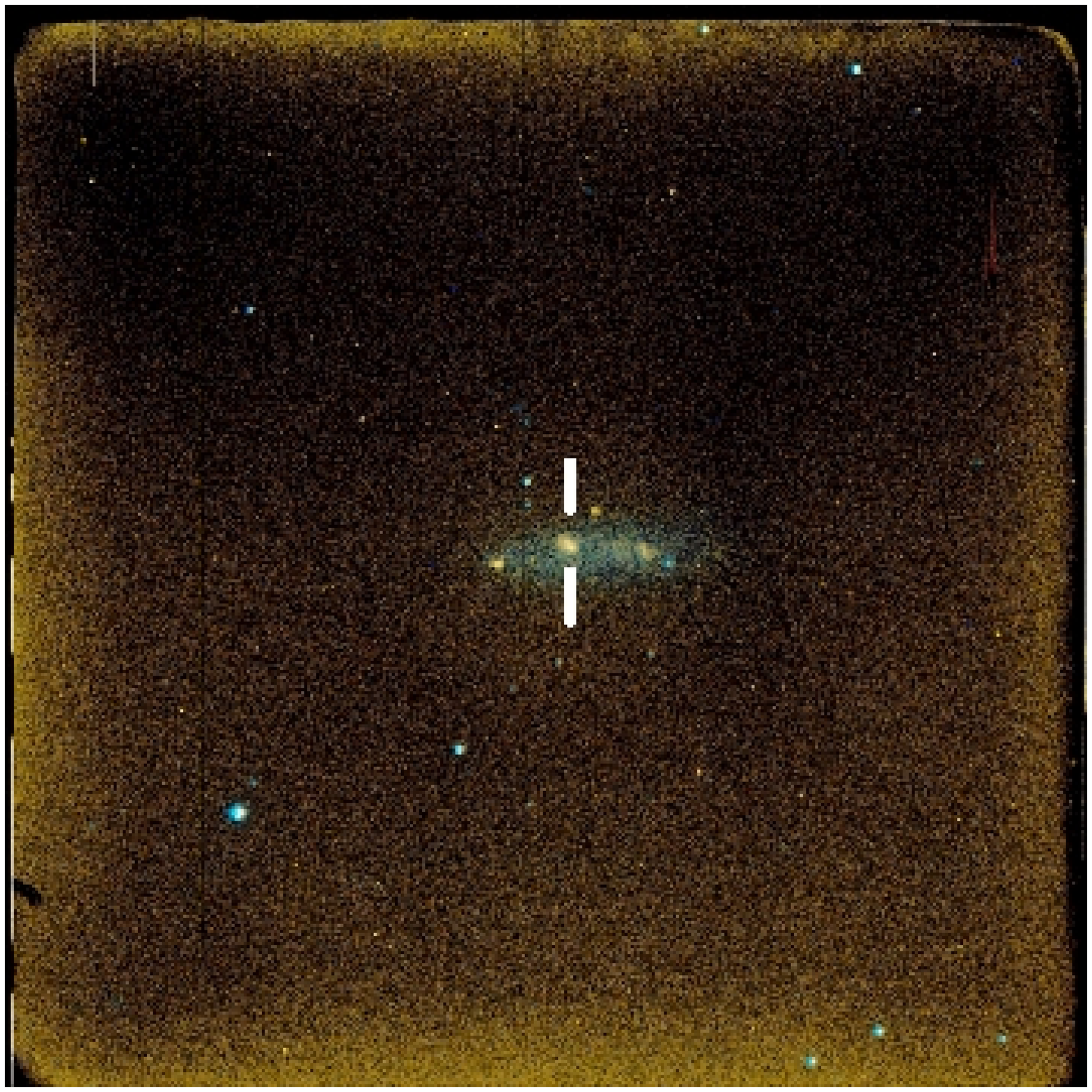}\tabularnewline
\hline
\end{tabular}
\par\end{centering}

\caption{From left to right, we show the morphology of the Sculptor group targets
in the optical (DSS), infrared $H$ (SOFI), and a false-color montage
of H$\alpha$ and R (EFOSC2) images. The reddish, diffuse nebulae
in the right image are \ion{H}{ii} regions, and the vertical
ticks mark those for which spectra have been obtained. The galaxy
luminosity decreases from top to bottom panels, with the exception
of ESO 294~G010, which was not observed in the NIR. In all images,
North is up and East to the left, and the field of view is $\sim5\arcmin\times5\arcmin$.\label{figcap:From-left-to}}

\end{figure}

\begin{figure}
\begin{centering}
\begin{tabular}{|c|c|c|}
\hline 
DSS  & $H$  & H$\alpha$,R\tabularnewline
\hline
\hline 
\multicolumn{3}{|c|}{ESO 348-G009}\tabularnewline
\hline 
\includegraphics[width=0.1\textheight]{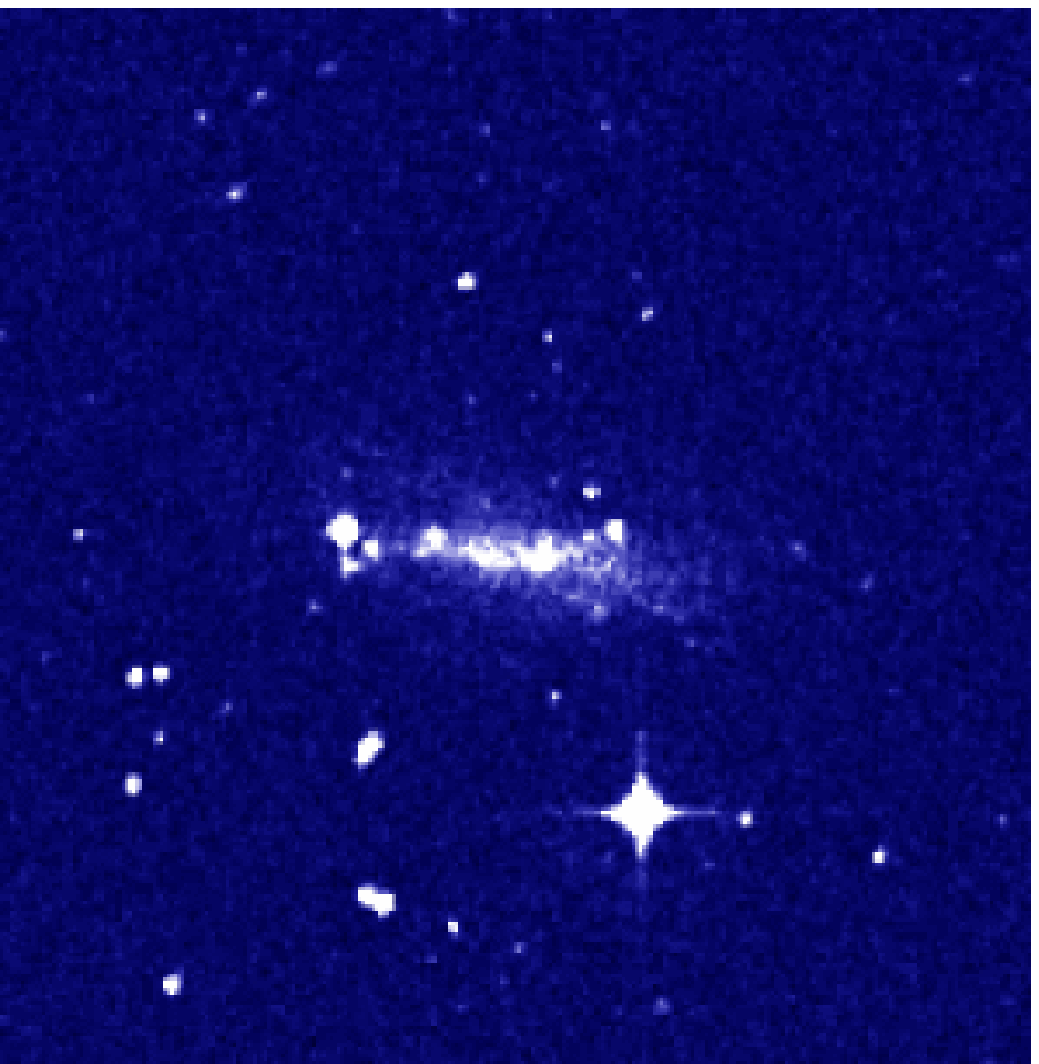}  & \includegraphics[width=0.1\textheight]{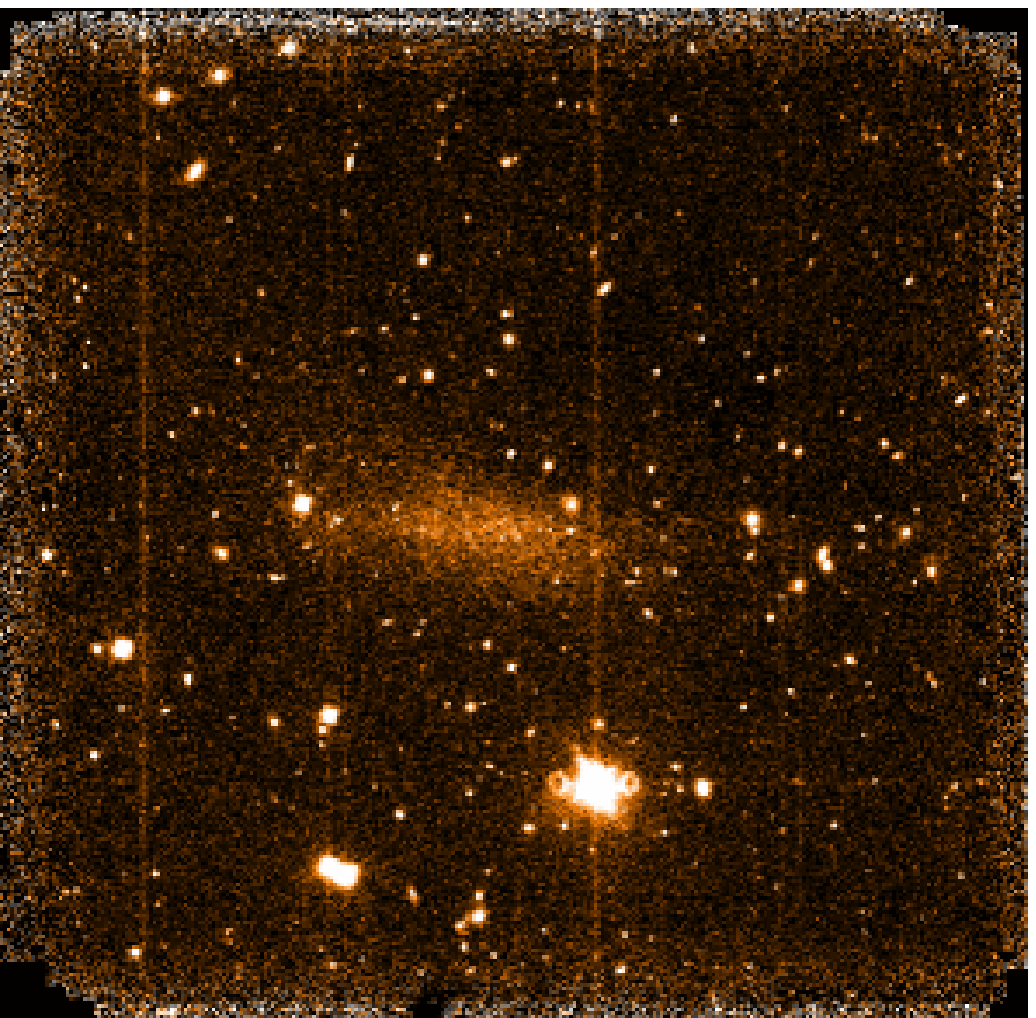}  & \tabularnewline
\hline 
\multicolumn{3}{|c|}{ESO 473-G024}\tabularnewline
\hline 
\includegraphics[width=0.1\textheight]{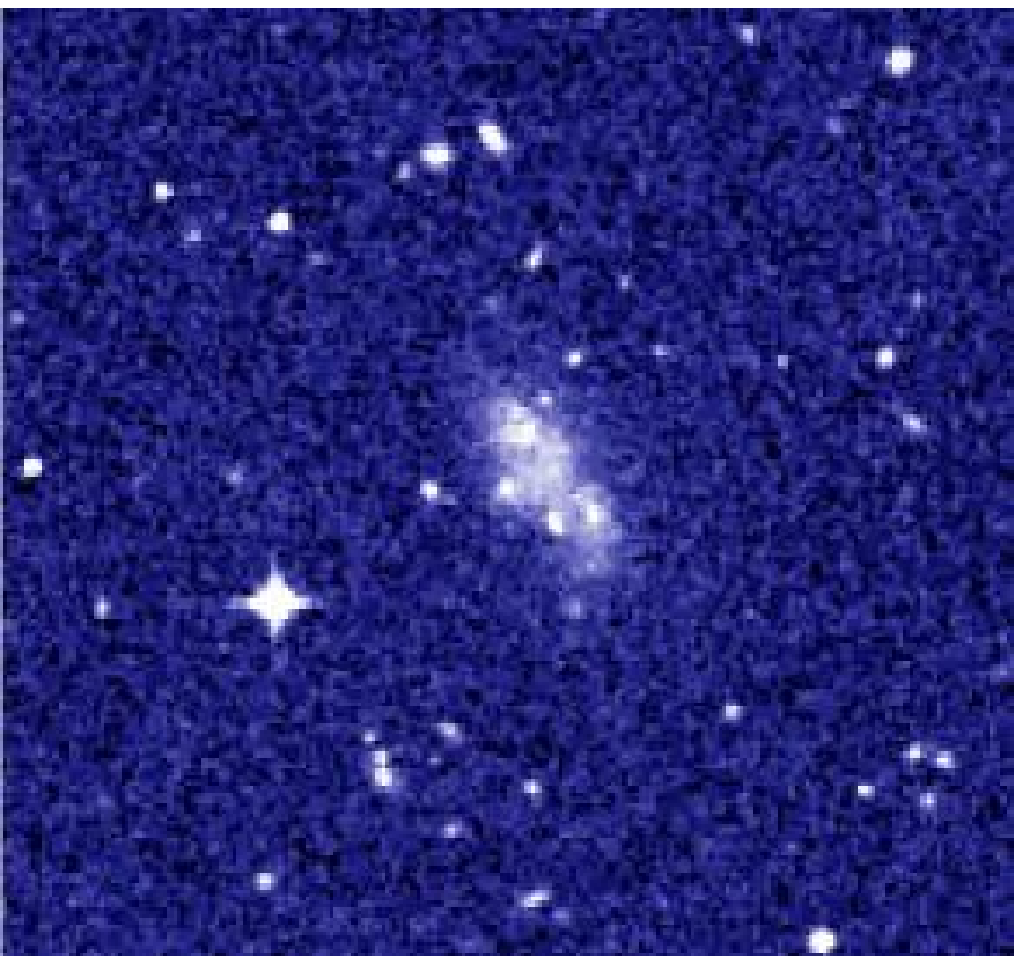}  & \includegraphics[width=0.1\textheight]{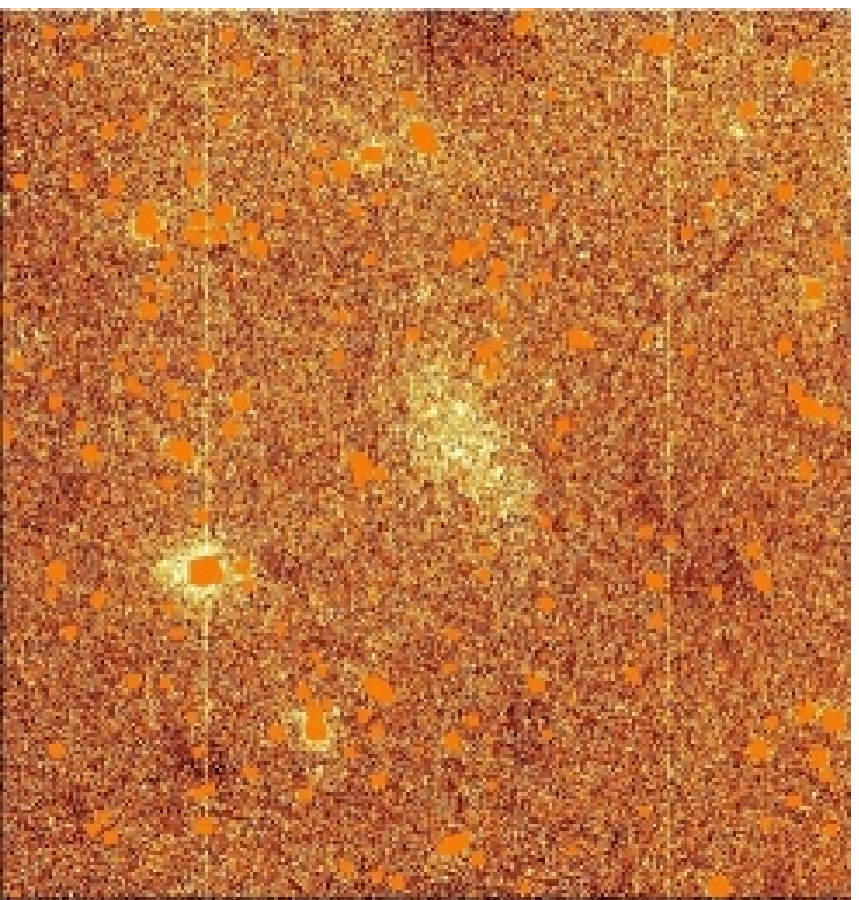}  & \includegraphics[width=0.1\textheight]{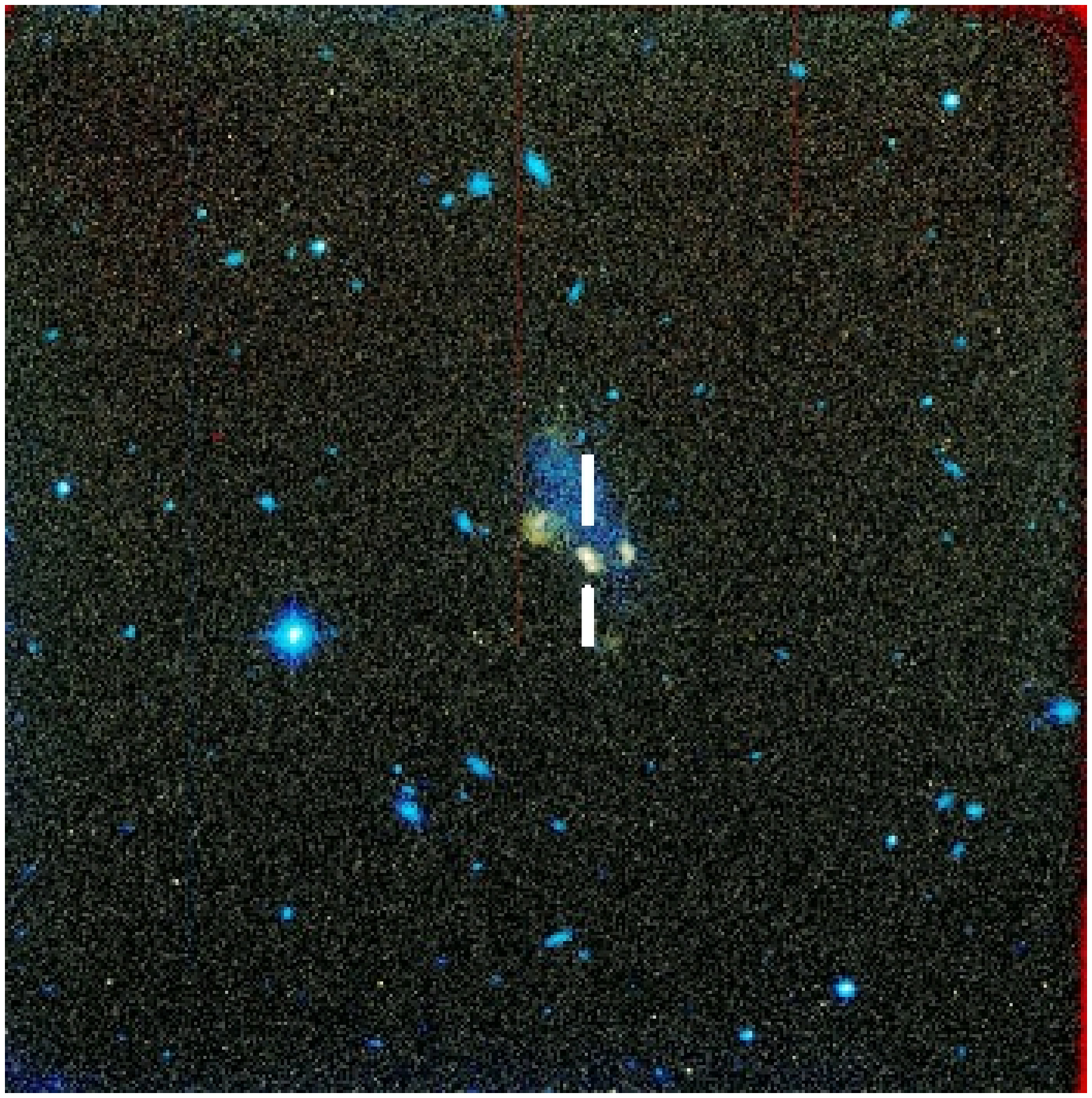}\tabularnewline
\hline 
\multicolumn{3}{|c|}{AM 106 382}\tabularnewline
\hline 
\includegraphics[width=0.1\textheight]{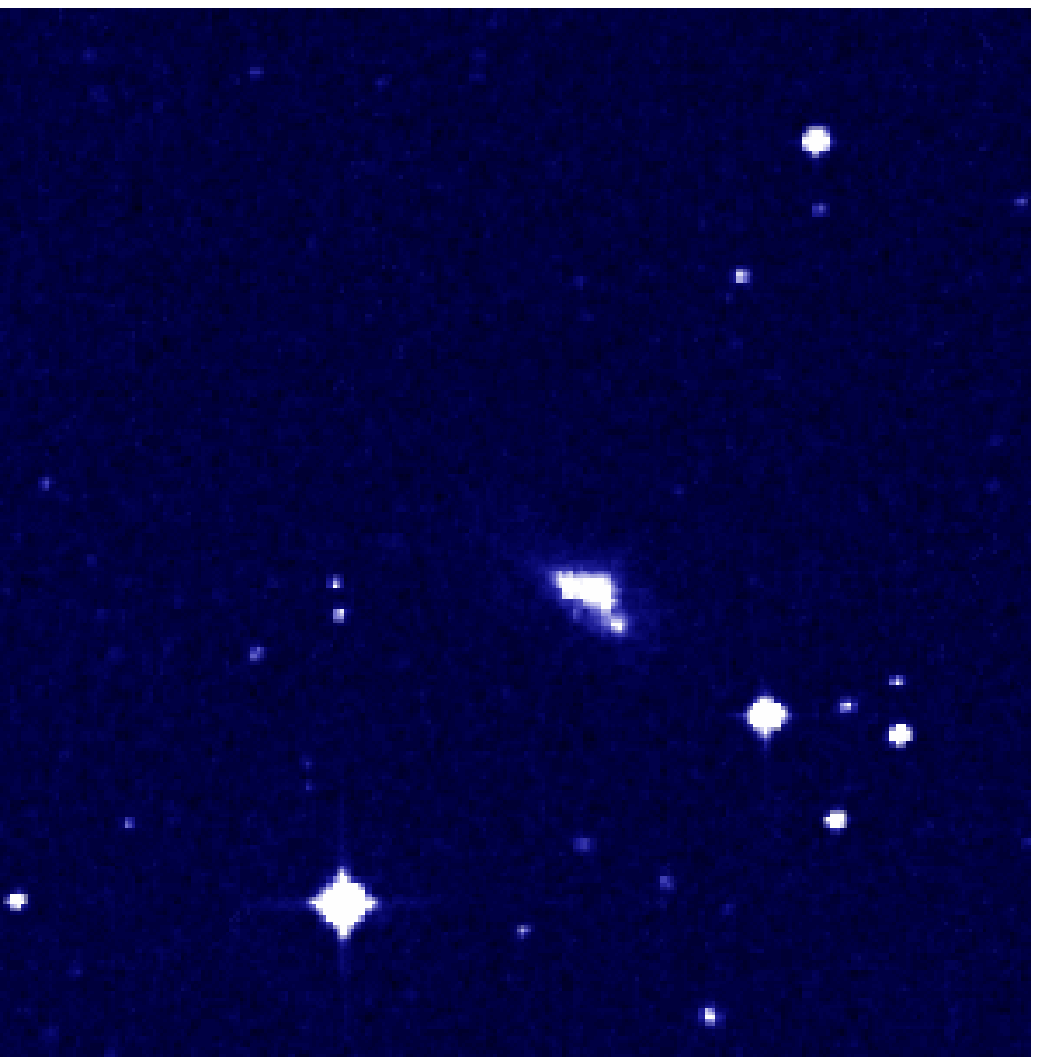}  & \includegraphics[width=0.1\textheight]{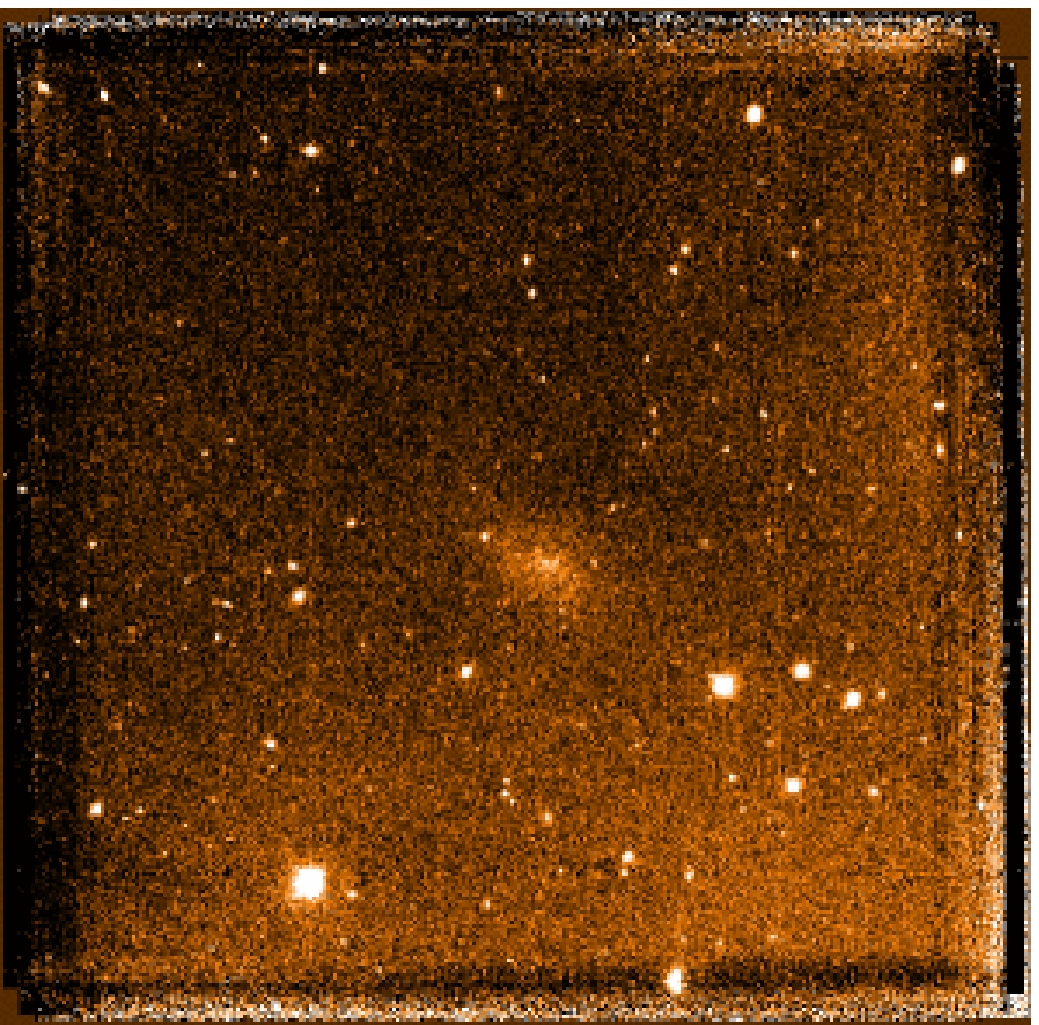}  & \includegraphics[width=0.1\textheight]{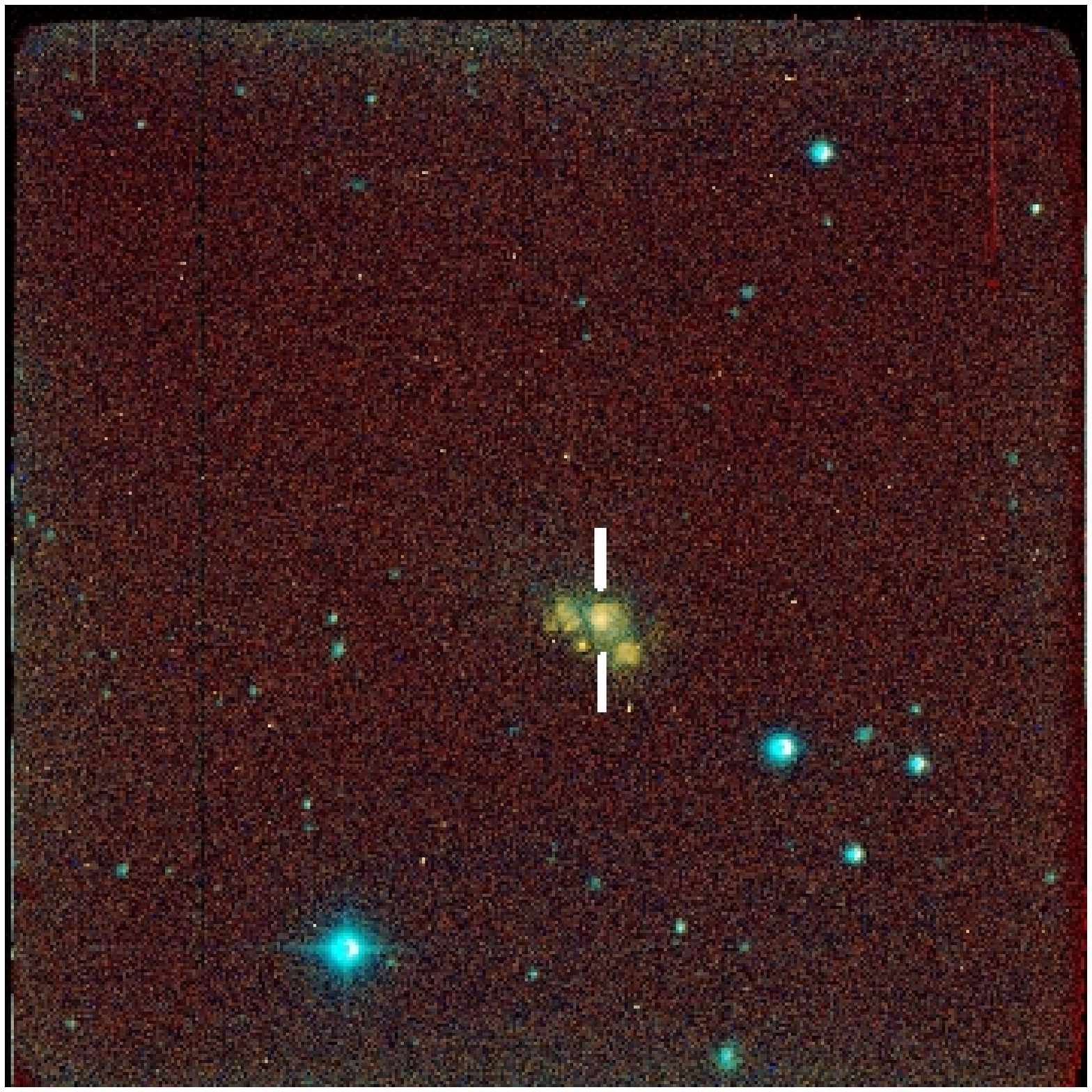}\tabularnewline
\hline 
\multicolumn{3}{|c|}{ESO 294-G010}\tabularnewline
\hline 
\includegraphics[width=0.1\textheight]{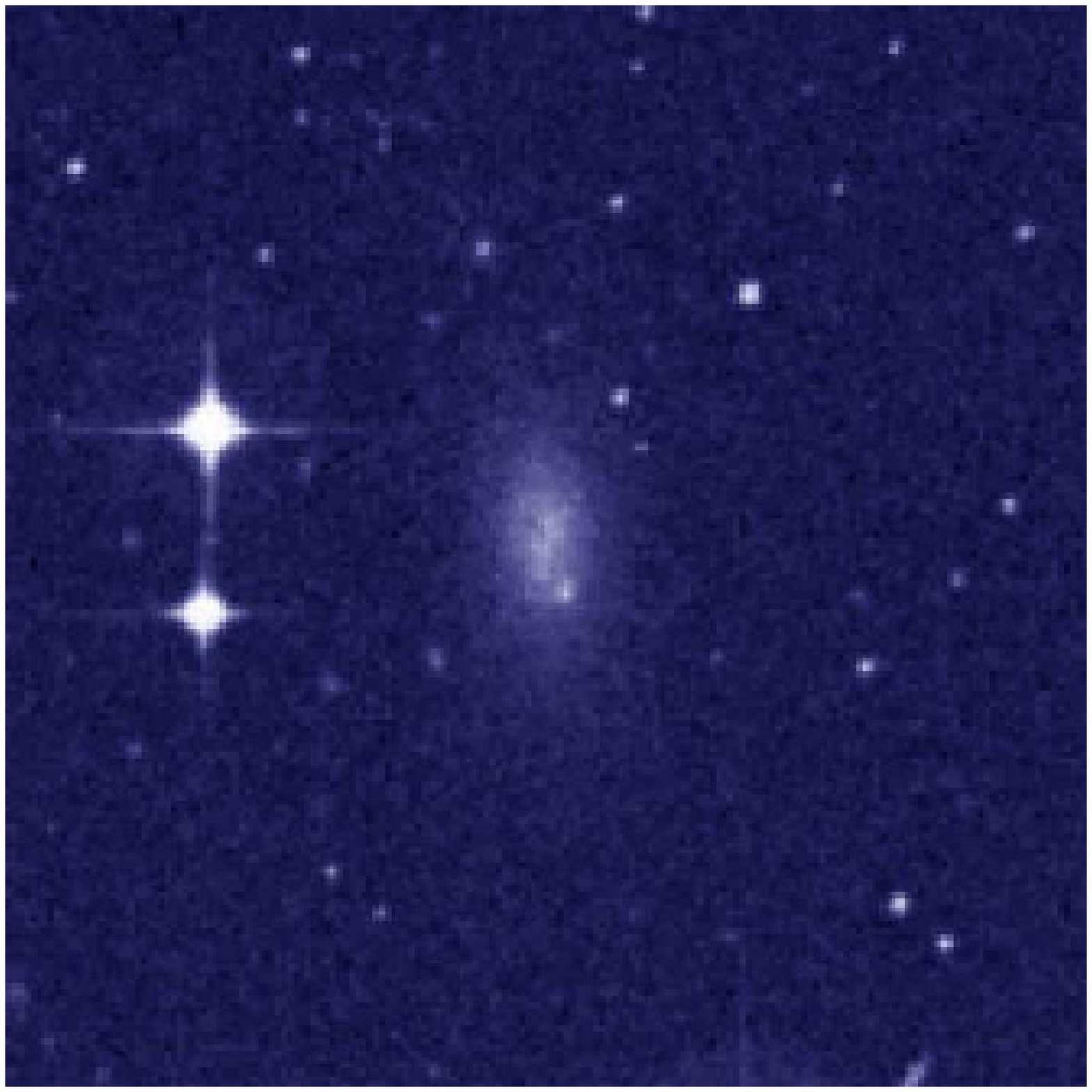}  &  & \includegraphics[width=0.1\textheight]{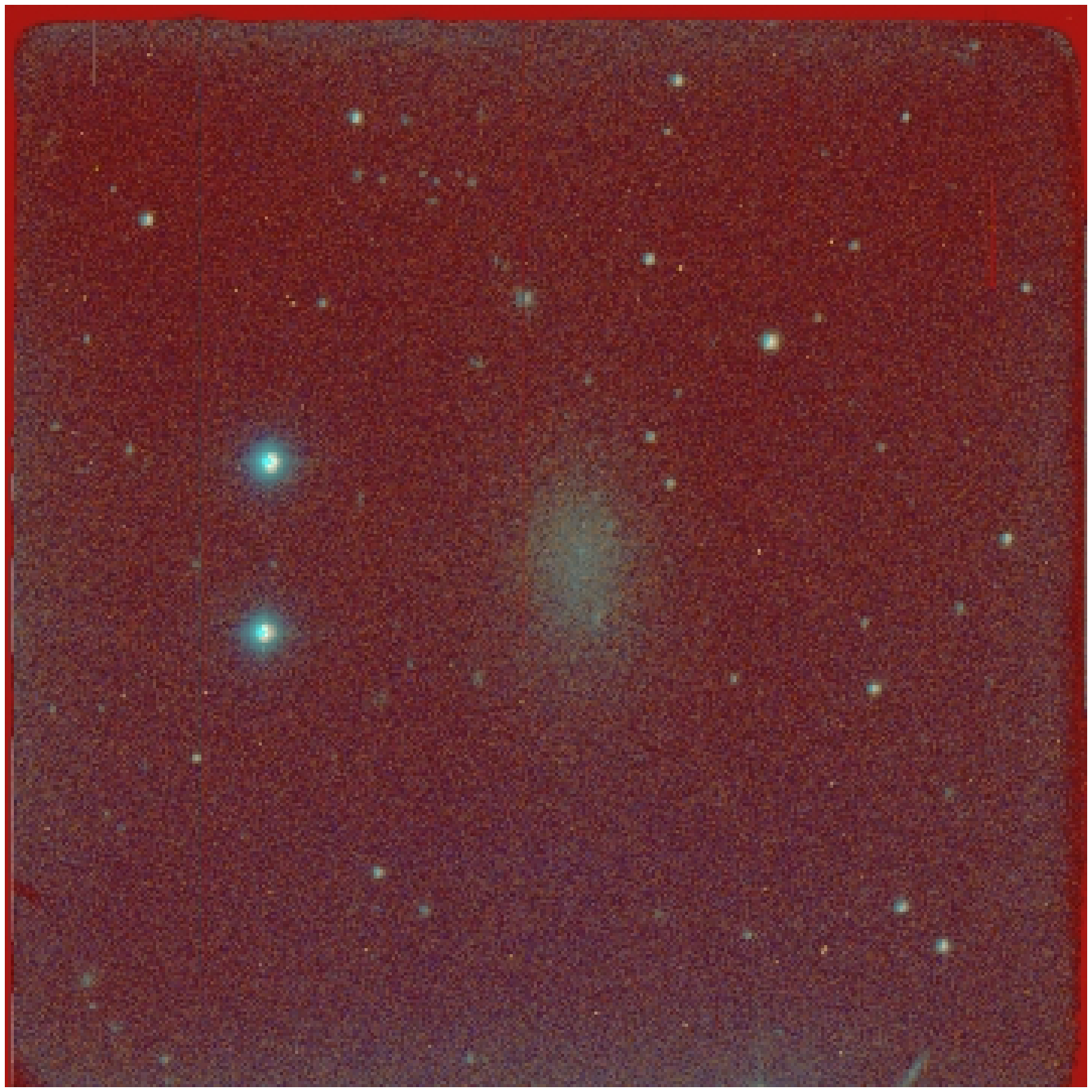}\tabularnewline
\hline
\end{tabular}
\par\end{centering}

{Figure 1.} (continued) \label{figcap:Same-as-Fig..} 
\end{figure}

\begin{table*}
\caption{Observations of Sculptor group targets. Additional data are the radial
velocity from Côté et al. (\cite{cote_etal97}), and the radial velocity
from the NASA extragalactic database. No emission was detected in
the H$\alpha$ pre-imaging for ESO~294-G010. \label{tabcap:Coordinates-and-exposure}}

\begin{centering}
\begin{tabular}{rrrcr|rrrrr}
\multicolumn{6}{r}{} &  &  &  & \tabularnewline
\hline
\hline 
\multicolumn{5}{c}{Target list} & \multicolumn{5}{c}{Journal of the observations}\tabularnewline
\multicolumn{1}{c}{Galaxy} & \multicolumn{1}{c}{R.A. } & \multicolumn{1}{c}{Dec. } & \multicolumn{2}{c}{$v_{{\rm r}}$ {[}km/sec] } & \multicolumn{2}{c}{EFOSC2} & \multicolumn{3}{c}{SOFI}\tabularnewline
 & \multicolumn{2}{c}{J2000} & \multicolumn{1}{c}{C97 } & \multicolumn{1}{c}{NED} & \multicolumn{1}{c}{Date} & t {[}s]  & Date  & \multicolumn{1}{c}{t {[}s]} & \multicolumn{1}{c}{airm.}\tabularnewline
\hline 
ESO 347-G017  & 23:26:56.0  & $-$37:20:49  & 659  & 692$\pm$ 4  & 13 Oct 2002  & $3\times1800$  & 15 Oct 2002  & 3420  & 1.10\tabularnewline
 &  &  &  &  & 14 Oct 2002  & $2\times1800$  & 16 Oct 2002  & 300  & 1.02\tabularnewline
UGCA 442  & 23:43:45.5  & $-$31:57:24  & 283  & 267$\pm$2  &  &  & 16 Oct 2002  & 3600  & 1.08\tabularnewline
ESO 348-G009  & 23:49:23.5  & $-$37:46:19  & 628  & 648$\pm$4  &  &  & 16 Oct 2002  & 3600  & 1.09\tabularnewline
NGC 59  & 00:15:25.1  & $-$21:26:40  & 357  & 382$\pm$60  & 13 Oct 2002  & $4\times1800$  & 15 Oct 2002  & 4080  & 1.05\tabularnewline
 &  &  &  &  &  &  & 16 Oct 2002  & 300  & 1.02\tabularnewline
ESO 294-G010  & 00:26:33.4  & $-$41:51:19  & 4450  & 117$\pm$5  & \ldots{}  & \ldots{}  &  &  & \tabularnewline
ESO 473-G024  & 00:31:23.1  & $-$22:46:02  & \ldots{}  & 541$\pm$1  & 19 Aug 2003  & $3\times1200$  & 10 Aug 2003  & 1560  & 1.10\tabularnewline
AM 0106-382  & 01:08:21.9  & $-$38:12:34  & 645  & 645$\pm$10  & 14 Oct 2002  & $4\times1800$  & 15 Oct 2002  & 3480  & 1.26\tabularnewline
 &  &  &  &  &  &  & 16 Oct 2002  & 1800  & 1.15\tabularnewline
NGC 625  & 01:35:04.2  & $-$41:26:15  & 415  & 396$\pm$1  & 14 Oct 2002  & $3\times900$  & 15 Oct 2002  & 1020  & 1.71\tabularnewline
 &  &  &  &  &  &  & 16 Oct 2002  & 900  & 1.78\tabularnewline
ESO 245-G005  & 01:45:03.7  & $-$43:35:53  & 389  & 391$\pm$2  & 13 Oct 2002  & $2\times1800$  & 16 Oct 2002  & 2460  & 1.32\tabularnewline
 &  &  &  &  & 14 Oct 2002  & $3\times1800$  &  &  & \tabularnewline
 &  &  &  &  & 14 Oct 2003  & $3\times1200$  &  &  & \tabularnewline
\hline
\end{tabular}
\par\end{centering}
\end{table*}

Sculptor group Irr galaxies were selected from the list of Côté et
al. (\cite{cote_etal97}; hereafter C97). The seven galaxies with
H$\alpha$ detection and radial velocities $v_{{\rm r}}<1000$~\kms,
were included in our target list shown in Table~\ref{tabcap:Coordinates-and-exposure}.
Note that the C97 catalog contains another set of galaxies with velocities
between $1000$ and $2000$~\kms, as well as sources with velocities
up to $\sim17000$~\kms. In addition to our main list, we included
ESO~294-G010, which has been classified as intermediate dS0/Im by
Jerjen et al. (\cite{jerjen_etal98}; hereafter J98) and, although
C97 measured an H$\alpha$ velocity $v_{{\rm r}}=4450$~\kms, a
much smaller velocity is quoted in NED. Moreover, J98 have detected
{[}\ion{O}{iii}] and H$\alpha$ emission, albeit very weak. Yet,
as no emission was detected during our H$\alpha$ pre-imaging, we
did not carry over further observations: it is possible that J98 detected
a PN with H$\alpha$ emission below our detection threshold. Bouchard
et al. (\cite{bouchard_etal05}) re-classified this galaxy as dSph/dIrr.

The main observational campaign was carried out in October 2002: NIR
imaging of the seven targets was collected, while bad weather undermined
the spectroscopic observing run and spectra could be collected for
five targets only. A few months after the completion of our observations,
Skillman et al. (\cite{skillman_sfr}; hereafter S03a) identified
other dwarfs showing H$\alpha$ emission despite a null detection
by C97, bringing the total number of actively star-forming dIrrs in
the Sculptor group up to twelve. In a companion paper, Skillman et
al. (\cite{skillman_scl_hii}; hereafter S03b) derived oxygen abundances
for five of these galaxies (reported in Table~\ref{tab:Data-for-the-LZ}),
two of them being in common with our original sample. So, to improve
the comparison with S03b, in August 2003 we complemented our data
set with NIR imaging and long-slit spectroscopy for ESO~473-G024.
Finally, ESO~245-G005 was re-observed in October 2003, in order to
measure an H\noun{~ii} region closer to the center of the galaxy.

In summary, with the observations presented here there are now eight
Sculptor group dwarfs with measured abundances, of which three are
in common between our study and S03b. NIR data are available for all
eight objects.

\subsubsection{M81 group}

\begin{figure}
\begin{centering}
\begin{tabular}{|c|c|}
\hline 
DSS  & $H$\tabularnewline
\hline
\hline 
\multicolumn{2}{|c|}{DDO 42}\tabularnewline
\hline 
\includegraphics[width=0.1\textheight]{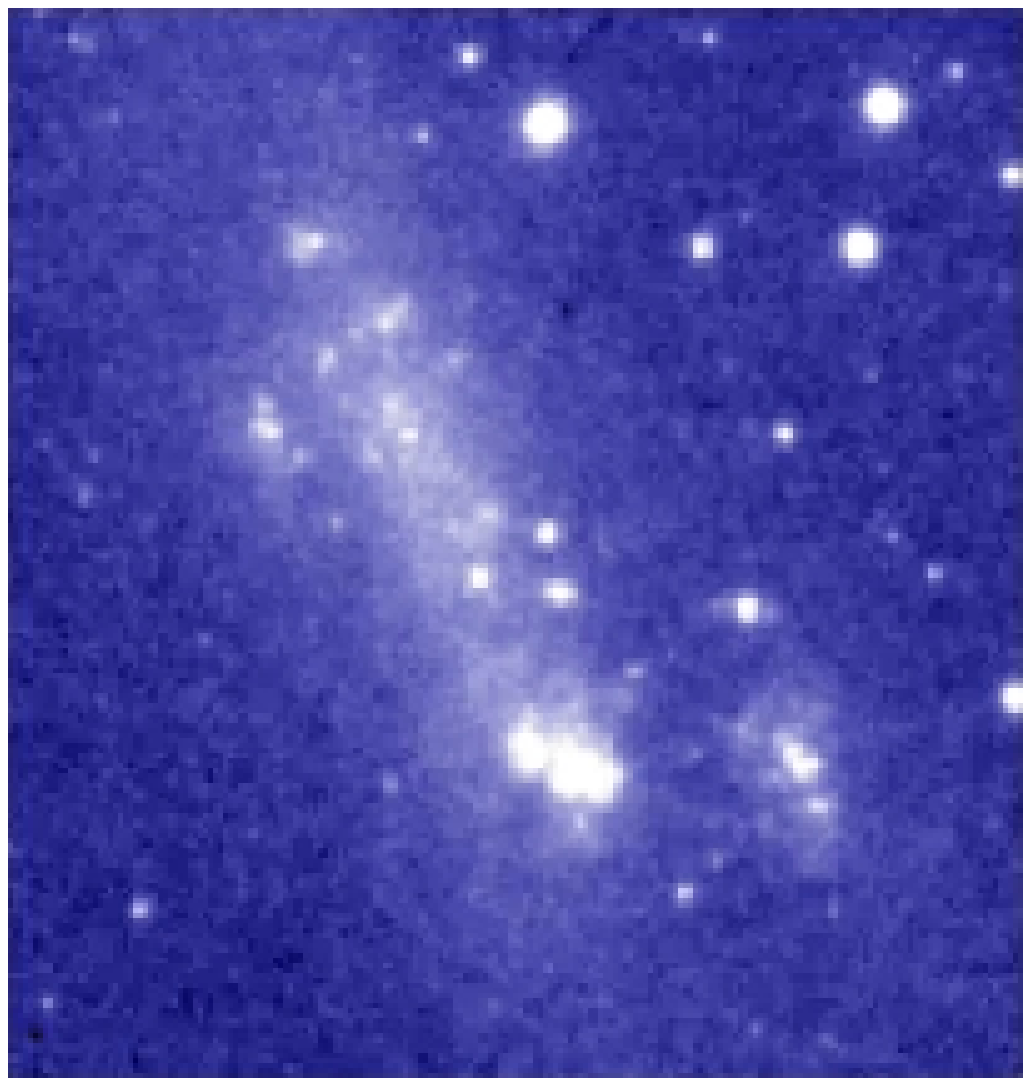}  & \includegraphics[width=0.1\textheight]{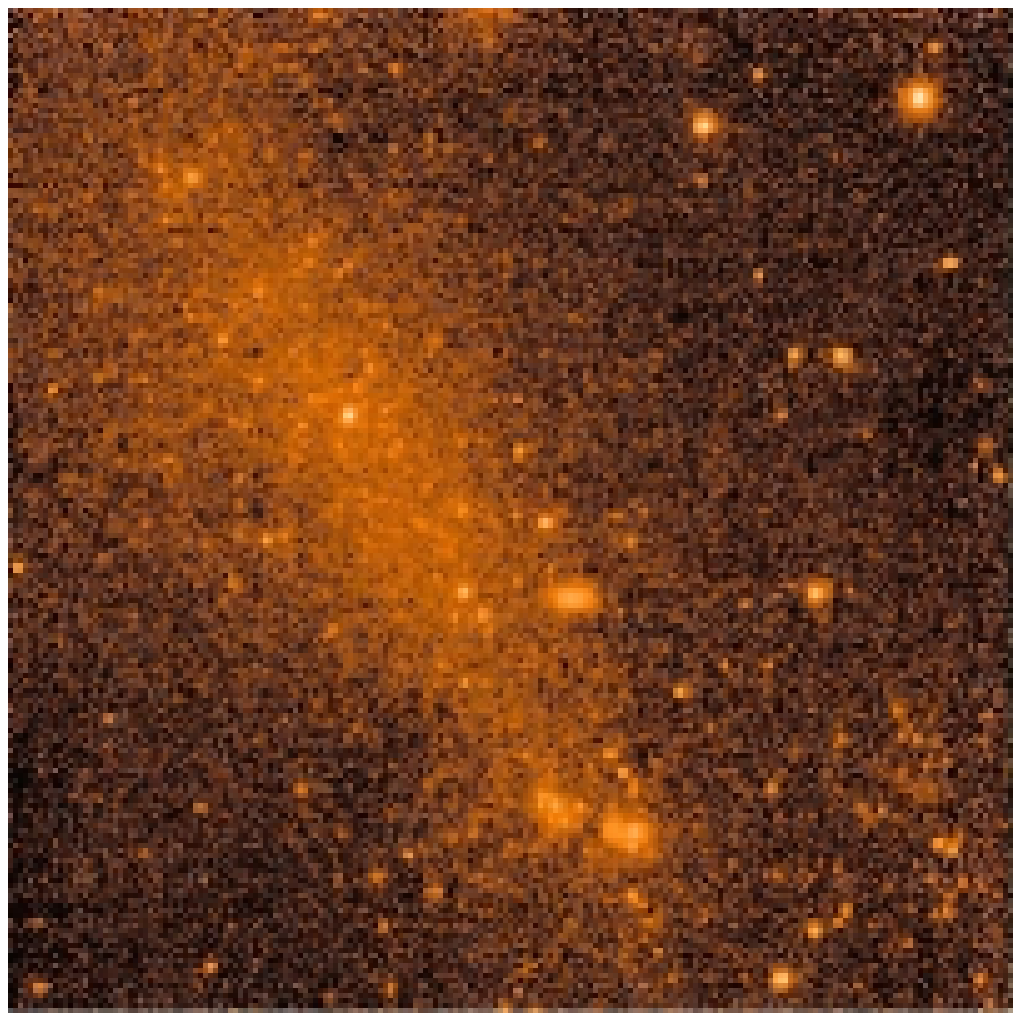}\tabularnewline
\hline 
\multicolumn{2}{|c|}{DDO 82}\tabularnewline
\hline 
\includegraphics[width=0.1\textheight]{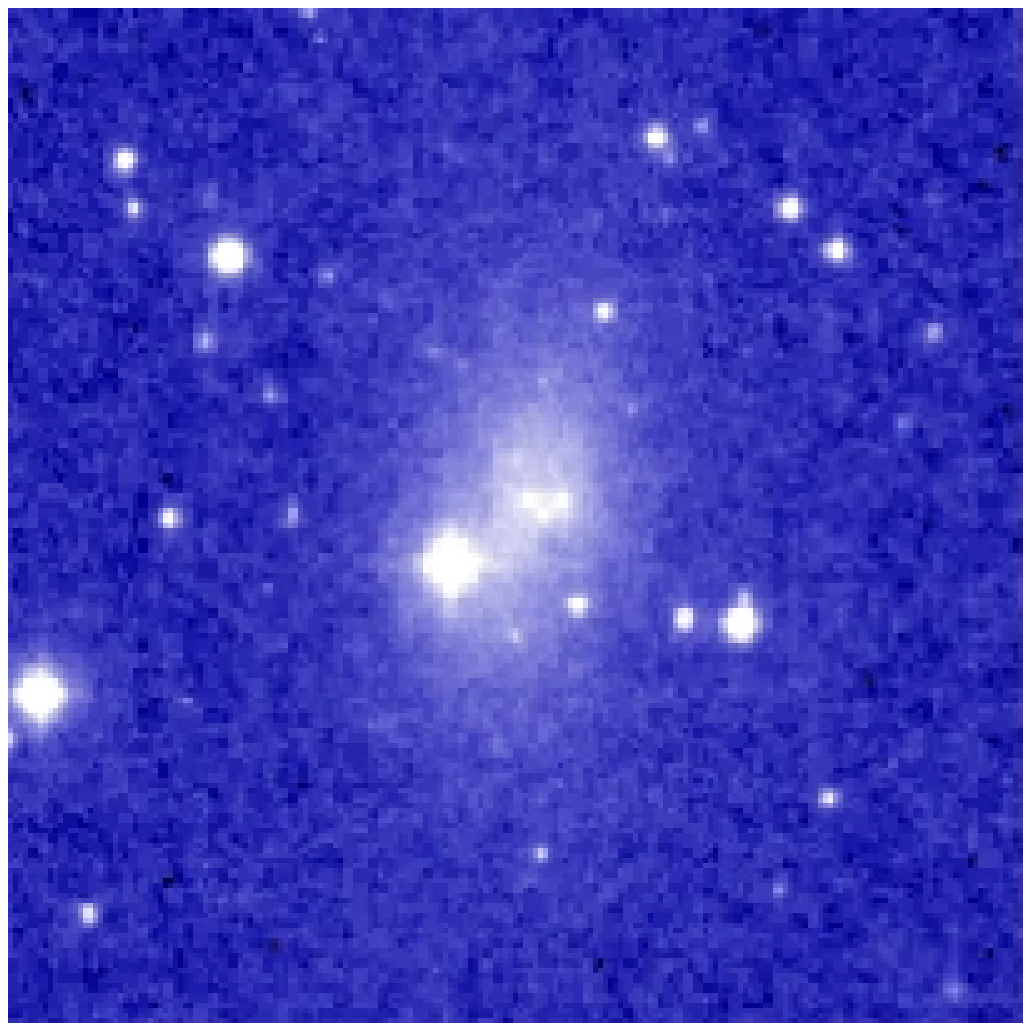}  & \includegraphics[width=0.1\textheight]{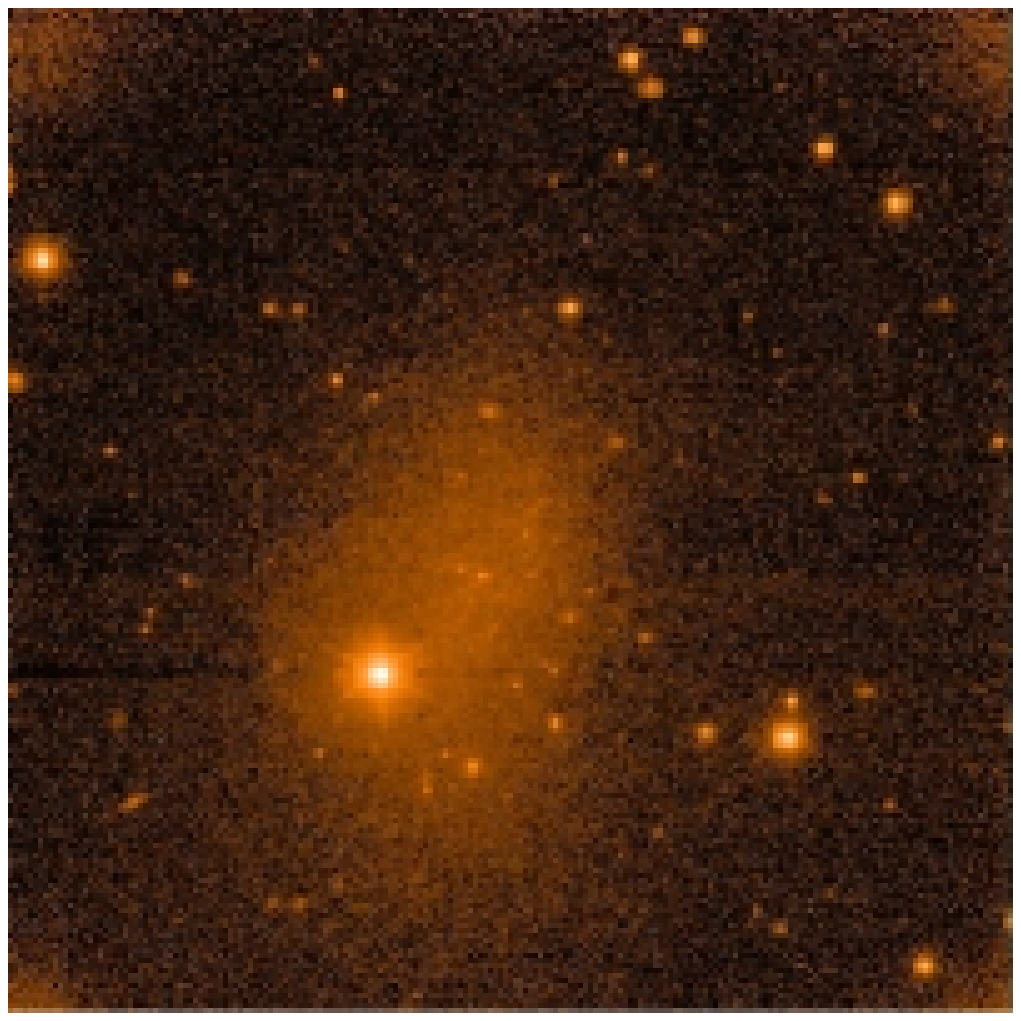}\tabularnewline
\hline 
\multicolumn{2}{|c|}{DDO 53}\tabularnewline
\hline 
\includegraphics[width=0.1\textheight]{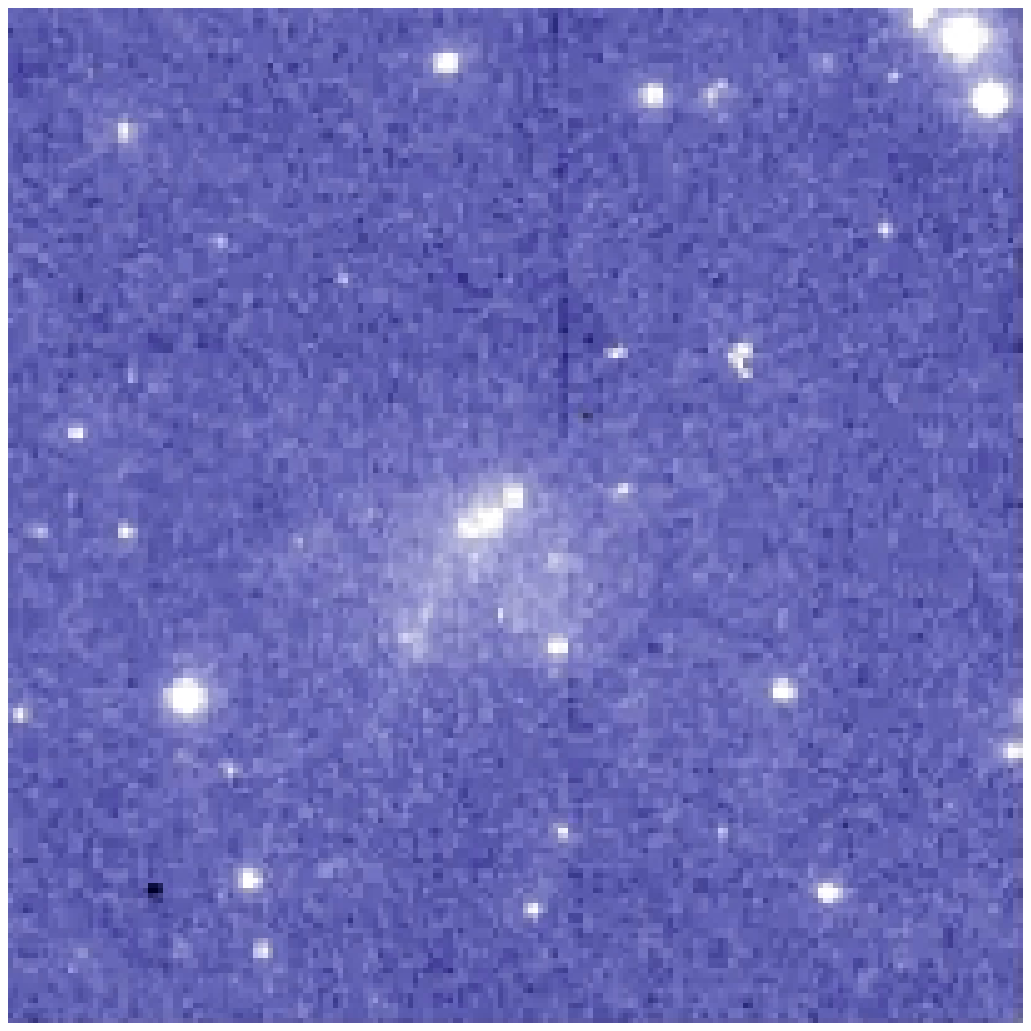}  & \includegraphics[width=0.1\textheight]{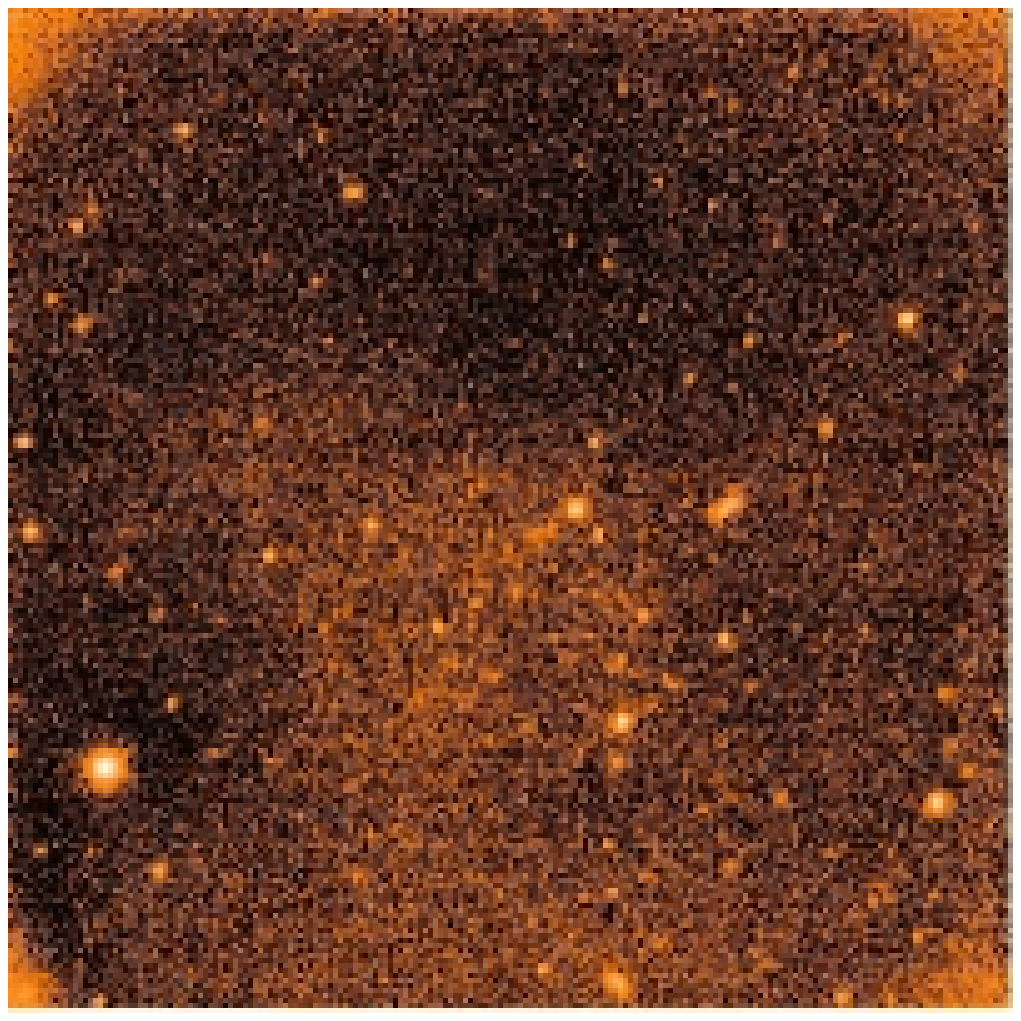}\tabularnewline
\hline 
\multicolumn{2}{|c|}{UGC 4483}\tabularnewline
\hline 
\includegraphics[width=0.1\textheight]{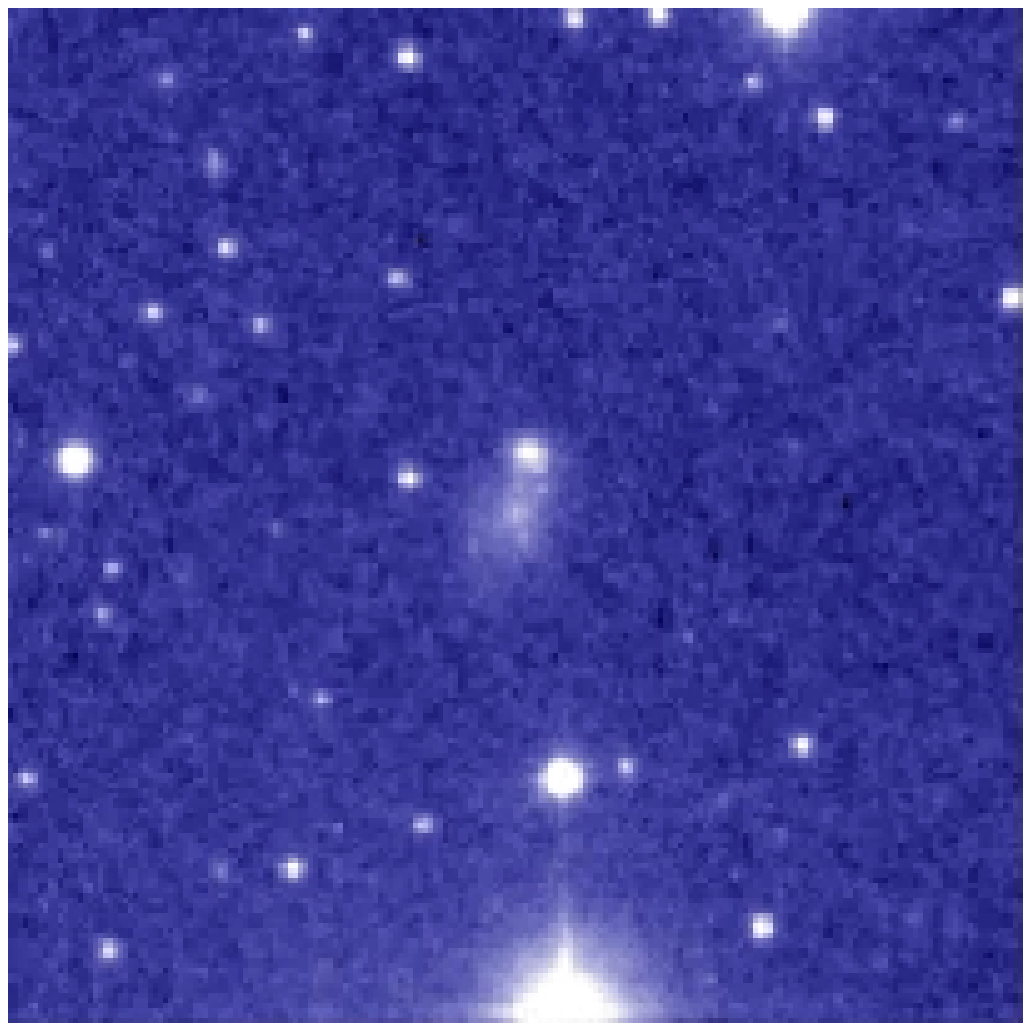}  & \includegraphics[width=0.1\textheight]{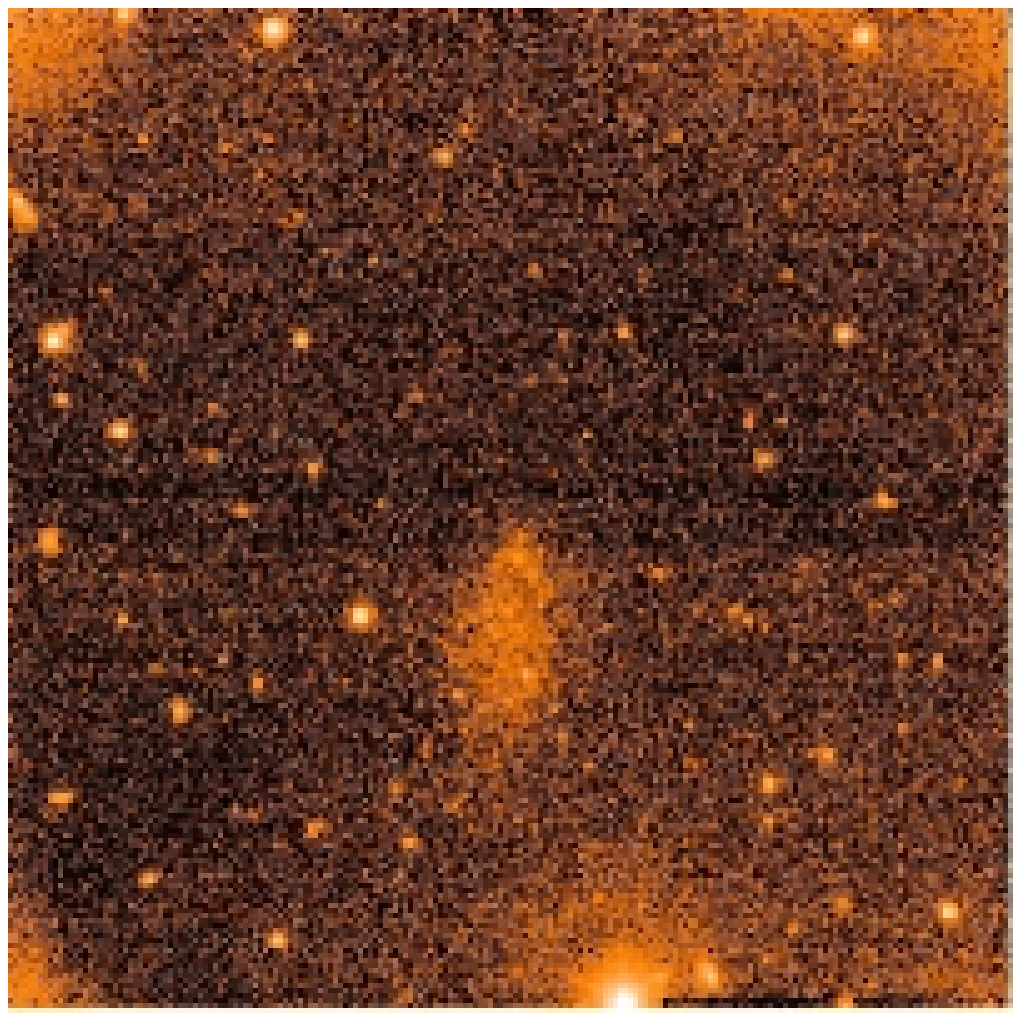}\tabularnewline
\hline
\end{tabular}
\par\end{centering}

\caption{The NIR, $H$-band, morphology of the M81 group dIrrs observed at
WHT with \noun{ingrid} (right column) is compared to their optical
appearance as seen in DSS images (left column). The galaxy luminosity
decreases from top to bottom panels. In all images, North is up and
East to the left, and the field of view is $\sim6\arcmin\times6\arcmin$
in the left column and $\sim5\arcmin\times5\arcmin$ in the right
column. \label{figcap:The-near-IR,-H}}

\end{figure}

\begin{table*}
\caption{Observations of M81 group targets. The group membership has been established
in K85, except for the candidate 5 which comes from Froebrich \& Meusinger
(\cite{fm00}). Additional data are the radial velocity from the NASA
extragalactic database. Exposure times for the \emph{Kast} spectrograph
observations are flagged with a code indicating the instrument setup:
B=blue arm with grism 830/3460; R=red arm with grating 300/4230; BR=both
arms with dichroic D46; F=red arm without dichroic; HR=red arm with
grating 1200/5000 centered around H$\alpha$. A `n.e.' in the INGRID
observation means that no emission was detected and so, no NIR imaging
had been performed. \label{tabcap:list-of-M81}}

\begin{centering}
\begin{tabular}{rccr|rlccc}
\multicolumn{9}{c}{}\tabularnewline
\hline
\hline 
\multicolumn{4}{c}{Target list} & \multicolumn{5}{c}{Journal of the observations}\tabularnewline
\multicolumn{1}{c}{Galaxy} & R.A.  & Dec.  & \multicolumn{1}{c}{$v_{{\rm r}}$ } & \multicolumn{2}{c}{\emph{Kast}} & \multicolumn{3}{c}{INGRID}\tabularnewline
\multicolumn{1}{c}{Main ID} & \multicolumn{2}{c}{J2000} & \multicolumn{1}{c||}{{[}km/sec]} &  & \multicolumn{1}{c}{t {[}s] } & \multicolumn{1}{c}{Date} & \multicolumn{1}{c}{t {[}s]} & \multicolumn{1}{c}{airm.}\tabularnewline
\hline 
DDO 42  & 07:28:53.7  & +69:12:54  & +80  & 17 Dec 2001  & {\tiny 600+2$\times$1200 RB}  & 15 Jan 2003  & 3600  & 1.36\tabularnewline
 &  &  &  & 17 Dec 2001  & {\tiny 1800 B + 1800 HR}  & 16 Jan 2003  & 300  & 1.46\tabularnewline
 &  &  &  & 17 Dec 2001  & {\tiny 600 F + 600 HR}  &  &  & \tabularnewline
 &  &  &  & 14 Jan 2002  & {\tiny 600+2$\times$1200 RB}  &  &  & \tabularnewline
 &  &  &  & 14 Jan 2002  & {\tiny 1800 B + 1800 HR}  &  &  & \tabularnewline
 &  &  &  & 15 Jan 2002  & {\tiny 1800 B + 1800 HR}  &  &  & \tabularnewline
KDG 52  & 08:23:56.0  & +71:01:45  & +113  & 17 Dec 2001  & {\tiny 2 $\times$300 RB}  & \multicolumn{3}{c}{n.e.}\tabularnewline
DDO 53  & 08:34:07.2  & +66:10:54  & +20  & 17 Dec 2001  & {\tiny 3 $\times$1800 RB}  & 16 Jan 2003  & 2160  & 1.38\tabularnewline
 &  &  &  & 17 Dec 2001  & {\tiny 1800 HR + 1800 B}  &  &  & \tabularnewline
UGC 4483  & 08:37:03.0  & +69:46:50  & +156  & 14 Jan 2002  & {\tiny 2 $\times$1200 RB}  & 15 Jan 2003  & 960  & 1.29\tabularnewline
 &  &  &  & 14 Jan 2002  & {\tiny 1800 B + 1800 HR}  & 16 Jan 2003  & 3210  & 1.34\tabularnewline
KDG 54  & 09:22:25.2  & +75:45:57  & +659  & 14 Jan 2002  & {\tiny 2 $\times$(1200+1800) RB}  & \multicolumn{3}{c}{higher redshift}\tabularnewline
candidate 5  & 09:39:02.0  & +69:25:01  & \ldots{}  &  &  & \multicolumn{3}{c}{n.e.}\tabularnewline
BK 3N  & 09:53:48.5  & +68:58:08  & $-$40  &  &  & \multicolumn{3}{c}{n.e.}\tabularnewline
DDO 66  & 09:57:30.1  & +69:02:52  & +46  &  &  & \multicolumn{3}{c}{n.e.}\tabularnewline
DDO 82  & 10:30:34.8  & +70:37:14  & +56  & 15 Jan 2002  & {\tiny 300 + 2$\times$600 RB}  & 16 Jan 2003  & 3600  & 1.41\tabularnewline
 &  &  &  & 15 Jan 2002  & {\tiny 3$\times$1200 RB}  &  &  & \tabularnewline
DDO~165  & 13:06:24.8  & +67:42:25  & +31  &  &  & \multicolumn{3}{c}{n.e.}\tabularnewline
\hline
\end{tabular}
\par\end{centering}
\end{table*}

A first list of M81 group members has been compiled by Karachentseva
et al. (\cite{kkb85}; hereafter K85), by inspecting photographic
plates taken at the 6-m telescope. It includes $14$ probable and
$11$ questionable members (on the basis of radial velocity and visual
appearance), all of them of dIrr type. Later, Karachentseva \& Karachentsev
(\cite{kk98}) added six new possible members in the dE class, by
searching POSS-II and ESO/SERC films. A search based on CCD imaging
was carried out by Caldwell et al. (\cite{cads98}), who, however,
did not publish their catalog. Finally, Froebrich \& Meusinger (\cite{fm00};
hereafter FM00) discovered six more candidates in a survey that employed
digitally stacked Schmidt plates. The candidates are equally divided
into dE and dIrr types. A new version of the catalog of M81 member
candidates was recently published by Karachentsev et al. (\cite{k_etal02};
hereafter K02), showing a few additions and a few deletions when compared
to the previous catalog.

Ten M81 group dwarfs have been observed from 2001 to 2002 (Table~\ref{tabcap:list-of-M81}).
Our first run was with the \emph{Kast} spectrograph at Lick: the instrument
offers limited imaging capabilities, so we selected the target H\noun{~ii}
regions with the following strategy. The slit was first placed on
the brightest nebulosity within a target dwarf and a short-exposure
spectrum was taken. If it showed a conspicuous H$\alpha$ line, then
the integration was continued, otherwise we moved to another candidate
region and tried again. If no emission was detected in two to three
regions (depending on the galaxy luminosity), we moved to the next
dwarf target. With this procedure, emission was detected in only half
of the galaxies, namely DDO~42, DDO~53, UGC~4483, DDO~82, and
KDG~54 (which turned out to be a background, higher redshift object).
No emission was detected in KDG~52, BK~3N, DDO~66, DDO~165, nor
in the candidate~5 of Froebrich \& Meusinger (\cite{fm00}). For
objects which did show emission lines, the spectroscopic observations
were followed-up by NIR imaging performed with the INGRID camera at
the ING/WHT telescope on La Palma.

After the completion of the spectroscopic observations, three H$\alpha$
surveys including M81 group galaxies were published by Hunter \& Elmegreen
(\cite{hunter_elmegreen04}), Gil de Paz et al. (\cite{gildepaz_etal03}),
and James et al. (\cite{james_etal04}). These surveys have detected
emission-line regions in, among others, DDO~42, DDO~53, UGC~4483,
DDO~66, DDO~82, and DDO~165. So they confirm our detections, but
they also unveil emission-line regions in DDO~66 and DDO~165, for
which our trial-and-error technique gave a null result.

\subsection{Optical spectrophotometry}

\subsubsection{Sculptor group}


%
\begin{figure}
\begin{centering}
\includegraphics[width=1\columnwidth]{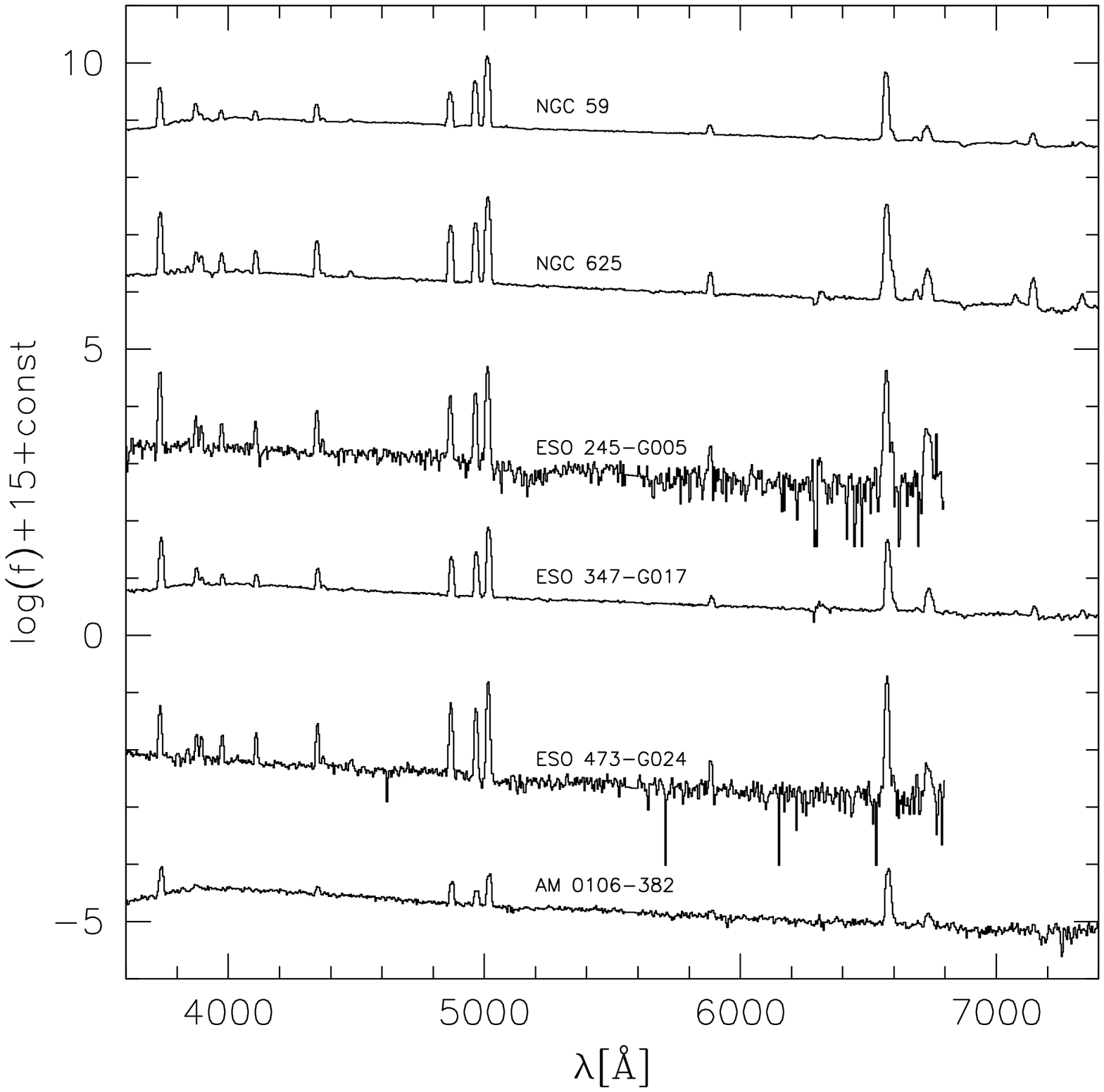} 
\par\end{centering}

\caption{For each dIrr in the Sculptor group sample we show the spectrum of
one of the \ion{H}{ii} regions that enter the \lz\ relation.
The spectra are masked near the strong night sky line {[}\ion{O}{i}]$\lambda5579$,
which leads to poor sky subtraction. The reddest, noisiest, part of
the spectrum is omitted for two objects. For display purposes, each
flux has been multiplied by $10^{15}$ and a constant has been added.
\label{fig:For-each-galaxy}}

\end{figure}

\begin{table*}
\caption{Reddening-corrected fluxes for all \ion{H}{ii} regions in the
Sculptor group dwarfs. \label{tabcap:Reddening-corrected-fluxes-for}}

\begin{centering}
\begin{sideways}
\begin{tabular}{rccccccccc}
 &  &  &  &  &  &  &  &  & \tabularnewline
\hline
\hline 
 & AM106-382  & ESO~245-G005  & ESO 245-G005  & ESO 347-G017  & ESO~473-G024  & NGC 59  & NGC 59  & NGC 625  & NGC 625\tabularnewline
Ion~~ $\lambda$ {[}\AA]  &  & external  & central  &  &  & \#1  & \#2  & \#1  & \#2\tabularnewline
\hline 
{[}\ion{O}{ii}] 3726.0  & 2.007 $\pm$ 0.041  & 1.508 $\pm$ 0.018  & 2.767 $\pm$ 0.087  & 2.191 $\pm$ 0.016  & 0.942 $\pm$ 0.023  & 1.080 $\pm$ 0.029  & 2.716 $\pm$ 0.092  & 1.454 $\pm$ 0.019  & 1.516 $\pm$ 0.009 \tabularnewline
H11 3770.6  & \ldots{}  & \ldots{}  & \ldots{}  & \ldots{}  & \ldots{}  & \ldots{}  & \ldots{}  & 0.049 $\pm$ 0.016  & 0.052 $\pm$ 0.008 \tabularnewline
H10 3797.9  & \ldots{}  & \ldots{}  & \ldots{}  & \ldots{}  & \ldots{}  & \ldots{}  & 0.622 $\pm$ 0.086  & 0.064 $\pm$ 0.016  & 0.063 $\pm$ 0.008 \tabularnewline
\ion{He}{i} 3819.6  & 0.143 $\pm$ 0.037  & \ldots{}  & \ldots{}  & \ldots{}  & \ldots{}  & \ldots{}  & 0.424 $\pm$ 0.084  & \ldots{}  & \ldots{} \tabularnewline
H9 3835.4  & \ldots{}  & \ldots{}  & \ldots{}  & \ldots{}  & \ldots{}  & \ldots{}  & \ldots{}  & 0.085 $\pm$ 0.016  & 0.077 $\pm$ 0.008 \tabularnewline
{[}\ion{Ne}{iii}] 3868.8  & 0.513 $\pm$ 0.037  & 0.146 $\pm$ 0.016  & 0.312 $\pm$ 0.071  & 0.348 $\pm$ 0.015  & 0.190 $\pm$ 0.021  & 0.374 $\pm$ 0.026  & 0.932 $\pm$ 0.082  & 0.274 $\pm$ 0.015  & 0.207 $\pm$ 0.008 \tabularnewline
{[}\ion{Ne}{iii}]+H8 3879.0  & \ldots{}  & \ldots{}  & \ldots{}  & \ldots{}  & \ldots{}  & \ldots{}  & \ldots{}  & \ldots{}  & \ldots{} \tabularnewline
H8 3889.1  & \ldots{}  & 0.151 $\pm$ 0.016  & 0.160 $\pm$ 0.070  & 0.117 $\pm$ 0.014  & 0.155 $\pm$ 0.020  & 0.217 $\pm$ 0.026  & \ldots{}  & 0.247 $\pm$ 0.015  & 0.186 $\pm$ 0.008 \tabularnewline
H$\epsilon$ 3970.1  & \ldots{}  & 0.138 $\pm$ 0.014  & 0.209 $\pm$ 0.066  & 0.163 $\pm$ 0.013  & 0.209 $\pm$ 0.018  & 0.250 $\pm$ 0.023  & \ldots{}  & 0.258 $\pm$ 0.015  & 0.208 $\pm$ 0.007 \tabularnewline
\ion{He}{i} 4026.2  & \ldots{}  & \ldots{}  & \ldots{}  & \ldots{}  & \ldots{}  & \ldots{}  & \ldots{}  & 0.024 $\pm$ 0.013  & \ldots{} \tabularnewline
{[}\ion{S}{ii}]blend 4072.0  & \ldots{}  & \ldots{}  & \ldots{}  & \ldots{}  & \ldots{}  & \ldots{}  & \ldots{}  & \ldots{}  & \ldots{} \tabularnewline
H$\delta$ 4101.7  & \ldots{}  & 0.212 $\pm$ 0.013  & 0.183 $\pm$ 0.057  & 0.172 $\pm$ 0.011  & 0.235 $\pm$ 0.016  & 0.250 $\pm$ 0.020  & \ldots{}  & 0.265 $\pm$ 0.013  & 0.248 $\pm$ 0.006 \tabularnewline
H$\gamma$ 4340.5  & 0.518 $\pm$ 0.032  & 0.442 $\pm$ 0.010  & 0.483 $\pm$ 0.042  & 0.438 $\pm$ 0.010  & 0.401 $\pm$ 0.012  & 0.466 $\pm$ 0.020  & 0.481 $\pm$ 0.047  & 0.465 $\pm$ 0.010  & 0.463 $\pm$ 0.005 \tabularnewline
{[}\ion{O}{iii}] 4363.2  & \ldots{}  & 0.032 $\pm$ 0.010  & 0.055 $\pm$ 0.037  & 0.049 $\pm$ 0.009  & 0.045 $\pm$ 0.011  & 0.038 $\pm$ 0.020  & 0.063 $\pm$ 0.045  & 0.028 $\pm$ 0.010  & 0.017 $\pm$ 0.004 \tabularnewline
\ion{He}{i} 4471.5  & \ldots{}  & \ldots{}  & \ldots{}  & \ldots{}  & 0.027 $\pm$ 0.008  & 0.033 $\pm$ 0.013  & \ldots{}  & 0.035 $\pm$ 0.008  & 0.046 $\pm$ 0.004 \tabularnewline
H$\beta$ 4861.3  & 1.000 $\pm$ 0.024  & 1.000 $\pm$ 0.008  & 1.000 $\pm$ 0.034  & 1.000 $\pm$ 0.009  & 1.000 $\pm$ 0.007  & 1.000 $\pm$ 0.014  & 1.000 $\pm$ 0.042  & 1.000 $\pm$ 0.009  & 1.000 $\pm$ 0.005 \tabularnewline
{[}\ion{O}{iii}] 4958.9  & 0.616 $\pm$ 0.020  & 0.588 $\pm$ 0.007  & 1.072 $\pm$ 0.036  & 1.267 $\pm$ 0.009  & 0.786 $\pm$ 0.006  & 1.629 $\pm$ 0.016  & 1.337 $\pm$ 0.052  & 1.421 $\pm$ 0.010  & 1.089 $\pm$ 0.005 \tabularnewline
{[}\ion{O}{iii}] 5006.8  & 1.782 $\pm$ 0.024  & 1.797 $\pm$ 0.010  & 3.223 $\pm$ 0.058  & 3.849 $\pm$ 0.013  & 2.473 $\pm$ 0.011  & 4.883 $\pm$ 0.023  & 2.930 $\pm$ 0.057  & 4.268 $\pm$ 0.017  & 3.304 $\pm$ 0.008 \tabularnewline
{[}\ion{Fe}{iii}] 5412.0  & \ldots{}  & 0.034 $\pm$ 0.008  & \ldots{}  & \ldots{}  & \ldots{}  & \ldots{}  & \ldots{}  & \ldots{}  & \ldots{} \tabularnewline
\ion{He}{i} 5875.7  & 0.159 $\pm$ 0.030  & 0.082 $\pm$ 0.012  & 0.104 $\pm$ 0.045  & 0.095 $\pm$ 0.011  & 0.062 $\pm$ 0.012  & 0.106 $\pm$ 0.019  & 0.152 $\pm$ 0.056  & 0.106 $\pm$ 0.011  & 0.093 $\pm$ 0.006 \tabularnewline
{[}\ion{S}{iii}] 6312.1  & \ldots{}  & 0.047 $\pm$ 0.014  & \ldots{}  & \ldots{}  & \ldots{}  & 0.039 $\pm$ 0.023  & \ldots{}  & 0.049 $\pm$ 0.012  & \ldots{} \tabularnewline
{[}\ion{O}{i}] 6363.8  & \ldots{}  & \ldots{}  & \ldots{}  & \ldots{}  & \ldots{}  & \ldots{}  & \ldots{}  & 0.006 $\pm$ 0.013  & \ldots{} \tabularnewline
H$\alpha$ 6562.8  & 2.850 $\pm$ 0.043  & 2.594 $\pm$ 0.024  & 2.849 $\pm$ 0.103  & 2.592 $\pm$ 0.020  & 2.849 $\pm$ 0.037  & 2.850 $\pm$ 0.032  & 2.850 $\pm$ 0.090  & 2.850 $\pm$ 0.020  & 2.850 $\pm$ 0.011 \tabularnewline
{[}\ion{N}{ii}] 6583.4  & \ldots{}  & \ldots{}  & 0.069 $\pm$ 0.103  & 0.064 $\pm$ 0.020  & 0.029 $\pm$ 0.037  & 0.075 $\pm$ 0.032  & \ldots{}  & \ldots{}  & \ldots{} \tabularnewline
\ion{He}{i} 6678.1  & \ldots{}  & \ldots{}  & \ldots{}  & 0.016 $\pm$ 0.016  & 0.033 $\pm$ 0.017  & 0.023 $\pm$ 0.025  & 0.031 $\pm$ 0.064  & 0.029 $\pm$ 0.015  & 0.030 $\pm$ 0.009 \tabularnewline
{[}\ion{S}{ii}] 6716.5  & 0.125 $\pm$ 0.040  & 0.134 $\pm$ 0.018  & 0.265 $\pm$ 0.068  & 0.208 $\pm$ 0.016  & 0.089 $\pm$ 0.019  & 0.129 $\pm$ 0.025  & 0.282 $\pm$ 0.064  & 0.130 $\pm$ 0.015  & 0.139 $\pm$ 0.009 \tabularnewline
{[}\ion{S}{ii}] 6730.8  & 0.045 $\pm$ 0.040  & 0.114 $\pm$ 0.018  & 0.164 $\pm$ 0.068  & 0.144 $\pm$ 0.017  & 0.058 $\pm$ 0.019  & 0.105 $\pm$ 0.025  & 0.318 $\pm$ 0.065  & 0.123 $\pm$ 0.015  & 0.097 $\pm$ 0.009 \tabularnewline
\ion{He}{i} 7065.3  & \ldots{}  & \ldots{}  & \ldots{}  & 0.021 $\pm$ 0.018  & \ldots{}  & 0.027 $\pm$ 0.028  & \ldots{}  & 0.027 $\pm$ 0.017  & 0.026 $\pm$ 0.011 \tabularnewline
{[}\ion{Ar}{iii}] 7135.8  & \ldots{}  & \ldots{}  & \ldots{}  & 0.070 $\pm$ 0.019  & \ldots{}  & 0.105 $\pm$ 0.029  & 0.137 $\pm$ 0.075  & 0.110 $\pm$ 0.018  & 0.098 $\pm$ 0.011 \tabularnewline
{[}\ion{O}{ii}]blend 7325.0  & \ldots{}  & \ldots{}  & \ldots{}  & \ldots{}  & \ldots{}  & \ldots{}  & \ldots{}  & \ldots{}  & \ldots{} \tabularnewline
C H$\alpha$/H$\beta$$^{a}$  & 0.002 $\pm$ 0.042  & 0.000 $\pm$ 0.019  & 0.192 $\pm$ 0.076  & 0.192 $\pm$ 0.076  & 0.191 $\pm$ 0.022  & 0.017 $\pm$ 0.028  & 0.302 $\pm$ 0.079  & 0.042 $\pm$ 0.018  & 0.004 $\pm$ 0.010 \tabularnewline
\hline
\end{tabular}
\end{sideways}
\par\end{centering}

Notes: {($^{a}$) The reddening constant is computed by the expression:
$I_{H\alpha}/I_{H\beta}=I_{H\alpha0}/I_{H\beta0}10^{-C[f(H\alpha)-f(H\beta)]}$,
where $I_{0}$ and $I$ denote the flux before and after extinction,
and $f(\lambda)$ is the extinction curve.}
\end{table*}

The observations summarized in
Table~\ref{tabcap:Coordinates-and-exposure} were carried out using
EFOSC2 (ESO Faint Object Spectrograph and Camera, 2; Buzzoni et
al. \cite{buzzoni_etal84}) at La Silla. Most spectra were taken on
October 13 and 14, 2002, an additional spectrum of ESO245-G005 was
collected on October 14, 2003, and finally ESO473-G024 was observed in
August 2003. EFOSC2 is a multi-mode instrument working at the Cassegrain
focus of the ESO/3.6m telescope, allowing both imaging and long-slit,
low-resolution spectroscopy. A camera images the aperture onto a
$2048\times2048$~px Loral/Lesser, thinned, and UV-flooded CCD. Its pixel
size is $15\mu$m, which corresponds to $0\farcs157$ on the sky, for a
total field of view of $5.4\times5.4$~\arcmin. Under normal operation,
the CCD is read out by the left amplifier at a gain of $1.3{\rm
e^{-}}/$ADU, and since numerical saturation occurs at $2^{16}$ADU, it
remains well below the full well capacity of $104,000{\rm
e^{-}px^{-1}}$.  The dark current is $7{\rm e^{-}px^{-1}hr^{-1}}$,
comparable to the readout noise of $\sim10\,{\rm e^{-}px^{-1}}$. The CCD
was binned at $2\times2$ both in imaging and spectroscopy mode. Indeed,
rebinning allows the line profile of the calibration arcs to be sampled
with more than four pixels, even with the highest dispersion grisms and
a $0\farcs5$ wide slit, so for spectroscopic observations there is no
real advantage in using the $1\times1$ binning mode.

For each galaxy, the target H\noun{~ii} regions were identified by
subtracting a $5$~min R image from a $15$~min H$\alpha$ image,
and the slit was then centered on the brightest region. Maps of the
H\noun{~ii} regions are shown in Fig.~\ref{figcap:From-left-to}.
During the night, two spectrophotometric standards (Feige 110 and
LTT~3218) from Oke (\cite{oke90}) and Hamuy et al. (\cite{hamuy92,hamuy94})
were observed. Other calibration frames were taken in the afternoon
preceding each observing night (bias, darks, HeAr arcs, and dome flatfields).
Grism \#11 was used without an order sorting filter, yielding a dispersion
of $4.2$~\AA~px$^{-1}$ and {giving a range} from $3400$\AA\ to
$7500$\AA. The second order spectrum (appearing at wavelengths longer
than $\sim6800$\AA) is only a few percent of the first order, so
the contamination is negligible for the target spectra. It is however
important for the determination of the response function. Indeed,
it can be seen in the spectrum of the extremely blue standard Feige~110
(spectral type D0), but it disappears already in the spectrum of LTT~3218,
although its spectral type is DA. The latter standard was then used
to compute the response function. Spectra of the target regions were
obtained with a $1\arcsec$ width slit, while the $5\arcsec$ width
slit, aligned along the parallactic angle, was used for observing
the spectrophotometric standards, in order to minimize slit losses.
The resolution with the $1\arcsec$ width slit is $\approx13$\AA\ FWHM
(from the HeAr lines near $5000$\AA). All slits cover the entire
EFOSC2 field of view.

The reduction and analysis of the spectroscopic data was performed
in a standard way, with the main steps: (a) 2D wavelength and distortion
calibration; (b) extraction of spectra using the wings of H$\alpha$
to define a good window; (c) flux calibration; (d) measurement of
line fluxes in the \textsc{midas/alice}%
\footnote{\noun{eso-midas} is the acronym for the European Southern Observatory
Munich Image Data Analysis System which is developed and maintained
by the European Southern Observatory%
} framework; (e) correction of fluxes for underlying absorption; (f)
correction for internal reddening of the region using the H$\alpha$/H$\beta$
ratio; (g) computation of $T_{e}$ using {[}\ion{O}{iii}] lines
$\lambda4363$ and $\lambda\lambda4959,5007$; and finally (h) computation
of nebular abundances using \textsc{iraf/ionic}%
\footnote{IRAF is distributed by the National Optical Astronomy Observatories,
which are operated by the Association of Universities for Research
in Astronomy, Inc., under cooperative agreement with the National
Science Foundation.%
}. In some cases the H$\alpha$ line profile showed more than one region:
in this case the two brightest regions were extracted (see Table~\ref{tabcap:Reddening-corrected-fluxes-for}).
The spectra of the brightest H~\noun{ii} regions that enter the \lz\ relation
for each galaxy are shown in Fig.~\ref{fig:For-each-galaxy}.

Full account of the data reduction is reported in Appendix~\ref{sec:Reduction-of-EFOSC2}.
The reddening corrected fluxes for each useful region are listed in
Table~\ref{tabcap:Reddening-corrected-fluxes-for}; for the regions
where {[}\ion{O}{iii}]$\lambda4363$ has been detected, we list
their physical parameters and abundances in Table~\ref{tabcap:Physical-parameters-and}.

\subsubsection{M81 group}

\begin{figure}
\begin{centering}
\includegraphics[width=1\columnwidth]{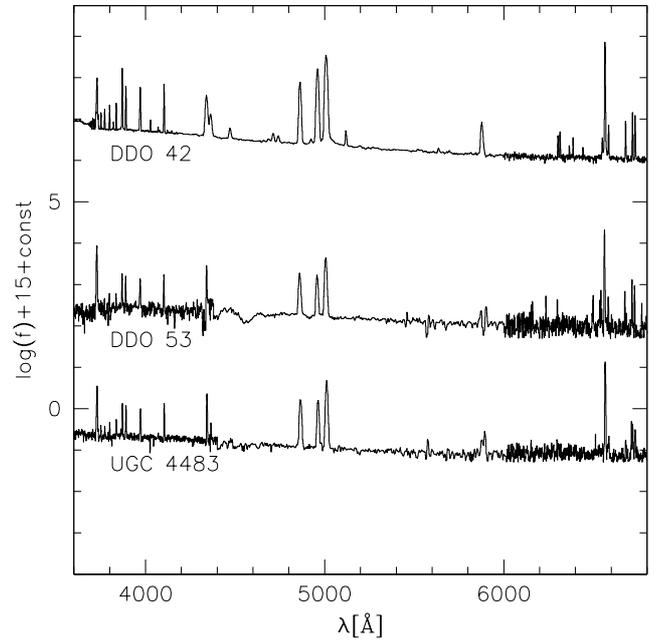} 
\par\end{centering}

\caption{For three galaxies in the M81 group sample the spectrum of the target
\ion{H}{II} region is shown, with a label identifying the host
galaxy. The spectra of other galaxies are of poor quality and have
not been used for abundance determinations. For display purposes,
each flux has been multiplied by $10^{15}$ and a constant has been
added. \label{figcap:For-three-galaxies-m81}}

\end{figure}

\begin{table*}
\caption{Reddening-corrected fluxes for all \ion{H}{ii} regions in M81
group galaxies. 
\label{tabcap:M81a-Reddening-corrected-fluxes-for}}

\begin{centering}
\begin{tabular}{rcccrccc}
 &  &  &  &  &  &  & \tabularnewline
\hline
\hline 
Ion $\lambda$ {[}\AA]  & DDO 42  & DDO 53  & UGC 4483  & Ion $\lambda$ {[}\AA]  & DDO 42  & DDO 53  & UGC 4483\tabularnewline
\hline 
{\tiny \ion{He}{i} 3613.6 }  & {\tiny 0.004 } $\pm$ {\tiny 0.009 }  & {\tiny \ldots{}}  & {\tiny \ldots{}}  & {\tiny \ion{He}{i} 4921.9 }  & {\tiny 0.010 } $\pm$ {\tiny 0.003 }  & {\tiny \ldots{}}  & {\tiny \ldots{}} \tabularnewline
{\tiny \ion{He}{i} 3634.3 }  & {\tiny 0.007 } $\pm$ {\tiny 0.009 }  & {\tiny \ldots{}}  & {\tiny \ldots{}}  & {\tiny $[$\ion{O}{iii}] 4958.9 }  & {\tiny 2.239 } $\pm$ {\tiny 0.011 }  & {\tiny 0.805 } $\pm$ {\tiny 0.029 }  & {\tiny 0.848 } $\pm$ {\tiny 0.014 }\tabularnewline
{\tiny H19 3686.8 }  & {\tiny 0.002 } $\pm$ {\tiny 0.009 }  & {\tiny \ldots{}}  & {\tiny \ldots{}}  & {\tiny {[}\ion{O}{iii}] 5006.8 }  & {\tiny 5.328 } $\pm$ {\tiny 0.017 }  & {\tiny 2.445 } $\pm$ {\tiny 0.045 }  & {\tiny 2.551 } $\pm$ {\tiny 0.021 }\tabularnewline
{\tiny H18 3691.6 }  & {\tiny 0.004 } $\pm$ {\tiny 0.009 }  & {\tiny \ldots{}}  & {\tiny \ldots{}}  & {\tiny ? 5118.0 }  & {\tiny 0.023 } $\pm$ {\tiny 0.003 }  & {\tiny \ldots{}}  & {\tiny \ldots{}} \tabularnewline
{\tiny H17 3697.2 }  & {\tiny 0.005 } $\pm$ {\tiny 0.009 }  & {\tiny \ldots{}}  & {\tiny \ldots{}}  & {\tiny {[}\ion{O}{i}] 5577.3 }  & {\tiny \ldots{}}  & {\tiny \ldots{}}  & {\tiny 0.041 } $\pm$ {\tiny 0.014 }\tabularnewline
{\tiny H16 3703.9 }  & {\tiny 0.012 } $\pm$ {\tiny 0.009 }  & {\tiny \ldots{}}  & {\tiny 0.061 } $\pm$ {\tiny 0.027 }  & {\tiny $^{{\rm (a)}}$5635.0 }  & {\tiny 0.005 } $\pm$ {\tiny 0.004 }  & {\tiny \ldots{}}  & {\tiny \ldots{}} \tabularnewline
{\tiny H15 3712.0 }  & {\tiny 0.010 } $\pm$ {\tiny 0.009 }  & {\tiny \ldots{}}  & {\tiny \ldots{}}  & {\tiny \ion{He}{i} 5875.7 }  & {\tiny 0.107 } $\pm$ {\tiny 0.007 }  & {\tiny \ldots{}}  & {\tiny 0.061 } $\pm$ {\tiny 0.019 }\tabularnewline
{\tiny \ion{S}{iii}+H14 3722.0 }  & {\tiny 0.025 } $\pm$ {\tiny 0.009 }  & {\tiny \ldots{}}  & {\tiny \ldots{}}  & {\tiny {[}\ion{O}{i}] 6300.3 }  & {\tiny 0.012 } $\pm$ {\tiny 0.007 }  & {\tiny \ldots{}}  & {\tiny \ldots{}} \tabularnewline
{\tiny {[}\ion{O}{ii}] 3726.0 }  & {\tiny 0.581 } $\pm$ {\tiny 0.009 }  & {\tiny 2.275 } $\pm$ {\tiny 0.076 }  & {\tiny 0.871 } $\pm$ {\tiny 0.028 }  & {\tiny {[}\ion{S}{iii}] 6312.1 }  & {\tiny 0.017 } $\pm$ {\tiny 0.007 }  & {\tiny \ldots{}}  & {\tiny \ldots{}} \tabularnewline
{\tiny H13 3734.4 }  & {\tiny 0.021 } $\pm$ {\tiny 0.009 }  & {\tiny \ldots{}}  & {\tiny \ldots{}}  & {\tiny {[}\ion{O}{i}] 6363.8 }  & {\tiny 0.005 } $\pm$ {\tiny 0.007 }  & {\tiny \ldots{}}  & {\tiny \ldots{}} \tabularnewline
{\tiny H12 3750.2 }  & {\tiny 0.031 } $\pm$ {\tiny 0.009 }  & {\tiny \ldots{}}  & {\tiny 0.064 } $\pm$ {\tiny 0.026 }  & {\tiny ? 6386.0 }  & {\tiny 0.006 } $\pm$ {\tiny 0.007 }  & {\tiny \ldots{}}  & {\tiny \ldots{}} \tabularnewline
{\tiny H11 3770.6 }  & {\tiny 0.042 } $\pm$ {\tiny 0.009 }  & {\tiny \ldots{}}  & {\tiny 0.072 } $\pm$ {\tiny 0.026 }  & {\tiny ? 6441.0 }  & {\tiny 0.003 } $\pm$ {\tiny 0.007 }  & {\tiny \ldots{}}  & {\tiny \ldots{}} \tabularnewline
{\tiny H10 3797.9 }  & {\tiny 0.055 } $\pm$ {\tiny 0.008 }  & {\tiny 0.104 } $\pm$ {\tiny 0.067 }  & {\tiny 0.076 } $\pm$ {\tiny 0.025 }  & {\tiny {[}\ion{N}{ii}] 6548.0 }  & {\tiny 0.013 } $\pm$ {\tiny 0.008 }  & {\tiny \ldots{}}  & {\tiny 0.015 } $\pm$ {\tiny 0.024 }\tabularnewline
{\tiny \ion{He}{i} 3819.6 }  & {\tiny 0.011 } $\pm$ {\tiny 0.008 }  & {\tiny \ldots{}}  & {\tiny \ldots{}}  & {\tiny H$\alpha$} {\tiny 6562.8 }  & {\tiny 2.849 } $\pm$ {\tiny 0.008 }  & {\tiny 2.849 } $\pm$ {\tiny 0.101 }  & {\tiny 2.851 } $\pm$ {\tiny 0.037 }\tabularnewline
{\tiny H9 3835.4 }  & {\tiny 0.075 } $\pm$ {\tiny 0.008 }  & {\tiny 0.111 } $\pm$ {\tiny 0.065 }  & {\tiny 0.099 } $\pm$ {\tiny 0.025 }  & {\tiny {[}\ion{N}{ii}] 6583.4 }  & {\tiny 0.027 } $\pm$ {\tiny 0.008 }  & {\tiny 0.051 } $\pm$ {\tiny 0.050 }  & {\tiny 0.034 } $\pm$ {\tiny 0.025 }\tabularnewline
{\tiny $[$\ion{Ne}{iii}] 3868.8 }  & {\tiny 0.632 } $\pm$ {\tiny 0.008 }  & {\tiny 0.294 } $\pm$ {\tiny 0.064 }  & {\tiny 0.183 } $\pm$ {\tiny 0.024 }  & {\tiny \ion{He}{i} 6678.1 }  & {\tiny 0.030 } $\pm$ {\tiny 0.008 }  & {\tiny 0.070 } $\pm$ {\tiny 0.052 }  & {\tiny 0.018 } $\pm$ {\tiny 0.026 }\tabularnewline
{\tiny H8 3889.1 }  & {\tiny 0.225 } $\pm$ {\tiny 0.008 }  & {\tiny 0.282 } $\pm$ {\tiny 0.062 }  & {\tiny 0.196 } $\pm$ {\tiny 0.024 }  & {\tiny {[}\ion{S}{ii}] 6716.5 }  & {\tiny 0.055 } $\pm$ {\tiny 0.008 }  & {\tiny 0.148 } $\pm$ {\tiny 0.056 }  & {\tiny 0.080 } $\pm$ {\tiny 0.026 }\tabularnewline
{\tiny H$\epsilon$ 3970.1 }  & {\tiny 0.353 } $\pm$ {\tiny 0.008 }  & {\tiny 0.343 } $\pm$ {\tiny 0.058 }  & {\tiny 0.207 } $\pm$ {\tiny 0.022 }  & {\tiny {[}\ion{S}{ii}] 6730.8 }  & {\tiny 0.045 } $\pm$ {\tiny 0.008 }  & {\tiny 0.104 } $\pm$ {\tiny 0.056 }  & {\tiny 0.059 } $\pm$ {\tiny 0.026 }\tabularnewline
{\tiny \ion{He}{i}4026.2 }  & {\tiny 0.018 } $\pm$ {\tiny 0.007 }  & {\tiny \ldots{}}  & {\tiny \ldots{}}  & {\tiny \ion{He}{i} 7065.3 }  & {\tiny 0.031 } $\pm$ {\tiny 0.010 }  & {\tiny \ldots{}}  & {\tiny 0.029 } $\pm$ {\tiny 0.030 }\tabularnewline
{\tiny $[$\ion{S}{ii}] 4068.6 }  & {\tiny 0.006 } $\pm$ {\tiny 0.007 }  & {\tiny \ldots{}}  & {\tiny 0.013 } $\pm$ {\tiny 0.020 }  & {\tiny $[$\ion{Ar}{iii}] 7135.8 }  & {\tiny 0.059 } $\pm$ {\tiny 0.010 }  & {\tiny 0.035 } $\pm$ {\tiny 0.058 }  & {\tiny 0.026 } $\pm$ {\tiny 0.031 }\tabularnewline
{\tiny $[$\ion{S}{ii}] 4076.4 }  & {\tiny 0.001 } $\pm$ {\tiny 0.007 }  & {\tiny \ldots{}}  & {\tiny \ldots{}}  & {\tiny \ion{He}{i} 7281.3 }  & {\tiny 0.004 } $\pm$ {\tiny 0.010 }  & {\tiny \ldots{}}  & {\tiny \ldots{}} \tabularnewline
{\tiny H$\delta$ 4101.7 }  & {\tiny 0.297 } $\pm$ {\tiny 0.007 }  & {\tiny 0.302 } $\pm$ {\tiny 0.050 }  & {\tiny 0.233 } $\pm$ {\tiny 0.019 }  & {\tiny \ion{O}{ii}] 7319.6 }  & {\tiny 0.009 } $\pm$ {\tiny 0.010 }  & {\tiny \ldots{}}  & {\tiny \ldots{}} \tabularnewline
{\tiny \ion{He}{i} 4120.9 }  & {\tiny 0.002 } $\pm$ {\tiny 0.006 }  & {\tiny \ldots{}}  & {\tiny \ldots{}}  & {\tiny \ion{O}{ii}] 7330.2 }  & {\tiny 0.013 } $\pm$ {\tiny 0.010 }  & {\tiny \ldots{}}  & {\tiny \ldots{}} \tabularnewline
{\tiny \ion{He}{i} 4143.8 }  & {\tiny 0.001 } $\pm$ {\tiny 0.006 }  & {\tiny \ldots{}}  & {\tiny \ldots{}}  & {\tiny $[$\ion{Ar}{iii}] 7751.1 }  & {\tiny 0.015 } $\pm$ {\tiny 0.012 }  & {\tiny \ldots{}}  & {\tiny \ldots{}} \tabularnewline
{\tiny H$\gamma$ 4340.5 }  & {\tiny 0.536 } $\pm$ {\tiny 0.008 }  & {\tiny 0.470 } $\pm$ {\tiny 0.035 }  & {\tiny 0.419 } $\pm$ {\tiny 0.015 }  & {\tiny ? 8058.0 }  & {\tiny 0.012 } $\pm$ {\tiny 0.013 }  & {\tiny \ldots{}}  & {\tiny \ldots{}} \tabularnewline
{\tiny $[$\ion{O}{iii}] 4363.2 }  & {\tiny 0.165 } $\pm$ {\tiny 0.008 }  & {\tiny \ldots{}}  & {\tiny 0.049 } $\pm$ {\tiny 0.013 }  & {\tiny ? 8083.0 }  & {\tiny 0.011 } $\pm$ {\tiny 0.013 }  & {\tiny \ldots{}}  & {\tiny \ldots{}} \tabularnewline
{\tiny \ion{He}{i} 4471.5 }  & {\tiny 0.048 } $\pm$ {\tiny 0.005 }  & {\tiny \ldots{}}  & {\tiny \ldots{}}  & {\tiny Pa16 8502.5 }  & {\tiny 0.003 } $\pm$ {\tiny 0.014 }  & {\tiny \ldots{}}  & {\tiny \ldots{}} \tabularnewline
{\tiny ? 4693.0 }  & {\tiny 0.004 } $\pm$ {\tiny 0.001 }  & {\tiny \ldots{}}  & {\tiny \ldots{}}  & {\tiny Pa15 8545.4 }  & {\tiny 0.006 } $\pm$ {\tiny 0.014 }  & {\tiny \ldots{}}  & {\tiny \ldots{}} \tabularnewline
{\tiny \ion{He}{i} 4713.2 }  & {\tiny 0.034 } $\pm$ {\tiny 0.003 }  & {\tiny \ldots{}}  & {\tiny \ldots{}}  & {\tiny Pa14 8598.4 }  & {\tiny 0.005 } $\pm$ {\tiny 0.014 }  & {\tiny \ldots{}}  & {\tiny \ldots{}} \tabularnewline
{\tiny \ion{C}{ii}4744.9 }  & {\tiny 0.026 } $\pm$ {\tiny 0.003 }  & {\tiny \ldots{}}  & {\tiny \ldots{}}  & {\tiny Pa13 8665.0 }  & {\tiny 0.016 } $\pm$ {\tiny 0.015 }  & {\tiny \ldots{}}  & {\tiny \ldots{}} \tabularnewline
{\tiny H$\beta$ 4861.3 }  & {\tiny 1.000 } $\pm$ {\tiny 0.008 }  & {\tiny 1.000 } $\pm$ {\tiny 0.028 }  & {\tiny 1.000 } $\pm$ {\tiny 0.014 }  & {\tiny C H/$\alpha$H$\beta$$^{b}$}  & {\tiny 0.020 } $\pm$ {\tiny 0.010 }  & {\tiny 0.216 } $\pm$ {\tiny 0.068}  & {\tiny 0.001 } $\pm$ {\tiny 0.030 }\tabularnewline
\hline
\end{tabular}
\par\end{centering}

Notes: ($^a$) \ion{S}{ii} $\lambda5640$? {($^{b}$) The reddening constant is computed by the expression:
$I_{H\alpha}/I_{H\beta}=I_{H\alpha0}/I_{H\beta0}10^{-C[f(H\alpha)-f(H\beta)]}$,
where $I_{0}$ and $I$ denote the flux before and after extinction,
and $f(\lambda)$ is the extinction curve.}
\end{table*}

The spectra of the M81 group dwarfs were obtained with the \emph{Kast}
Double Spectrograph at the Cassegrain focus of the Shane 3m Telescope,
at Lick observatory. The instrument consists of two separate spectrographs
-- one optimized for the red, and the other for the blue --. Dichroic
beam-splitters and separate CCD detectors allow simultaneous observation.
Both arms use a UV-flooded Reticon 1200$\times$400 px device, with
$2.7{\rm \mu m}$ pixels covering $0\farcs8$ on the sky; the ADC
sets a maximum ADU count of 32000, with a gain of $3.8$~e$^{-}$/ADU,
and the read-out noise is $6$~e$^{-}$. The CCDs have a 40\% quantum
efficiency at $3200$\AA.We used a $1\farcs5$ wide slit. The observations
were carried out simultaneously with the blue and the red arms, using
dichroic D46. In one occasion we also obtained a spectrum with the
red arm alone (without dichroic) to check the flux normalization in
the overlap region of the red and blue spectra. Table~\ref{tabcap:list-of-M81}
summarizes the observations, carried out in December 2001 and January
2002. Due to poor weather conditions, and the overheads of the trial-and-error
technique, only six galaxy targets were observed in three nights,
of which one is the higher redshift object KDG~54.

The reduction and analysis of the spectroscopic data is similar to
that of EFOSC2 (see Appendix~\ref{sec:Reduction-of-EFOSC2}), although
more steps are needed to correct for the large flexures and to ensure
a proper normalization of the spectra in the two arms (see Appendix~\ref{sec:Reduction-of-KAST}).
Spectra with adequate $S/N$ ratio could be obtained for three galaxies,
they are shown in Fig.~\ref{figcap:For-three-galaxies-m81}.

The reddening corrected fluxes for each useful region are listed in
Table~\ref{tabcap:M81a-Reddening-corrected-fluxes-for}, and for
the regions where {[}\ion{O}{iii}]$\lambda4363$ was detected,
their derived physical parameters and abundances are provided in Table~\ref{tabcap:Physical-parameters-and}.

\subsection{NIR imaging}

The imaging part of the project was carried out with the 4m William
Herschel Telescope (WHT) and the 3.5m New Technology Telescope (NTT).
The instrument INGRID (Isaac Newton Group Red Imaging Device) was
used at the former, and SOFI (Morwood et al. \cite{morwood_etal98})
at the latter. Both instruments make use of 1024x1024 HgCdTe Hawaii
arrays, and the chosen scales were respectively of $0\farcs238$~px$^{-1}$
and $0\farcs288$~px$^{-1}$.

The observations were performed with the typical pattern for infrared
imaging: iterating every 1-2 minutes between the object and a nearby
patch of clear sky, usually $3$-$5$\arcmin ~away from the galaxy.
Each image was averaged over a number of short integrations, usually
10-30 sec, to avoid saturation of the array from the combination of
the sky background and any bright foreground objects. Small offsets
-- of order of $5$-$10$\arcsec -- were introduced between successive
object or sky images to ensure that the target was placed on different
locations of the array. This helps removing cosmetic defects and improves
the flat-fielding. A log of the observations is presented in Table~\ref{tabcap:Coordinates-and-exposure}
and in Table~\ref{tabcap:list-of-M81}.

Detailed account of the data reduction and calibration is provided
in Appendix~\ref{sec:Reduction-of-near-IR}. The main steps of the
NIR surface photometry are: (a) accurate flat-fielding with polynomial
fitting of the background; (b) masking of foreground stars; (c) deriving
the light-profile by integration in elliptical annuli. Errors on the
total magnitudes have been estimated by changing the background level
by $\pm1\sigma$, integrating the two new growth curves, and then
computing the difference with the magnitude obtained from the average
growth curve. In turn, the r.m.s. fluctuation of the background was
estimated by measuring the background itself in $10$--$20$ independent
areas. 
%
%
The total magnitudes of all
galaxies are listed in Table~\ref{tab:Data-for-the-LZ}, except for
DDO~82, dropped out because no estimate of its abundance could be
obtained; its magnitude is $H$$=10.83\pm0.37$~mag.

\section{Derivation of the chemical abundances}

\begin{table*}
\caption{Physical parameters and abundances for \ion{H}{ii} regions in
the dIrrs of the Sculptor and M81 groups, for which the electron temperature
could be determined. The ${\rm O}^{++}/{\rm H}^{+}$ and ${\rm S}^{+}/{\rm H}^{+}$
abundances are the average of the two values obtained from the two
lines of the doublet separately ($\lambda\lambda5007,4959$, and $\lambda\lambda6716,6731$,
respectively) \label{tabcap:Physical-parameters-and} }

\begin{centering}
\begin{tabular}{rccccc}
 &  &  &  &  & \tabularnewline
\hline
\hline 
\multicolumn{1}{c}{ Parameter } & \multicolumn{1}{c}{ESO 245-G005} & \multicolumn{1}{c}{ESO 245-G005} & \multicolumn{1}{c}{ESO 347-G017} & \multicolumn{1}{c}{ ESO 473-G024 } & DDO 42 \tabularnewline
 & \multicolumn{1}{c}{external} & \multicolumn{1}{c}{central} & \multicolumn{1}{c}{\#1} &  & \tabularnewline
\hline 
$T$ (\ion{O}{iii}) (K)  & $14500{}_{-2020}^{+1970}$  & $14200_{-4680}^{+4290}$  & $12660{}_{-970}^{+933}$  & $14770{}_{-1690}^{+1670}$  & $18500{}_{-542}^{+548}$\tabularnewline
$n$(\ion{S}{ii}) (cm$^{-3}$)  & $605\pm547$  & $674\pm615$  & $164\pm100$  & $545\pm487$  & $682\pm636$\tabularnewline
${\rm O}^{++}/{\rm H}^{+}$ ($\times10^{5}$)  & $2.27$ $\pm$ $0.09$  & $7.75\pm0.16$  & $6.63\pm0.24$  & $2.84$ $\pm$ $0.18$  & $3.82\pm0.53$\tabularnewline
${\rm O}^{+}/{\rm H}^{+}$ ($\times10^{5}$)  & $1.90$ $\pm$ $0.96$  & $8.67\pm7.45$  & $3.54\pm1.03$  & $1.06$ $\pm$ $0.47$  & $0.31\pm0.06$\tabularnewline
${\rm O/{\rm H}}$ ($\times10^{5}$)  & $4.17$ $\pm$ $0.96$  & $16.42\pm7.45$  & $10.17\pm1.06$  & $3.90$ $\pm$ $0.50$  & $4.13\pm0.53$\tabularnewline
$12+\log{\textrm{(O/H)}}$  & $7.61$ $\pm$ $0.10$  & $8.17\pm0.21$  & $8.00\pm0.05$  & $7.59$ $\pm$ $0.06$  & $7.61\pm0.06$\tabularnewline
${\rm N}^{+}/{\rm O}^{+}$ ($\times10^{2}$)  & \ldots{}  & $2.24\pm4.15$  & $2.18\pm1.62$  & $3.30$ $\pm$ $4.77$  & $6.77\pm5.50$\tabularnewline
$\log{\textrm{(N/O)}}$  & \ldots{}  & $-1.65\pm0.46$  & -$1.66\pm0.24$  & $-1.48$ $\pm$ $0.39$  & $-1.17\pm0.26$\tabularnewline
${\rm N}/{\rm H}$ ($\times10^{6}$)  & \ldots{}  & $1.94\pm5.27$  & $0.77\pm0.80$  & $0.35$ $\pm$ $0.66$  & $0.21\pm0.21$\tabularnewline
${\rm S}^{+}/{\rm H}^{+}$ ($\times10^{7}$)  & $3.29$ $\pm$ $0.38$  & $9.73\pm3.68$  & $5.06\pm0.48$  & $1.81$ $\pm$ $0.52$  & $0.72\pm0.18$\tabularnewline
${\rm S}^{++}/{\rm H}^{+}$ ($\times10^{7}$)  & $38.1$ $\pm$ $24.6$  & \ldots{}  & \ldots{}  & \ldots{}  & $5.29\pm2.58$\tabularnewline
ICF  & $1.32$ $\pm$ $0.31$  & \ldots{}  & \ldots{}  & \ldots{}  & $1.97\pm0.24$\tabularnewline
${\rm S/H}$ ($\times10^{6}$)  & $5.44$ $\pm$ $2.46$  & \ldots{}  & \ldots{}  & \ldots{}  & $1.19\pm0.26$\tabularnewline
$12+\log{\textrm{(S/H)}}$  & $6.69$ $\pm$ $0.21$  & \ldots{}  & \ldots{}  & \ldots{}  & $6.06\pm0.10$\tabularnewline
${\rm S/O}$ ($\times10^{2}$)  & $1.30$ $\pm$ $0.89$  & \ldots{}  & \ldots{}  & \ldots{}  & $0.29\pm0.10$\tabularnewline
$\log{\textrm{(S/O)}}$  & $-1.89$ $\pm$ $0.23$  & \ldots{}  & \ldots{}  & \ldots{}  & $-2.54\pm0.13$\tabularnewline
 &  &  &  &  & \tabularnewline
\hline
\hline 
\multicolumn{1}{c}{Parameter } & \multicolumn{1}{c}{NGC 625} & \multicolumn{1}{c}{NGC 625 } & \multicolumn{1}{c}{NGC 59} & \multicolumn{1}{c}{NGC 59} & UGC 4483\tabularnewline
 & \multicolumn{1}{c}{\#1} & \multicolumn{1}{c}{\#2} & \multicolumn{1}{c}{\#1 } & \multicolumn{1}{c}{\#2} & \tabularnewline
\hline 
$T$ (\ion{O}{iii}) (K)  & $10100{}_{-1220}^{+1060}$  & $9400{}_{-758}^{+678}$  & $10700{}_{-2020}^{+1670}$  & $15200{}_{-5640}^{+5340}$  & $15000{}_{-1900}^{+1890}$\tabularnewline
$n$(\ion{S}{ii}) (cm$^{-3}$)  & $622\pm549$  & $135\pm63$  & $736\pm667$  & $3120\pm3070$  & $2000\pm1940$\tabularnewline
${\rm O}^{++}/{\rm H}^{+}$ ($\times10^{5}$)  & $16.42$ $\pm$ $0.45$  & $15.41\pm0.54$  & $19.01\pm0.48$  & $8.11$ $\pm$ $1.69$  & $2.92\pm0.08$\tabularnewline
${\rm O}^{+}/{\rm H}^{+}$ ($\times10^{5}$)  & $7.37$ $\pm$ $4.01$  & $8.48\pm3.04$  & $6\pm4.29$  & $19.03$ $\pm$ $6.93$  & $1.21\pm0.72$\tabularnewline
${\rm O/{\rm H}}$ ($\times10^{5}$)  & $23.79$ $\pm$ $4.04$  & $23.89\pm3.09$  & $25.01\pm4.32$  & $27.14$ $\pm$ $7.13$  & $4.13\pm0.72$\tabularnewline
$12+\log{\textrm{(O/H)}}$  & $8.37$ $\pm$ $0.07$  & $8.37\pm0.06$  & $8.39\pm0.08$  & $8.42$ $\pm$ $0.12$  & $7.61\pm0.08$\tabularnewline
${\rm N}^{+}/{\rm O}^{+}$ ($\times10^{2}$)  & \ldots{}  & \ldots{}  & $3.10\pm4.47$  & \ldots{}  & $5.12\pm8.17$\tabularnewline
$\log{\textrm{(N/O)}}$  & \ldots{}  & \ldots{}  & $-1.51\pm0.39$  & \ldots{}  & $-1.29\pm0.41$\tabularnewline
${\rm N}/{\rm H}$ ($\times10^{6}$)  & \ldots{}  & \ldots{}  & $1.86\pm4.01$  & \ldots{}  & $0.62\pm1.36$\tabularnewline
${\rm S}^{+}/{\rm H}^{+}$ ($\times10^{7}$)  & $7.49$ $\pm$ $0.44$  & $6.79\pm0.51$  & $7.32\pm1.62$  & $21.33$ $\pm$ $6.80$  & $2.34\pm0.81$\tabularnewline
${\rm S}^{++}/{\rm H}^{+}$ ($\times10^{7}$)  & $153$ $\pm$ $98$  & \ldots{}  & $156\pm139$  & \ldots{}  & \tabularnewline
ICF  & $1.43$ $\pm$ $0.37$  & \ldots{}  & $1.95\pm0.86$  & \ldots{}  & \tabularnewline
${\rm S/H}$ ($\times10^{6}$)  & $23.06$ $\pm$ $9.87$  & \ldots{}  & $31.80\pm14.02$  & \ldots{}  & \tabularnewline
$12+\log{\textrm{(S/H)}}$  & $7.32$ $\pm$ $0.20$  & \ldots{}  & $7.46\pm0.21$  & \ldots{}  & \tabularnewline
${\rm S/O}$ ($\times10^{2}$)  & $0.97$ $\pm$ $0.58$  & \ldots{}  & $1.27\pm0.78$  & \ldots{}  & \tabularnewline
$\log{\textrm{(S/O)}}$  & $-2.01$ $\pm$ $0.20$  & \ldots{}  & $-1.90\pm0.21$  & \ldots{}  & \tabularnewline
\hline
\end{tabular}
\par\end{centering}
\end{table*}

Since the main purpose of this study is to use oxygen abundances in
the search for a correlation with the galaxy luminosity, we defer
a complete discussion of the elements other than oxygen to a future
paper. Yet, the abundances of two other important elements (nitrogen
and sulfur) have been derived as described in this section, and are
listed in Table~\ref{tabcap:Physical-parameters-and}. The abundances
have been computed with the so-called direct method (e.g. Osterbrock
\cite{osterbrock}).

The \ion{O}{iii} region electron temperatures $T_{{\rm e}}$,
electron densities $n_{{\rm e}}$, and abundances have been computed
using tasks within the \noun{nebular} package of \noun{iraf.} Since
temperature and density cannot be derived independently, we start
with suitable initial values and iterate the task \noun{temden} until
the values stabilize within the errors: the temperature is computed
using the {[}\ion{O}{iii}] line ratio $(\lambda4959+\lambda5007)/\lambda4363$,
and the density is derived from the {[}\ion{S}{ii}] line ratio
$\lambda6716/\lambda6731$. Using these values for $T_{{\rm e}}$
and $n_{{\rm e}}$, for every ion the abundance is computed with the
\noun{ionic} task, with central wavelengths, line ratios, and errors
listed in Table~\ref{tabcap:Reddening-corrected-fluxes-for} and
Table~\ref{tabcap:M81a-Reddening-corrected-fluxes-for} for the Sculptor
and M81 groups, respectively. The tolerance on the central wavelength
is generally taken as $1$\AA, except in cases where the line is
actually an unresolved doublet or blend. This happens for the following
multiplets: {[}\ion{O}{i}]$\lambda6364$ (tolerance=$2$\AA),
{[}\ion{O}{ii}]$\lambda3727$ ($4$\AA), {[}\ion{O}{ii}]$\lambda7325$
blend ($10$\AA), and the {[}\ion{S}{ii}]$\lambda4072$ blend
($7$\AA).

The total oxygen abundance is computed as ${\rm O/H=O^{++}/H^{+}+O^{+}/H^{+}}$,
i.e. neglecting the usually small contribution from ${\rm O^{+3}}$
(Skillman \& Kennicutt \cite{skillman_kennicutt93}). On the other
hand, higher ionization stages of sulfur do contribute to the total
abundance (Dufour et al. \cite{dufour_etal88}), so an ionization
correction factor (ICF) has to be computed and taken into account.
This is extensively discussed in Garnett (\cite{garnett89}), who
computed a series of photoionization models and compared them with
the relation ${\rm (S^{+}+S^{++})}/{\rm S=[1-(O^{+}/O)^{\alpha}]^{1/\alpha}}$
proposed by Stasinska (\cite{stasinska78}) with $\alpha=3$ and French
(\cite{french81}) with $\alpha=2$. Neither choice of $\alpha$ gives
an entirely satisfactory result, however we have found that for $\alpha=2.6$
the relation reproduces Garnett's models in the range ${\rm O^{+}/O>0.2}$,
which corresponds to the case of all our \ion{H}{ii} regions.
Thus, sulfur abundances are computed as ${\rm S/H=ICF\times(S^{+}+S^{++})}$,
with ${\rm ICF=[1-(1-O^{+}/O)^{2.6}]^{-1/2.6}}$; the correction factors
are listed in Table~\ref{tabcap:Physical-parameters-and}. Finally
nitrogen abundances are computed assuming ${\rm (N/O)=(N^{+}/O^{+})}$,
and then using the nitrogen to oxygen ratio in the equation ${\rm (N/H)=(N/O)}\times{\rm (O/H)}$.

\subsection{Comparison with previous abundance determinations\label{sub:Comparison-with-previous}}

\begin{figure}
\begin{centering}
\includegraphics[width=1\columnwidth]{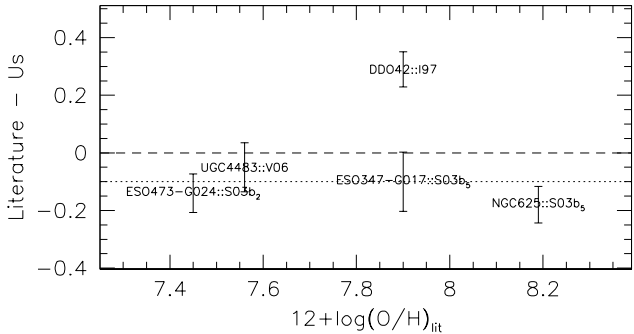} 
\par\end{centering}

\caption{Comparison of our abundance determinations to literature values, for
\ion{H}{ii} regions in galaxies with independent -- and direct
-- determinations of $12+\log{\rm (O/H)}$. The dotted line marks
the median deviation of $-0.1$~dex. The galaxy name is followed
by the reference acronym, and its subscript is the region number in
the reference paper. \label{fig:Comparison-of-our-oh}}

\end{figure}

A summary of our measurements is given in Table~\ref{tab:Data-for-the-LZ},
and, for each galaxy, the best literature direct measurement of the
oxygen abundance is reported as well. If no direct measurement exists,
then we list the best indirect measurement (i.e. the temperature cannot
be estimated, so abundance estimates rely upon semi-empirical relations
making use of different line ratios). There are five galaxies in our
sample for which the abundances can be compared to values already
published: this is shown in Fig.~\ref{fig:Comparison-of-our-oh}.
There is a general agreement except for DDO~42, which exhibits an
internal metallicity gradient and is discussed in more detail in Sect.~\ref{sec:The-near-IR-luminosity-metallicity}.
Excluding this galaxy, the dispersion in the value differences is
only $0.06$~dex. However, the median of the literature values is
$0.1$~dex lower than ours (dotted line in the figure). We tried
to understand this discrepancy by examining NGC~625, which shows
the largest difference. As a first check, we compared our fluxes to
those of S03b for the same nebula in this galaxy: the two studies
are fairly consistent, our fluxes being only $3\%$ smaller on average
and with a dispersion of $14\%$, and no trend with wavelength. The
exception is the {[}\ion{S}{iii}]~$\lambda6312$ line, for which
our flux is $\sim2$ times larger than that quoted in S03b. With respect
to the oxygen lines, our fluxes are $17\%$ and $25\%$ smaller for
the {[}\ion{O}{ii}]~$\lambda3727$ and {[}\ion{O}{iii}]~$\lambda4363$
lines, respectively, while the {[}\ion{O}{iii}]~$\lambda4959$
and $\lambda5007$ line fluxes are only $7\%$ and $6\%$ smaller,
respectively. Such flux discrepancies are entirely within the errors
we quote, yet they are responsible for most of the abundance differences.
Indeed, using the fluxes listed in S03b we obtain 12 + log(O/H)$=8.26$~dex
for NGC~625 (region 5), so the difference is reduced from $0.18$~dex
to $0.07$~dex, i.e. it is now within our $1\sigma$ error on the
abundance. 

Since the measured green {[}\ion{O}{iii}] line fluxes are very
close to those of S03b, we discuss only the blue {[}\ion{O}{iii}]~$\lambda4363$
and {[}\ion{O}{ii}] lines here. The difference in flux for the
{[}\ion{O}{ii}]~$\lambda3727$ line could be due to the fact
that for $\lambda<4000$\AA\ the EFOSC2 response function starts
dropping steadily, and large corrections have to be applied. On the
other hand, comparing the response functions obtained at different
epochs shows that the uncertainty on our response curve is at the
level of a few $0.1\%$. S03b do not show their response function,
but since they are using an order sorting filter with $50\%$ cut
at $3600$\AA\, it is very likely that their correction to the {[}\ion{O}{ii}]
line is quite large as well (almost a factor of two). In any case,
adopting the S03b {[}\ion{O}{ii}] line flux leads only to a minor
\emph{increase} in abundance of $0.03$~dex. Most of the abundance
dependence is then in the {[}\ion{O}{iii}]~$\lambda4363$ line,
and indeed changing its flux and adopting the S03b value we get ${\rm 12+\log(O/H)=8.19}$.
In this case, the flux quoted in S03b might be more reliable than
ours. Indeed, we have to deblend this line from H$\gamma$, while
on the spectra published by S03b the line is well separated from H$\gamma$.
Since {[}\ion{O}{iii}]~$\lambda4363$ becomes weaker as the
oxygen abundance increases -- hence the temperature decreases -- this
could explain why the most deviant abundance value is that corresponding
to NGC~625, the most metal-rich galaxy in our sample.

\section{The NIR luminosity-metallicity relation \label{sec:The-near-IR-luminosity-metallicity}}

\begin{table*}
\caption{Data for the \lz\ relation of Sculptor and M81 dIrrs, taken both
from the literature and from the present work. The abundances, computed
with the method identified in the third column, have been extracted
from the publication identified by its acronym in the fourth column
(see the bibliography); S08 is this paper. The abundance methods are
D=direct, M=McGaugh, P=Pilyugin, and I=indirect. Notes on abundances
are: (a) average of 3 regions; (b) average of two methods; (c) central
region; (d) external region. Distances have been computed with the
method identified in the sixth column, and have been extracted from
the publication identified by its acronym in the seventh column (see
the bibliography). The distance methods are J98, assuming an error
of $20\%$ and RGBT -- red giant branch tip --. Total apparent $H$-band
magnitudes are from this work, and the $H$-band extinctions are from
Schlegel et al. (\cite{schlegel_etal98}). \label{tab:Data-for-the-LZ}}

\begin{centering}
\begin{tabular}{rrrrrrrrrr}
 &  &  &  &  &  &  &  &  & \tabularnewline
\hline
\hline 
Name  & 12+log(O/H)  & Method  & Ref  & D {[}Mpc]  & Method  & Ref  & ${\rm A}_{{\rm H}}$  & $m_{{\rm H,tot}}$  & \multicolumn{1}{c}{$M_{H}$}\tabularnewline
\hline 
 &  &  &  &  &  &  &  &  & \tabularnewline
\multicolumn{9}{c}{Sculptor group} & \tabularnewline
{\tiny ESO 347-G017 }  & {\tiny $8.00\pm0.05$}  & {\tiny D}  & {\tiny S08\#1}  & {\tiny $6.99\pm1.40$}  & {\tiny J98}  & {\tiny S03a}  & {\tiny 0.010 }  & {\tiny $11.79\pm0.50$ }  & {\tiny $-17.40\pm0.66$ }\tabularnewline
 & {\tiny $7.87\pm0.20$}  & {\tiny P}  & {\tiny S08\#1}  &  &  &  &  &  & \tabularnewline
 & {\tiny $7.90\pm0.09$}  & {\tiny D}  & {\tiny S03b\#5}  &  &  &  &  &  & \tabularnewline
{\tiny UGCA 442 }  & {\tiny \ldots{}}  & {\tiny \ldots{}}  & {\tiny S08}  & {\tiny $4.27\pm0.52$}  & {\tiny RGBT}  & {\tiny K03}  & {\tiny 0.010 }  & {\tiny $11.70\pm0.50$}  & {\tiny $-16.45\pm0.56$ }\tabularnewline
 & {\tiny $7.72\pm0.03$}  & {\tiny D}  & {\tiny S03b}  &  &  &  &  &  & \tabularnewline
{\tiny ESO 348-G009 }  & {\tiny \ldots{}}  & {\tiny \ldots{}}  & {\tiny S08}  & {\tiny $6.52\pm1.30$}  & {\tiny J98}  & {\tiny S03a}  & {\tiny 0.008 }  & {\tiny $12.80\pm0.61$}  & {\tiny $-16.24\pm0.75$ }\tabularnewline
 & {\tiny $7.89\pm0.03$}  & {\tiny M}  & {\tiny S03b, L03}  &  &  &  &  &  & \tabularnewline
 & {\tiny $8.11\pm0.05$}  & {\tiny P}  & {\tiny S03b, L03}  &  &  &  &  &  & \tabularnewline
{\tiny NGC 59 }  & {\tiny $8.39\pm0.08$}  & {\tiny D}  & {\tiny S08\#1}  & {\tiny $4.39\pm0.87$}  & {\tiny J98}  & {\tiny S03a}  & {\tiny 0.012 }  & {\tiny $10.21\pm0.37$ }  & {\tiny $-17.97\pm0.57$ }\tabularnewline
 & {\tiny $7.74\pm0.20$}  & {\tiny P}  & {\tiny S08\#1}  &  &  &  &  &  & \tabularnewline
{\tiny ESO 473-G024}  & {\tiny $7.59\pm0.06$}  & {\tiny D}  & {\tiny S08}  & {\tiny $5.94\pm1.19$}  & {\tiny J98}  & {\tiny S03a}  & {\tiny 0.011 }  & {\tiny $13.99\pm0.81$}  & {\tiny $-14.85\pm0.92$ }\tabularnewline
 & {\tiny $7.45\pm0.20$}  & {\tiny P}  & {\tiny S08}  &  &  &  &  &  & \tabularnewline
 & {\tiny $7.45\pm0.03$}  & {\tiny D}  & {\tiny S03b\#2}  &  &  &  &  &  & \tabularnewline
{\tiny AM 0106-382 }  & {\tiny $7.74\pm0.20$}  & {\tiny P}  & {\tiny S08}  & {\tiny $6.13\pm1.23$}  & {\tiny J98}  & {\tiny S03a}  & {\tiny 0.007 }  & {\tiny $14.38\pm0.65$}  & {\tiny $-14.52\pm0.78$ }\tabularnewline
 & {\tiny $7.58\pm0.04$}  & {\tiny M}  & {\tiny L03(a)}  &  &  &  &  &  & \tabularnewline
 & {\tiny $7.62\pm0.08$}  & {\tiny P}  & {\tiny L03(a)}  &  &  &  &  &  & \tabularnewline
{\tiny NGC 625 }  & {\tiny $8.37\pm0.06$}  & {\tiny D}  & {\tiny S08\#2}  & {\tiny $3.89\pm0.22$}  & {\tiny RGBT}  & {\tiny C03}  & {\tiny 0.010 }  & {\tiny $9.01\pm0.26$ }  & {\tiny $-18.95\pm0.28$ }\tabularnewline
 & {\tiny $7.69\pm0.20$}  & {\tiny P}  & {\tiny S08\#2}  &  &  &  &  &  & \tabularnewline
 & {\tiny $8.19\pm0.02$}  & {\tiny D}  & {\tiny S03b\#5}  &  &  &  &  &  & \tabularnewline
{\tiny ESO 245-G005 }  & {\tiny $8.17\pm0.21$}  & {\tiny D}  & {\tiny S08(c)}  & {\tiny $4.43\pm0.45$}  & {\tiny RGBT}  & {\tiny K03}  & {\tiny 0.009 }  & {\tiny $10.99\pm0.50$ }  & {\tiny $-17.24\pm0.54$ }\tabularnewline
 & {\tiny $7.61\pm0.10$}  & {\tiny D}  & {\tiny S08(d)}  &  &  &  &  &  & \tabularnewline
 & {\tiny $7.96\pm0.20$}  & {\tiny P}  & {\tiny S08}  &  &  &  &  &  & \tabularnewline
 & {\tiny $7.65$-- $8.20$}  & {\tiny M}  & {\tiny M96, H03}  &  &  &  &  &  & \tabularnewline
 &  &  &  &  &  &  &  &  & \tabularnewline
\multicolumn{9}{c}{M81 group} & \tabularnewline
{\tiny DDO 42 }  & {\tiny $7.61\pm0.06$}  & {\tiny D}  & {\tiny S08}  & {\tiny $3.20\pm0.41$}  & {\tiny RGBT}  & {\tiny K02}  & {\tiny 0.021 }  & {\tiny $10.13\pm1.00$}  & {\tiny $-17.40\pm1.03$ }\tabularnewline
 & {\tiny $7.73\pm0.20$}  & {\tiny P}  & {\tiny S08}  &  &  &  &  &  & \tabularnewline
 & {\tiny $7.9\pm0.01$}  & {\tiny D}  & {\tiny I97}  &  &  &  &  &  & \tabularnewline
{\tiny DDO 53 }  & {\tiny $7.83\pm0.20$}  & {\tiny P}  & {\tiny S08}  & {\tiny $3.57\pm0.25$}  & {\tiny RGBT}  & {\tiny K02}  & {\tiny 0.021 }  & {\tiny $12.64\pm1.30$ }  & {\tiny $-15.14\pm1.31$ }\tabularnewline
 & {\tiny $8.00\pm0.23$}  & {\tiny i}  & {\tiny HH99(b)}  &  &  &  &  &  & \tabularnewline
{\tiny UGC 4483}  & {\tiny $7.61\pm0.08$}  & {\tiny D}  & {\tiny S08}  & {\tiny $3.40\pm0.20$}  & {\tiny RGBT}  & {\tiny IT02}  & {\tiny 0.020 }  & {\tiny $13.68\pm0.56$ }  & {\tiny $-13.99\pm0.57$ }\tabularnewline
 & {\tiny $7.43\pm0.20$}  & {\tiny P}  & {\tiny S08}  &  &  &  &  &  & \tabularnewline
 & {\tiny $7.56\pm0.03$}  & {\tiny D}  & {\tiny V06}  &  &  &  &  &  & \tabularnewline
\hline
\end{tabular}
\par\end{centering}
\end{table*}

\begin{table*}
\caption{Abundances and NIR luminosities for the fiducial sample of galaxies.
Distances are expressed in Mpc, and columns 6 and 7 indicate the method
used to compute the distance, and its reference. Our abundances have
been corrected by $-0.1$~dex. {For comparison we add also
I~Zw~18 ($H=16.03\pm0.15$; Thuan \cite{thuan83}) and SBS~0335-052
($H=16.09\pm0.03$, $15\arcsec$ aperture; Vanzi et al. \cite{vanzi_etal00})}
\label{tab:Fiducial Abundances-and-near-IR}}

\begin{centering}
\begin{tabular}{lcrcrccc}
 &  &  &  &  &  &  & \tabularnewline
\hline
\hline 
\multicolumn{1}{c}{Galaxy} & Group  & \multicolumn{1}{c}{$12+\log{\rm (O/H)}$} & Ref.  & \multicolumn{1}{c}{Distance} & Met.  & Ref.  & $M_{H}$\tabularnewline
\hline 
ESO 347-G017  & Scl  & $7.90\pm0.05$  & S08\#1  & $6.99\pm1.40$  & J98  & S03a  & $-17.40\pm0.66$ \tabularnewline
NGC 59  & Scl  & $8.29\pm0.08$  & S08\#1  & $4.39\pm0.87$  & J98  & S03a  & $-17.97\pm0.57$ \tabularnewline
ESO 473-G024  & Scl  & $7.49\pm0.06$  & S08  & $5.94\pm1.19$  & J98  & S03a  & $-14.85\pm0.92$ \tabularnewline
NGC 625  & Scl  & $8.27\pm0.06$  & S08\#2  & $3.89\pm0.22$  & RGBT  & C03  & $-18.95\pm0.28$ \tabularnewline
ESO 245-G005  & Scl  & $8.07\pm0.21$  & S08c  & $4.43\pm0.45$  & RGBT  & K03  & $-17.24\pm0.54$ \tabularnewline
UGC 4483  & M81  & $7.51\pm0.08$  & S08  & $3.40\pm0.20$  & RGBT  & IT02  & $-13.99\pm0.57$ \tabularnewline
UGCA 442  & M81  & $7.72\pm0.03$  & S03b  & $4.27\pm0.52$  & RGBT  & K03  & $-16.45\pm0.56$ \tabularnewline
DDO 42  & M81  & $7.90\pm0.01$  & I97  & $3.20\pm0.41$  & RGBT  & K02  & $-17.40\pm1.03$ \tabularnewline
 &  &  &  &  &  &  & \tabularnewline
{I ZW 18 -- SE}  & \ldots{}  & {$7.26\pm0.05$ } & {SK03}  & $18.2\pm1.5$  & {RGBT}  & {A08}  & $-15.27\pm0.23$\tabularnewline
{I ZW 18 -- NW } & \ldots{}  & {$7.17\pm0.04$ } & {SK03}  & $19.3\pm2.3$  & {Ceph}.  & {A08}  & $-15.40\pm0.28$\tabularnewline
{SBS 0335-052 } & \ldots{}  & {$7.29\pm0.02$ } & {I06}  & $54.8\pm3.8$  & {Hub.fl. } & {NED}  & $-17.60\pm0.15$\tabularnewline
\hline
\end{tabular}
\par\end{centering}
\end{table*}

\begin{figure}
\begin{centering}
\includegraphics[width=1\columnwidth]{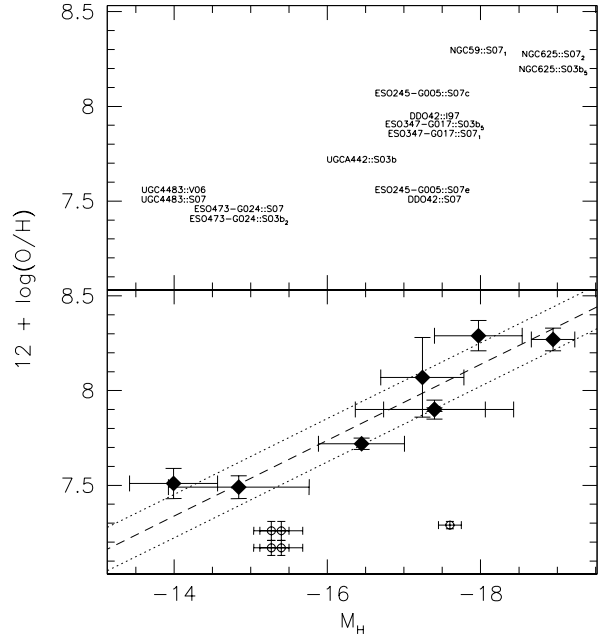} 
\par\end{centering}

\caption{Metallicity of \ion{H}{ii} regions vs. absolute $H$ luminosity
of galaxies in the Sculptor and M81 groups. The upper panel shows
all direct oxygen measurements listed in Table~\ref{tab:Data-for-the-LZ},
while in the lower panel only selected regions are plotted with filled
diamonds (see text). The dashed line is a weighted least-squares fit
to the data, and the dotted lines feature the $\pm1\sigma$ dispersion
around the fit. The galaxy IDs in the upper panel are followed by
the acronym of the paper where the abundances have been published
(see Table~\ref{tab:Data-for-the-LZ}), and the subscript is the
region number in that paper. To separate the labels, small random
vertical offsets have been introduced. {The open circles at
$M_{H}\sim-15.5$ represent the two components of I~Zw~18 for the
two distances listed in Table~\ref{tab:Data-for-the-LZ}, while the
open square represents SBS~0335-052.}\label{fig:lz}}

\end{figure}

Table~\ref{tab:Data-for-the-LZ} summarizes the galaxy parameters
which have been used to create the \lz\ relation. Besides oxygen
abundances and NIR luminosities, the table reports on distances and
total extinctions in the $H$ band, taken from Schlegel et al. (\cite{schlegel_etal98}).
Distances based on standard candles are available for only a fraction
of the targets, so in most cases we have used approximate values.
In particular for the Sculptor group, distances were taken from S03a
(computed with the method of J98). This method relies on the recession
velocity, and is claimed to provide distances within $20\%$ of the
direct measurement. The three galaxies with distances computed with
the red-giant branch tip (RGBT) {luminosity} {enable
a test of this method}: they are UGCA~442, NGC~625, and ESO~245-G005,
and their distances obtained with the J98 method are respectively
$18\%$ smaller, $10\%$ larger, and $15\%$ smaller than the ones
quoted in Table~\ref{tab:Data-for-the-LZ}. For the M81 group, all
four galaxies have distances obtained with the RGBT technique, and
the DDO~42 distance has been measured by both Karachentsev et al.
(\cite{k_etal02}; K02) and Thuan \& Izotov (\cite{thuan_izotov05};
TI05). Although the TI05 distance ($3.42\pm0.15$ Mpc) has a smaller
formal error, we adopt the K02 distance, as, doing so, we have consistent
distances for three galaxies in our sample.

All galaxies having direct abundance determinations are plotted in
the upper panel of Fig.~\ref{fig:lz}. A correction of $-0.1$~dex
has been applied to our data, in order to be consistent with Skillman
et al. (\cite{skillman_scl_hii}). There appears an obvious trend
of increasing metallicity with $H$ luminosity, except for the two
galaxies at $12+\log{\rm (O/H)\approx7.5}$ and $M_{H}\approx-17$~mag.
These are ESO~245-G005 and DDO~42: the latter has an independent
oxygen abundance from Izotov et al. (\cite{izotov_etal97}), which
is higher than our value and consistent with that of ESO~347-G017
at comparable luminosity. Since the work of Roy et al. (\cite{roy_etal96}),
it is known that DDO~42 displays a range of abundances consistent
with a shallow gradient, so our low value probably reflects the external
location of our target \ion{H}{ii} region. In the following we
will then adopt the Izotov et al. value which corresponds to a central
region, and we note that a key point in the definition of an \lz\ relation
is that, for each galaxy, \emph{regions close to their centers should
be selected}. The case of ESO~245-G005 is similar to that of DDO~42:
in this galaxy, we measured both an external and a central region,
finding an abundance difference of $\approx0.5$~dex. Our direct
abundances (the first ones for this galaxy) confirm the metallicity
gradient found by Miller (\cite{miller96}; M96) with an indirect
method: our range is $12+\log{\rm (O/H)=}7.61$ to $8.17$ while M96
finds a range $7.65$ to $8.20$. A similar result has been obtained
by Hidalgo-Gámez et al. (\cite{hidalgo_gamez_etal03}). Again, we
adopt the central abundance as the reference for this galaxy.

From the sample plotted in the upper panel of Fig.~\ref{fig:lz}
we extracted the galaxies listed in Table~\ref{tab:Fiducial Abundances-and-near-IR},
which are the fiducial objects defining our \lz\ relation. Whenever
possible we select abundances from our work, except for two galaxies:
UGCA~442 was not observed spectroscopically, and our observed DDO~42
region was excluded for the reason discussed above. Our fiducial \lz\ relation
is plotted in the lower panel of Fig.~\ref{fig:lz}. The obvious
abundance trend with luminosity is confirmed by a correlation coefficient
$r=-0.93$: a weighted least-squares fit yields the following relation:\[
12+\log{\rm (O/H)}=-0.20\pm0.03\, M_{H}+4.54\pm0.49\]
 with a dispersion of $0.11$~dex around the fit.

{Table~\ref{tab:Fiducial Abundances-and-near-IR} includes
also data for the blue-compact dwarf (BCD) galaxies I~Zw~18 and
SBS 0335-052. These are possibly the most metal-poor galaxies in the
local Universe, so it is worthwhile to see how they compare with our
sample of dwarf irregular galaxies. This is done in Fig.~\ref{fig:lz},
where the two objects are plotted with open symbols. They fall outside
the trend defined by dIrr galaxies, in the sense that their oxygen
abundances are too low for their luminosities. There might be several
explanations for this fact: the first possibility is that the gas
enrichment of BCD galaxies is slower than that of dIrr galaxies, either
because they produce less metals per stellar generation, or because
they lose metals into the IGM more easily. So, as the luminosity of
a BCD increases, its oxygen abundance does not increase as much as
that of a dIrr of comparable mass. A second possibility is that we
are not measuring the central abundances of the two BCDs. We have
shown above that gradients as large as $0.5$~dex are detected in
dwarf galaxies, so the low abundance of the \ion{H}{ii} regions
in I~Zw~18 and SBS~0335-052 might be explained by their non-central
location. Indeed the \ion{H}{i} maps presented in van Zee et
al. (\cite{vanzee_etal98}) and Hirashita et al. (\cite{hirashita_etal02}),
and compared to the optical and H$\alpha$ maps, show that the star
forming regions are slightly off-center. However a counter-argument
for I~Zw~18 is that the galaxy is too small to have appreciable
metallicity gradients. Finally in the case of SBS~0335-052, one might
think that its distance is overestimated: if the distance were $\sim4$
times smaller, then its luminosity would agree with the faintest objects
in our sample. However at the relatively large recession velocity
of the galaxy ($\sim4000$~km~sec$^{-1}$) peculiar motions cannot
induce such a factor of four uncertainty in the distance (e.g., Tonry
\cite{tonry_etal00}). The conclusion is then that these two BCD galaxies
must have had different evolutionary paths than dIrr galaxies. }

{This remains true even when the two galaxies are compared
to the general population of BCDs. Shi et al. (\cite{shi_etal05})
have shown that the \lz\ relation of BCD galaxies is similar to
that of dIrr galaxies, so at the luminosity of I~Zw~18 and SBS~0335-052
a typical BCD has higher oxygen abundances than these two objects.
On the other hand the Shi et al. \lz\ relation shows a very large
scatter: this is not quantified by the authors, but in their Fig.~3
one can see object-to-object abundance differences of up to 1~dex.
Shi et al. propose that the scatter might be due to differences in
SFH or the evolutionary status of the current starburst, or to different
starburst-driven outflows or gas infall rates. The abundances of I~Zw~18
and SBS~0335-052 are then particularly low, but still compatible
with those of BCDs in general, and could be due to one of these processes. }

\section{Discussion}

\subsection{Origin of the luminosity-metallicity relation}

\begin{figure}
\begin{centering}
\includegraphics[width=1\columnwidth]{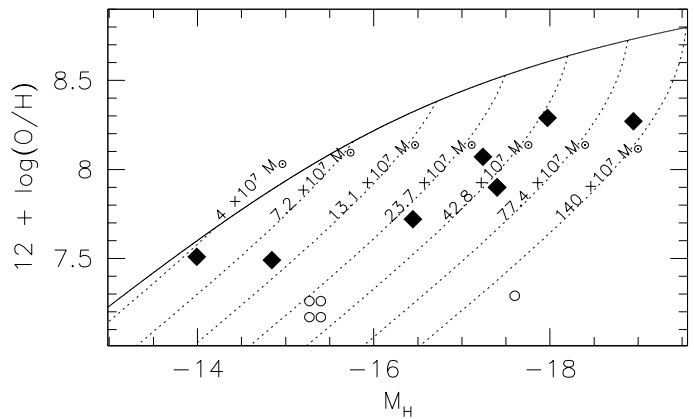} 
\par\end{centering}

\caption{Comparison of the luminosities and abundances of our fiducial galaxies
(filled diamonds) with the predictions of a set of closed-box models
of increasing total masses (dotted line curves). For each model the
evolution has been truncated when the remaining gas mass drops below
$3\times10^{7}\,\mathcal{M}_{\odot}$, and the sequence of the end-points
is outlined by the solid curve. {The open symbols are the same
as in Fig.~\ref{fig:lz}}. \label{fig:7617fig37.ps}}

\end{figure}

As recalled in the introduction, in the case of dE/dSph galaxies the
\lz\ relation is interpreted in the scenario of mass-loss through
galactic winds. The gas escapes more easily from low mass galaxies,
so their chemical evolution is {truncated} before that of more
massive galaxies. Some previous studies have proposed that dSphs and
dIrrs share the same \lz\ relation, thus extending the mass-loss
scenario to irregulars (e.g. Skillman, Kennicutt, \& Hodge \cite{skh89}).
However, recently it has been shown that the abundances of dIrrs are
lower than those of dSphs of comparable luminosity, whether we consider
their gas content (e.g. Mateo \cite{mateo98}) or their RGB stars
(e.g. Grebel \cite{grebel04}). Therefore, if we want to retain the
mass-loss scenario for both classes of dwarfs, then we must assume
not only that this process leaves behind an \lz\ relation after
the evolution has stopped (dSph), but also that it induces a (possibly
different) relation very early in the history of galaxies (dIrr),
and is able to maintain it throughout their lives.

More input on the issue of the chemical evolution of dIrrs has been
provided recently by Skillman et al. (\cite{skillman_scl_hii}) and
Pilyugin et al. (\cite{pilyugin_etal_04}). Both studies conclude
that the chemical evolution of dIrrs can be approximated by a closed-box,
provided that an effective oxygen yield $\approx1/3$ of the theoretical
value (e.g., Maeder \cite{maeder92}) is adopted. A low yield can
be the signature of gas exchange with the environment, and this is
indeed observed in a few cases, as discussed in the introduction.
If this conclusion is valid, then it is hard to imagine how mass-loss
can produce the \lz\ relation, because the galaxy mass is not a
parameter of the closed-box model. Conversely, one would expect that
the effective yield does depend on the mass of the galaxy, since the
efficiency of mass-loss via galactic winds should depend on the depth
of the potential well. Note also that Lee et al. (\cite{lee_etal06})
have found a few dwarf galaxies with large yields, a result difficult
to reconcile with the galactic wind hypothesis. In conclusion, there
does not appear to be any obvious interpretation of the observational
facts and the situation remains confusing.

Yet, pushing further the hypothesis of a low and universal yield,
we can gain some insight by looking at the dotted curves in Fig.~\ref{fig:7617fig37.ps},
which are in fact closed-box models computed with constant yield $p_{{\rm (O/H)}}=1.6\times10^{-4}$,
and total masses $m_{{\rm tot}}=m_{{\rm gas}}+m_{{\rm stars}}$ varying
between $4\times10^{7}\,\mathcal{M}_{\odot}$ and $1.4\times10^{9}\,\mathcal{M}_{\odot}$.
We also assumed $\mathcal{M}/L_{H}=1.2$ (typical of a globular cluster-like
population, e.g. Bruzual \& Charlot \cite{bc03}), and $M_{H,\odot}=3.1$~mag
(\emph{Astrophysical Quantities}); the only parameter varying along
a track is the gas mass fraction $\mu=m_{{\rm gas}}/m_{{\rm tot}}$.
If one assumes that all galaxies were born at the same time, then
this plot is telling us that more massive galaxies evolve faster along
their tracks, i.e. $d\mu/dt$ is larger for more massive galaxies.
If the galaxies were formed at different times, then larger galaxies
had more time to build up their metals. Skillman et al. (\cite{skillman_sfr})
computed SFRs for the Sculptor group galaxies: if their data are plotted
in the form $d\mu/dt={\textrm{SFR}}/(m_{{\rm gas}}+m_{{\rm stars}})$
vs. $(m_{{\rm gas}}+m_{{\rm stars}})$, then a general trend of $d\mu/dt$
increasing with total mass can be seen. This confirms that larger
galaxies are more effective in converting their gas into stars.

The conclusion is that, while a galaxy is still evolving (such as
a dIrr) the \lz\
relation could be a combined effect of mass-loss and of more efficient
gas processing. A similar conclusion was reached by Pilyugin \& Ferrini
(\cite{pilyugin_ferrini}): according to them, the \lz\ relation
is a combined effect of smaller gas-loss and higher astration level
as the mass increases. Another prediction of this scenario is that
the slope of the \lz\ relation should increase with time, and the
effect would be even larger if the gas is removed at earlier stages
in smaller galaxies. The reason why astration becomes more efficient
as the galaxy mass increases, remains to be understood. On the other
hand, we should recall that one of the most accepted facts about star
formation is that it depends on the gas density (Schmidt \cite{schmidt59}),
a conclusion also suggested by recent simulations (e.g., Chiosi \&
Carraro \cite{chiosi_carraro02}). So it might be that as its mass
increases, a galaxy becomes more effective at compressing its gas.

{The open symbols in Fig.~\ref{fig:7617fig37.ps} represent again
the two BCD galaxies I~Zw~18 and SBS~0335-052. In Sect.~\ref{sec:The-near-IR-luminosity-metallicity}
we concluded that they do not follow the trend defined by dIrr galaxies
because their chemical evolution must be different. Our simplified
approach would tell us that I~Zw~18 is on the track of a $\sim3\times10^{8}\mathcal{M}_{\odot}$
galaxy, while SBS~0335-052 is on that of a $\sim2\times10^{9}\mathcal{M}_{\odot}$
object, and that -- compared to a dIrr of similar total mass -- they
did not convert much of their gas into stars, thus having very low
metallicities. However it is also possible that their evolutionary
paths are different: for example their compactness might imply higher
densities, hence SFRs, which in turn would trigger more powerful stellar
winds and remove metals more easily than in dIrr galaxies. Or they
might be unusually young}.

The closed-box scenario itself does not make any prediction about
the final abundance a galaxy can reach: actually it predicts arbitrarily
large values of $Z$ as the gas mass fraction tends to zero, since
$Z=p\,\ln(1/\mu)$ (Searle \& Sargent \cite{ss72}). However, if we
assume that the mass of each new generation of stars, $m_{{\rm SF}}$,
is roughly constant, then the maximum metallicity we can measure will
be $Z_{{\rm max}}=p\,\ln(m_{{\rm tot}}/m_{{\rm SF}})\propto p\,\ln m_{{\rm tot}}$,
i.e. it will be proportional to the mass of the galaxy. This is shown
in Fig.~\ref{fig:7617fig37.ps}, where we assumed $m_{{\rm SF}}=3\times10^{7}\mathcal{M}_{\odot}$,
a mass of stars that can be created in $\sim10^{8}\textrm{\, yr}$
for a typical SFR~$\sim0.1~\,\mathcal{M}_{\odot}\textrm{y}r^{-1}$.
This formulation seems to provide a natural upper limit to the measurable
metallicities, and it works even for large galaxy masses. Indeed,
the dotted line in the top panel of Fig.~\ref{fig:dichot} represents
the same relation extended to $m_{{\rm tot}}=2\times10^{11}\,\mathcal{M}_{\odot}$,
and with the exception of a couple of objects, the metallicities of
star-forming galaxies are always lower than this limit. It is quite
surprising that an upper limit to the metallicity is provided by such
a simple expression, as it is an oversimplification of galactic chemical
evolution. Still, it might suggest that a characteristic mass is involved
in the star formation process.

\subsection{Comparison with other galaxies \label{subsec:Comparison-with-other}}

\begin{table*}
\caption{A compilation of luminosities and abundances for a sample of nearby
dwarf star-forming galaxies. The abundances listed in column 6 have
been taken from the publication the number of which is given in column
8 and have been computed with the method identified in column 7: 1=
direct method but temperature not computed, assumed $T_{{\rm e}}=10,000{\rm \, K}$;
2= direct method; 3= unknown. The publication numbers correspond to
the following papers: (1) Hunter \& Hoffman \cite{hunter_hoffman99};
(2) Hidalgo-Gàmez \& Olofsson (\cite{hidalgo-gamez_olofsson02});
(3) Vigroux et al. (\cite{vigroux_etal87}); (4) Storchi-Bergmann
et al. (\cite{storchi-bergmann_etal94}); (5) Martin (\cite{martin97});
(6) Kobulnicky \& Skillman (\cite{kobulnicky_skillman96}); (7) average
$\pm1\,\sigma$ of abundances from the following papers: Izotov \&
Thuan (\cite{izotov_thuan04}), Hunter \& Hoffman (\cite{hunter_hoffman99}),
Thuan \& Izotov (\cite{thuan_izotov05}), Alloin et al. (\cite{alloin_etal79}),
Shi et al. (\cite{shi_etal05}); (8) Izotov \& Thuan (\cite{izotov_thuan98});
(9) Shi et al. (\cite{shi_etal05}); (10) Thuan \& Izotov (\cite{thuan_izotov05});
(11) van Zee, Haynes, \& Salzer (\cite{vanzee_etal97}); (12) Izotov
\& Thuan (\cite{izotov_thuan98}); (13) Izotov \& Thuan (\cite{izotov_thuan97}).
Most distances were taken from Hunter \& Elmegreen (\cite{hunter_elmegreen06};
HE06): in column 10 the label `h' means that these authors computed
the distance from the recession velocity, while a number identifies
their source publication for the distance. If no distance is given
in HE06, then we derived it using the NED velocity and $H_{\circ}=72\,{\rm km}\,{\rm sec}^{-1}\,{\rm Mpc}^{-1}$.
Finally the DDO~43 distance is from Sharina (2004, private communication;
cited in Karachentsev et al. \cite{karachentsev_etal04}). \label{tab:A-compilation-of}}

\begin{sideways} \begin{tabular}{rrrrrrrrrrcr}
 &  &  &  &  &  &  &  &  &  &  & \tabularnewline
\hline
\hline 
\multicolumn{1}{c}{Name} & \multicolumn{1}{c}{Alt. name} & \multicolumn{1}{c}{R.A.} & \multicolumn{1}{c}{Dec.} & \multicolumn{1}{c}{$v_{{\rm r}}$ (NED)} & \multicolumn{1}{c}{12+log(O/H) } & \multicolumn{1}{c}{M} & \multicolumn{1}{c}{Ref.} & \multicolumn{1}{c}{$D$} & \multicolumn{1}{c}{source} & \multicolumn{1}{c}{${\rm A}_{{\rm tot}}$} & \multicolumn{1}{c|}{${\rm H}_{{\rm tot}}$}\tabularnewline
 &  & \multicolumn{2}{c}{J2000} & \multicolumn{1}{c}{{[}km~sec$^{-1}$]} &  &  &  & \multicolumn{1}{c}{{[}Mpc]} & \multicolumn{1}{c}{of $D$} & {[}mag]  & \multicolumn{1}{c|}{{[}mag]}\tabularnewline
\hline 
MRK 600  &  & 02:51:04.6  & +04:27:14  & $1008\pm1$  & $7.83\pm0.01$  & 2  & 12  & $15.7$  & h  & $0.038$  & $14.38\pm0.19$\tabularnewline
NGG 1156  &  & 02:59:42.2  & +25:14:14  & $375\pm1$  & $8.23$$\pm$\ldots{}  & 3  & 3  & $7.8$  & 25  & $0.129$  & $9.83\pm0.01$\tabularnewline
NGG 1569  &  & 04:30:49.0  & +64:50:53  & $-104\pm4$  & $8.37$$\pm$\ldots{}  & 2  & 4  & $2.5$  & 41  & $0.403$  & $9.40\pm0.02$\tabularnewline
DDO 43  & UGC 03860  & 07:28:17.4  & +40:46:11  & $354\pm1$  & $8.30\pm0.09$  & 1  & 1  & $7.8$  & rg  & $0.034$  & $13.00\pm0.17$\tabularnewline
Haro 23  & UGCA 201  & 10:06:18.1  & +28:56:40  & 1$363\pm7$  & $8.35$$\pm$\ldots{}  & 2  & 7  & $20.2$  & h  & $0.014$  & $11.77\pm0.13$\tabularnewline
MRK 33  & UGC 05720  & 10:32:31.9  & +54:24:04  & $1430\pm4$  & $8.35$$\pm$\ldots{}  & 2  & 9  & \ldots{}  & n  & $0.007$  & $10.70\pm0.04$\tabularnewline
 &  &  &  &  &  &  &  &  &  &  & $10.84\pm0.03$\tabularnewline
Haro 3  & NGG 3353  & 10:45:22.4  & +55:57:37  & $944\pm5$  & $8.3\pm0.10$  & 2  & 7  & $15.5$  & h  & $0.004$  & $10.71\pm0.03$\tabularnewline
 &  &  &  &  &  &  &  &  &  &  & $10.83\pm0.02$\tabularnewline
 &  &  &  &  &  &  &  &  &  &  & $10.74\pm0.06$\tabularnewline
Haro 4  & UGCA 225  & 11:04:58.5  & +29:08:22  & $646\pm5$  & $7.81\pm0.02$  & 2  & 8  & $9.5$  & h  & $0.017$  & $14.03\pm0.03$\tabularnewline
 &  &  &  &  &  &  &  &  &  &  & $13.75\pm0.36$\tabularnewline
MRK 178  & UGC 06541  & 11:33:28.9  & +49:14:14  & $250\pm1$  & $7.93\pm0.02$  & 2  & 10  & $3.9$  & 27  & $0.010$  & $12.47\pm0.15$\tabularnewline
NGG 3738  &  & 11:35:48.8  & +54:31:26  & $229\pm4$  & $8.23\pm0.01$  & 2  & 5  & $4.9$  & 27  & $0.006$  & $9.63\pm0.04$\tabularnewline
MRK 1307  & UGC 06850  & 11:52:37.3  & $-$02:28:10  & $1055\pm12$  & $7.95\pm0.01$  & 2  & 8  & \ldots{}  & n  & $0.011$  & $12.87\pm0.02$\tabularnewline
NGG 4214  &  & 12:15:39.2  & +36:19:37  & $291\pm3$  & $8.25\pm0.05$  & 2  & 6  & $2.9$  & 33  & $0.013$  & $8.26\pm0.03$\tabularnewline
Haro 8  & UGC 07354  & 12:19:09.9  & +03:51:21  & $1526\pm4$  & $8.12$$\pm$\ldots{}  & 2  & 9  & $22.2$  & h  & $0.010$  & $11.97\pm0.14$\tabularnewline
Haro 29  & UGCA 281  & 12:26:16.0  & +48:29:37  & $281\pm4$  & $7.82\pm0.01$  & 2  & 10  & $5.4$  & h  & $0.009$  & $13.54\pm0.22$\tabularnewline
 &  &  &  &  &  &  &  & $4.72$  & 100  &  & \tabularnewline
LeoI HH  & DDO 168  & 13:14:27.9  & +45:55:09  & $192\pm1$  & $7.5\pm0.20$  & 2  & 2  & $3.5$  & h  & $0.009$  & $11.91\pm0.12$\tabularnewline
 &  &  &  &  &  &  &  & $4.33$  & 101  &  & \tabularnewline
Haro 38  & UGC 08578  & 13:35:35.6  & +29:13:01  & $853\pm7$  & $8.02$$\pm$\ldots{}  & 2  & 9  & $13.7$  & h  & $0.007$  & $14.03\pm0.17$\tabularnewline
Haro 43  &  & 14:36:08.8  & +28:26:59  & $1912\pm10$  & $8.17\pm0.10$  & 2  & 11  & $30.5$  & h  & $0.011$  & $13.66\pm0.17$\tabularnewline
MRK 829  & UGC 09560  & 14:50:56.5  & +35:34:18  & $1180\pm2$  & $7.69$$\pm$\ldots{}  & 2  & 9  & \ldots{}  & n  & $0.007$  & $13.07\pm0.08$\tabularnewline
 &  &  &  &  &  &  &  &  &  &  & $13.05\pm0.04$\tabularnewline
II Zw 71  & UGC 09562  & 14:51:14.4s  & +35:32:32  & $1292\pm18$  & $8.24$$\pm$\ldots{}  & 2  & 9  & \ldots{}  & n  & $0.007$  & $11.89\pm0.05$\tabularnewline
 &  &  &  &  &  &  &  &  &  &  & $12.14\pm0.03$\tabularnewline
MRK 487  & UGCA 410  & 15:37:04.2s  & +55:15:48  & $665\pm7$  & $8.10\pm0.04$  & 2  & 13  & \ldots{}  & n  & $0.008$  & $13.35\pm0.11$\tabularnewline
 &  &  &  &  &  &  &  &  &  &  & $13.25\pm0.03$\tabularnewline
MRK 297  & NGC 6052  & 16:05:12.9s  & +20:32:32  & $4716\pm4$  & $8.34$$\pm$\ldots{}  & 2  & 9  & \ldots{}  & n  & $0.044$  & $10.92\pm0.06$\tabularnewline
\hline
\end{tabular}\end{sideways} 
\end{table*}

\begin{figure}
\begin{centering}
\includegraphics[width=1\columnwidth]{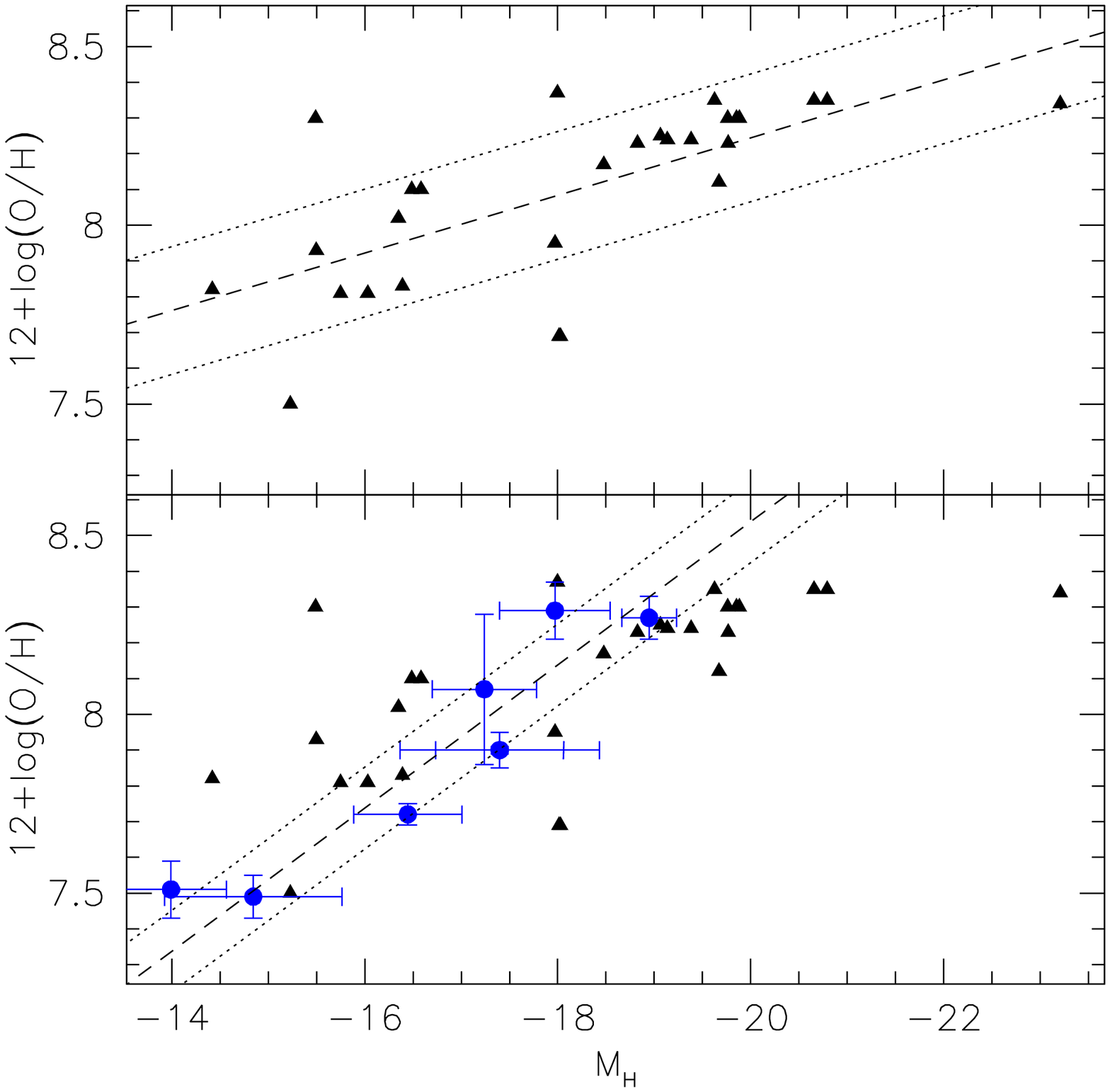} 
\par\end{centering}

\caption{Abundance vs. absolute $H$ luminosity for galaxies with data taken
from the literature (filled triangles) and for our targets (filled
circles). In the upper panel, the dashed line is an unweighted fit
to the literature data, while our fiducial relation is shown in the
lower panel. In both panels dotted lines outline the r.m.s. dispersion
around the fit. \label{fig:Abundance-vs.-absolute}}

\end{figure}

The literature offers only a few O/H measurements for galaxies as
faint as those in our sample, so a direct consistency check of our
results is not possible. However, one can find a few dIrrs with oxygen
abundances and total infrared magnitudes allowing us to verify our
relation statistically.

In order to minimize any systematic errors, we limited ourselves to
using the infrared magnitudes from three sources: the 2MASS Extended
Source Catalog (Jarrett et al. \cite{jarret_etal00}), Cairós et al.
(\cite{cairos_etal03}), and Hunter \& Elmegreen (\cite{hunter_elmegreen06}).
No corrections for the different filter systems were applied. Furthermore,
we used only oxygen abundances derived with the temperature sensitive
method. The only exception is DDO\,43, where T$_{e}$=10000\,K was
assumed instead of being measured. Finally, the distances for galaxies
closer than $\sim$$10$~Mpc were based on direct measurements rather
than on the Hubble flow. The data used for the comparison are listed
in Table~\ref{tab:A-compilation-of}. More than one O/H measurement
per galaxy is listed, whenever available. Radial velocities are taken
from the NASA extragalactic database. If the literature sources do
not list the O/H uncertainty, $0.1$\,dex was adopted. The distances
flagged with {}``h'' are based on the Hubble flow (${\rm H}_{0}$=72~km
s$^{-1}$ Mpc$^{-1}$). The $H$-band extinction A$_{H}$ has been
taken from Schlegel et al. (\cite{schlegel_etal98}). The table data
are presented in graphical form in Fig.~\ref{fig:Abundance-vs.-absolute}
and compared to our \lz\ relation. It appears that the properties
of our targets, dIrrs in the Sculptor and M81 groups, are generally
similar to those of nearby dIrr galaxies, but the \lz\ relations
yielded by the two sets of data are radically different. The slope
of an \lz\ relation based on literature data is $2.5$ times smaller
than the one we obtain ($-0.08\pm0.02$ vs. $-0.20\pm0.03$), and
its dispersion is $0.18$~dex, i.e. $58\%$ larger than the scatter
around our relation ($0.11$~dex).

The large scatter in the literature data {might be} ascribed
to the uncertain distances, and it is also likely related to the random
location of the \ion{H}{ii} regions considered inside each galaxy.
Moreover, this large scatter seems to affect faint galaxies more than
luminous ones, since all objects brighter than $M_{H}\simeq-18$ exhibit
a smaller scatter. Note that the literature compilation includes galaxies
in diverse environments, another factor which might account for the
larger spread in the literature \lz\
relation in comparison with our \lz\ relation for the group galaxies.

{Finally, although we tried to use a sample that is as homogeneous
as possible, and the method for the abundance calculation is formally
the same in all these works, yet systematic differences are still
possible (as demonstrated in Sect.~\ref{sub:Comparison-with-previous}),
and they can explain part of the scatter. A detailed discussion of
all the single galaxies contained in Table~\ref{tab:A-compilation-of}
is beyond the scope of this paper, but the conclusion is clear: any
quantitative interpretation of the \lz\ relation based on literature
data should be performed with great caution. However, this conclusion
might be revised as more and more large and homogeneous data samples
appear in the literature. }

\subsection{A dwarf vs. giant galaxy dichotomy? \label{sec:dichot}}

\begin{figure*}
\begin{centering}
\includegraphics[width=0.8\textwidth]{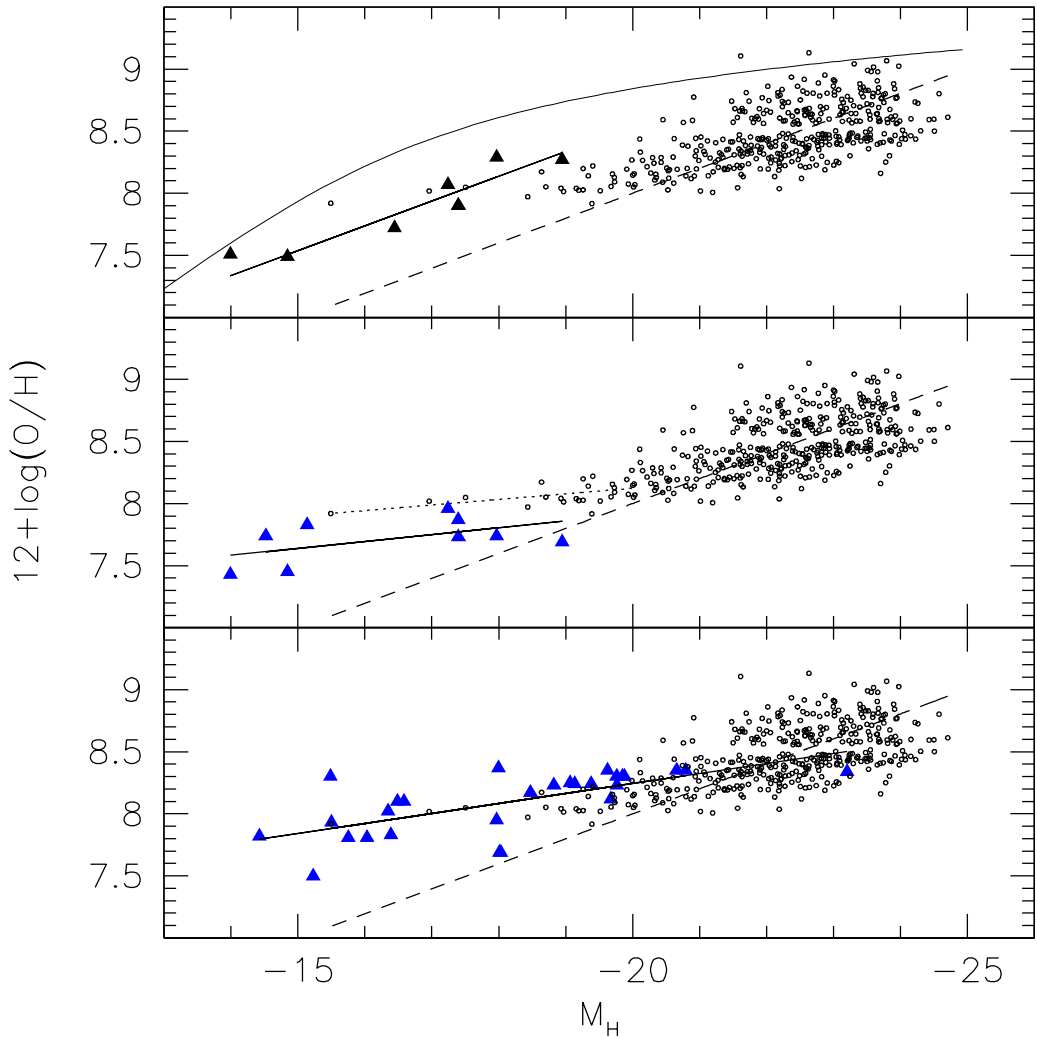}
\par\end{centering}

\caption{Oxygen abundances vs. $H$ luminosity for the Salzer et al. (\cite{salzer05})
emission-line galaxies (KBG calibration), our dIrrs, and data compiled
from the literature. In all panels the dashed line is the fit obtained
by Salzer et al. to their data points (small open circles). In the
top panel, the thin solid line is the same curve as that shown in
Fig.~\ref{fig:7617fig37.ps}, while the filled triangles and solid line
correspond to our fiducial \lz\ relation. In the central panel,
the dotted line is a weighted fit to Salzer et al. galaxies fainter
than $M_{H}=-20$, while the filled triangles and solid line relate
to our \lz\ relation obtained using abundances re-computed by us
with the empirical, indirect method. In the bottom panel, we show
the \lz\ relation from literature data (filled triangles and solid
line). \label{fig:dichot}}

\end{figure*}

The unique study of the \lz\ relation for giant galaxies that includes
NIR, $H$-band magnitudes is that by Salzer et al. (\cite{salzer05};
hereafter S05). They obtained spectra of emission-line galaxies in
fixed apertures of $1\farcs5$ or $2\arcsec$, and computed oxygen
abundances with a reduced number of emission lines, because of the
limited spectral coverage of their data: using additional spectra,
they first derived metallicities for a subsample of galaxies using
both the direct and the empirical methods, and then calibrated them
against {[}\ion{N}{ii}]$\lambda$6583/H$\alpha$ and {[}\ion{O}{iii}]$\lambda$5007/H$\beta$.
The empirical method they used is the Pilyugin (\cite{pilyugin00-pmethod};
P00) calibration for the lower branch of the $12+{\log({\rm O/H})}$
vs. $R_{23}$ relation, and three calibrations for the upper branch:
Edmunds \& Pagel (\cite{edmunds_pagel84}; EP), Kennicutt, Bresolin,
\& Garnett (\cite{kbg03}; KBG), and Tremonti et al. (\cite{tremonti_etal04}).
In Fig.~\ref{fig:dichot} we have plotted our data together with Salzer's
et al. data. The figure reveals that the scatter in the \lz\ relation
for giant galaxies is very large, compared to that of our dIrrs, perhaps
due to fixed-aperture effects and uncertainties specific to the
empirical methods (see also the discussion in S05). The straight lines
in the upper panel are the fits obtained by S05 and by us: taken at face
value, this plot would suggest a well-defined offset between the \lz\
relation of giants and dwarfs. The slope of the S05 \lz\ relation is in
fact close to what we find for dwarfs alone, namely $-0.215\pm0.003$ and
$-0.201\pm0.004$ for abundances obtained with the EP and KBG
calibrations, respectively. Finally, to be consistent with S05, in the
central panel we have plotted our abundances computed with the same P00
calibration of the empirical method, allowing two galaxies to be added
(see abundances identified by {}``P'' in
Table~\ref{tab:Data-for-the-LZ}). The offset between galaxies in the two
mass ranges seems to disappear: it is replaced by what looks like
different \lz\ relations for dwarf and giant galaxies, with a separation
at $M_{H}\approx-20$. A fit to S05 galaxies fainter than this limit
yields the dotted line shown in the panel, while the solid line is a fit
to our dwarfs. The two relations have a similar slope, and a
$\sim0.3$~dex offset, with the dwarfs of our sample being on average
more metal poor than those of S05. This might result from the two-step
calibration of S05, which introduces additional uncertainties in the
empirical method -- a method which, in any case, does not guarantee
abundance precisions better than $0.2$~dex --. The most important thing
to notice, however, is that extremely different \lz\ relations are
obtained for the dwarfs of our sample, if one uses abundances computed
by the two methods, direct or empirical. The solid line in the upper
panel of Fig.~\ref{fig:dichot} (corresponding to direct abundances in
the region of dwarf galaxies) has a slope of $-0.05$~dex~mag$^{-1}$ and
an r.m.s. dispersion around this fit of $0.14$~dex. The scatter is
larger, and the slope is four times smaller when empirical abundances
are used for the same sample of dwarfs (central panel). Striving for
abundances obtained with the direct method requires a much larger
observational effort, but the effort is wholly justified by the better
constraints it provides on galaxy evolution.

Interestingly, using direct abundances from the literature yields
an \lz\ relation similar to that obtained by fitting the S05 dwarfs.
This can be seen by comparing the solid line in the bottom panel of
Fig.~\ref{fig:dichot}, to the dotted line in the central panel.
The zero points are similar, and the two slopes are $-0.08\pm0.02$~dex~mag$^{-1}$
and $-0.04\pm0.02$~dex~mag$^{-1}$, respectively. The conclusion
is that the use of direct abundances is not a sufficient condition:
distances must be know with great accuracy as well. In that respect,
let us point out that selecting a sample of dIrr galaxies in groups
has been a winning strategy. Even assuming just an average distance
for all galaxies in a group has allowed us to obtain essentially the
same \lz\ relation (Saviane et al. \cite{sidney,diablerets}).

The question of a dwarf-giant dichotomy is then open: moving from
the empirical to the direct abundances, the \lz\ relation for dwarfs
becomes much better defined, parallel to the one obtained for more
massive galaxies, and with a substantial offset. It remains to be
seen what would happen to the \lz\ relation of giant galaxies if
one uses direct abundances: an important extension of this project
would be to measure direct abundances for at least some of the S05
galaxies, and check whether the observed offset is confirmed. Since
the S05 sample includes star-forming galaxies, one might suspect that
some of the giant galaxies are shifted in luminosity because of the
presence of a star-burst: however, Lee et al. (\cite{lee_etal04})
have found that this shift is a few tenths of magnitude in the $B$-band,
compared to quiescent galaxies (however see also Bica et al. \cite{bica_etal90}).
We expect that the effect in the IR must be even lower, and certainly
not comparable to the $\sim2$ magnitudes we observe.

{A dwarf-giant galaxy dichotomy has been discussed, e.g., in
Garnett (\cite{garnett02}; hereafter G02), where the trend of effective
yields $y_{{\rm eff}}=Z/\ln(\mu^{-1})$ vs. galaxy luminosity or
rotational velocity is studied. It is shown that dwarf galaxies (those
with $v_{{\rm rot}}\lesssim100$~km~s$^{-1}$) have lower effective yields
than massive spiral galaxies, which is interpreted as due to gas
outflows in dwarfs. A similar conclusion has been reached by Tremonti et
al. (\cite{tremonti_etal04}). In G02 the stellar mass that enters the
gas mass fraction $\mu$ was computed from blue luminosities, assuming a
color-$M/L$ relation.  Since more accurate stellar masses can be
computed with NIR luminosities, the availability for massive spiral
galaxies of abundances obtained with temperature-sensitive methods would
allow more accurate effective yields to be computed as well, and
therefore to update the G02 study. In particular it would be interesting
to check whether indeed dwarf galaxies can lose up to $90\%$ of their
metals and be the main source of enrichment of the inter-galactic
medium.}

\section{Summary and conclusions}

We have measured oxygen abundances in \ion{H}{ii} regions located
inside a number of dIrr galaxies, belonging to the nearby Sculptor
and M81 groups. The abundances were measured with the temperature-sensitive
(direct) method, based on the ratio of the auroral line {[}\ion{O}{iii}]$\lambda4363$
to the nebular lines {[}\ion{O}{iii}]$\lambda\lambda4959,5007$,
and were complemented by direct abundances from the literature for
two additional galaxies. The weak {[}\ion{O}{iii}]$\lambda4363$
line could be measured in \ion{H}{ii} regions belonging to five
galaxies of the Sculptor group and two galaxies of the M81 group.
Metallicity gradients were detected in ESO~245-G005 and DDO~42,
confirming earlier findings, so only the central highest-metallicity
regions were considered for the derivation of the luminosity-metallicity
relation. This forced us to discard our DDO~42 measurement, and adopt
a literature value. Our fiducial sample then includes six dIrr galaxies
of the Sculptor group and two galaxies of the M81 group.

For these eight galaxies we have obtained deep NIR, $H$-band, imaging
which allowed us to perform surface photometry and compute their total
luminosities. Thanks to the availability of distances with $\leq15\%$
errors, the galaxies could be placed in the $12+\log{\rm (O/H)}$
vs. $M_{H}$ diagram, revealing a clear \lz\ relation with a small
$0.11$~dex scatter around the average trend. The scatter is smaller
than that of relations obtained at optical wavelengths (e.g. $0.161$~dex
using $B$-band data, Lee et al. \cite{lee_etal06}), and is comparable
to the one obtained by Lee et al. (\cite{lee_etal06}) with their
MIR {[}4.5\micron]-band data ($0.12$~dex). Assuming the existence
of a fundamental mass-metallicity relation, the improved definition
of the \lz\ relation at NIR luminosities must be due to the fact
that it better traces the underlying relation with mass, since the
NIR mass-to-light ratio is more sensitive to the \emph{integrated}
star formation history of a galaxy. On the contrary, the blue mass-to-light
ratio is more sensitive to the instantaneous SFR, which has a large
galaxy-to-galaxy scatter (Saviane et al. \cite{sidney}; Salzer et
al. \cite{salzer05}).

Our work and that of Lee et al. (\cite{lee_etal06}) are the ones
that managed to obtain the best defined \lz\ relation, and as an
additional advantage, our standard NIR band allows an easy comparison
to other relations obtained in independent studies. Indeed we compared
our relation to that of Salzer et al. (\cite{salzer05}), who are
the only authors providing an $H$-band relation, at the same time
extending it to giant galaxies. Unfortunately a direct comparison
with our accurate abundances is not possible, since the abundances
of S05 are obtained with the so-called strong-line method (indirect
or empirical method). To be consistent with S05 the abundances of
our dwarf sample were also computed with the empirical method, and
although the new \lz\ relation has more scatter than the one using
direct abundances, in this way our dwarf galaxies can be placed in
a same graph together with the galaxies analyzed by S05. The slope
and zero-point of the relation for our dwarfs, using indirect abundances,
are different than those of the relation obtained with direct abundances,
and, when compared to S05 giants, a possible break-point appears at
$M_{H}\approx-20$. While this could suggest a different gas-consumption
mechanism for dwarf and giant galaxies, a solid conclusion cannot
be established at this stage. Indeed, the version of the empirical
method employed by S05 is different from that classically used, and
so a slight inconsistency between our abundances and those of S05
may still exist. The best way for giving the final word on this possible
giant-dwarf dichotomy would be to obtain direct abundances for a good
number of the KISS galaxies. And precisely, such a project has been
started recently by some of us (see Saviane et al. \cite{saviane_etal07}).

One of the motivations for the current study was to remedy the lack
of homogeneous abundances and luminosities in the literature, but
still it was interesting to make the experiment of assembling an \lz\ relation
based on literature data, and see whether our approach really turned
out to be superior. We concluded that indeed mixing data from a variety
of sources builds an \lz\
relation affected by large uncertainties, and very different from
the one obtained with a controlled sample and controlled methods.
Inferences based on literature data are thus to be taken as qualitative
at best, and should be discarded, if possible.

Limiting ourselves to our sample of dIrrs, we attempted to explain
the \lz\ relation assuming that the chemical evolution of these
objects is similar to that of a closed-box model with an effective
yield which is $1/3$ its value in the solar vicinity (Skillman et
al. \cite{skillman_scl_hii}, Pilyugin et al. \cite{pilyugin_etal_04}).
If this approximation holds true, and if we assume that the chemical
evolution has started at the same time irrespective of mass, then
the conclusion is that more massive galaxies have faster chemical
evolutions (gas-consumption rates). Alternatively, more massive galaxies
had more time to build up their metal content.

Due to a number of reasons (mainly adverse weather conditions) we
could assemble a large database only for the Sculptor group of galaxies,
while only two galaxies of the M81 group enter our \lz\ relation.
Although these two dwarfs seem to follow the same relation as that
defined by the Sculptor group galaxies, more M81 dIrrs need to be
measured, before one can understand whether the higher density in
that group is influencing the chemical evolution of its members.

{Indeed in general terms our \lz\ relation is based on a
small number of galaxies, therefore more objects need to be added
to Fig.~\ref{fig:lz} in order to confirm our \lz\ relation with
a larger sample. For example looking at the number of dIrr presented
in Lee et al. (\cite{lee_etal06}), we might be able to add $\sim20$
galaxies to our database. And 10m-class facilities will have to be
used to reach this goal in a reasonable time.} In the short term we
plan to keep collecting data for the M81 group, and start a similar
study for the Centaurus~A group. A natural extension of the project
would be to investigate clusters of galaxies, but this is a difficult
task from the observational point of view, since the two nearest galaxy
clusters are Virgo and Fornax, at $22$ and $24$~Mpc respectively
(Ferguson \& Sandage \cite{ferguson_sandage90}; hereafter FS90).
A more viable alternative would be to observe the Leo group of galaxies.
At a distance of $18.2$~Mpc (FS90), this group is the fourth nearest
one, and has a relatively high velocity dispersion of $250\,{\rm km\, sec}^{-1}$.
Its properties (density, number of galaxies, dwarf-to-giant galaxy
ratio, etc.) are intermediate between those of nearby groups and those
of clusters of galaxies (see e.g. Figure~1 and following in Ferguson
\& Sandage \cite{ferguson_sandage91}). It would then allow us to
begin the exploration of a cluster-like environment, with NIR cameras
at 10m-class telescopes.

\begin{acknowledgements}
This research has made use of the NASA/IPAC Extragalactic Database
(NED) which is operated by the Jet Propulsion Laboratory, California
Institute of Technology, under contract with the National Aeronautics
and Space Administration. We thank John Salzer for providing the data
of KISS galaxies in tabular form, and the referee for providing useful
comments that improved the presentation of our work. Regina Riegerbauer,
Nadia Millot, and Richard Whitaker helped with the data reductions
at different stages of this project. This paper is dedicated to the
memory of Maria Saviane. 
\end{acknowledgements}

\appendix

\section{Reduction of EFOSC2 long-slit spectroscopy \label{sec:Reduction-of-EFOSC2}}

\subsection{Extraction of spectra}

All nights were treated separately, and reductions were carried out
using a customized version of the EFOSC2 quick-look tool (\texttt{http://www.ls.eso.org/}),
which is based on the \noun{eso-midas} data reduction system. The
main reduction steps are summarized as follows.

All bias frames were inspected to ensure that their level was the
same, and an average bias was then computed and subtracted from the
rest of the frames. The master flatfield was computed by taking the
median of five normalized frames and then subtracting the spectral
signature of the quartz lamp. The latter was computed by first averaging
the flux along the spatial direction, then fitting a polynomial of
degree 6 along the dispersion, and finally expanding it back into
a bidimensional frame. The wavelength calibration was computed based
on the spectrum of the internal Helium-Argon lamp, taken with the
$1\arcsec$ wide slit. Since the projection of the $5\arcsec$ wide
slit on the CCD is in the same position, the calibration was used
both for the spectra of the \ion{H}{ii} regions and the spectrophotometric
standards. A transformation of the form $(x,y)\rightarrow(\lambda,s)$
was found, thus correcting also for the distortion of the spectra
introduced by the optical system. The transformation was modeled with
polynomials of degree $3$ and $2$ along the dispersion and the spatial
direction, respectively. After applying the calibration, the spectra
were linearly rebinned with a constant step of $\Delta\lambda=4.23$\AA.
The bidimensional sky frame was created by first sampling the sky
spectrum in two windows flanking the target spectrum, and then fitting
the spatial gradient with a polynomial of degree $1$. After subtracting
the sky frame, the extraction window of the emission-line spectra
was decided by looking at the spatial profile of the H$\alpha$ line
and making some experiments until the best S/N was found. On the other
hand, the flux of the spectrophotometric standard stars was summed
over almost the entire spatial profile, leaving out just the two sky
windows. As it was recalled above, the response function for the flux
calibration was computed using the star LTT~3218. To do this, the
instrumental spectrum was corrected for atmospheric extinction, divided
by the exposure time, and then divided by the published spectrum,
and the ratio was fitted with a polynomial of degree $12$. The spectrophotometric
standard was observed at several airmasses, and we checked that the
extinction-corrected and normalized spectra are almost identical,
thus confirming the extinction curve for La Silla provided by \noun{midas}.
The response function was then computed as the median of the five
functions found for the two nights. The pre-processed, wavelength-calibrated,
sky-subtracted and extracted spectra were thus finally flux-calibrated
using this response function.

\subsection{Flux measurements \label{sec:Flux-measurements}}

The fluxes were measured within the \noun{alice} \noun{contex} in
\noun{midas}. A fourth-degree continuum was fitted to the line-free
spectra, and then for each line the flux was summed within two wavelengths
bracketing the line. The blends H$\gamma$+{[}\ion{O}{iii}]$\lambda4363$,
H$\alpha$+{[}\ion{N}{ii}]$\lambda6583$ and the {[}\ion{S}{ii}]
doublet $\lambda\lambda6716,6731$ were deconvolved using Gaussian
fits, and we ensured that the sum of the single line fluxes equaled
that of their blend. The measured line ratios were corrected for the
effects of reddening, measured by comparing the observed and the expected
hydrogen line ratios. The theoretical H line ratios of Hummer \& Storey
(\cite{hummer}) were used, assuming an electron temperature $T_{\textrm{e}}\simeq10000$
K. We set $R=I_{\lambda}/I_{H\beta}$, $R_{0}=I_{\lambda,0}/I_{H\beta,0}$
and $\varphi(\lambda)=[f(\lambda)-f(H\beta)]$ so that $C=[\log R_{0}/R]/\varphi(\lambda)$.
The reddening constant was then computed using the Cardelli et al.
(\cite{cardelli}) extinction law, where $f(\lambda)=<A(\lambda)/A(V)>$,
and we assumed $R_{V}=3.1$. We obtained the value of $C$ both from
the ratio H${\alpha}$/H${\beta}$ and from H${\beta}$/H${\gamma}$.
These line ratios are affected by the unknown stellar absorption under
the H emission lines: we then corrected the fluxes assuming a constant
EW of the absorption lines, and adopted the EW that yielded the closest
values for the two $C$. In some cases we were not able to resolve
H$\alpha$ from the {[}\ion{N}{ii}] lines; however, the flux
of the latter lines is at most a few percent of that of H$\alpha$,
so we assumed that this correction is negligible. Since the H${\alpha}$/H${\beta}$
ratio is the more accurate, the value from this ratio was adopted
to correct the observed fluxes, which were finally normalized to H$\beta$
whose flux was set to $1$.

\section{Reduction of \emph{KAST} long-slit spectroscopy \label{sec:Reduction-of-KAST}}

\subsection{Extraction of spectra}

\begin{figure}
\begin{centering}
\includegraphics[width=1\columnwidth]{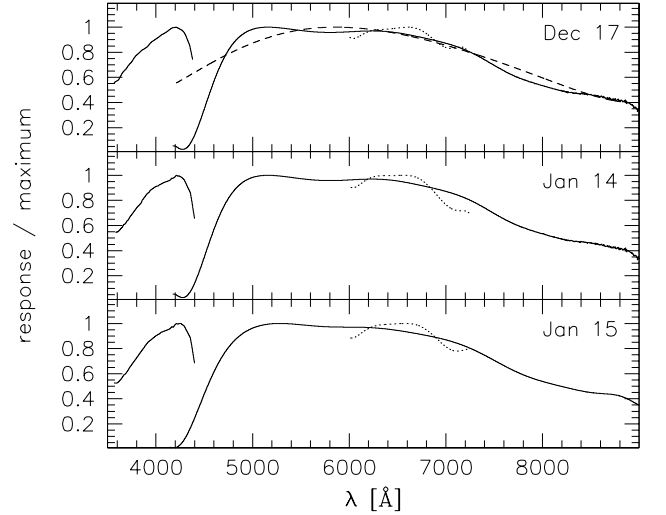} 
\par\end{centering}

\caption{Response functions for each date of the Lick run: the solid curves
are the functions for the blue and red spectra, the dashed curve is
the full spectrum function (no dichroic), and the dotted lines are
the functions for the high resolution spectra. The functions have
been normalized to the peak response. \label{figcap:Response-functions-for}}

\end{figure}

\begin{table}
\caption{Dispersions and wavelength range for each date and \emph{Kast} configuration
\label{tab:Dispersions-and-wavelength}}

\begin{centering}
\begin{tabular}{lccc}
 &  & \multicolumn{2}{c}{}\tabularnewline
\hline
\hline 
\multicolumn{1}{c}{\emph{Kast} setup} & disp.  & \multicolumn{2}{c}{wavelength range}\tabularnewline
\multicolumn{1}{c}{Arm/dichroic/grating|grism} & {[}\AA/px]  & \multicolumn{2}{c}{}\tabularnewline
\hline 
 &  & \multicolumn{2}{c}{}\tabularnewline
\multicolumn{4}{c}{Dec 17, 2001}\tabularnewline
blue/D46/830\_3460  & 1.1  & 3198.0  & 4532.2\tabularnewline
red/D46/1200\_5000  & 1.2  & 5856.5  & 7240.8 \tabularnewline
red/D46/300\_4230  & 4.6  & 4216.2  & 9099.2 \tabularnewline
red/none/300\_4230  & 4.6  & 3688.1  & 9097.7 \tabularnewline
 &  &  & \tabularnewline
\multicolumn{4}{c}{Jan 14, 2002}\tabularnewline
blue/D46/830\_3460  & 1.1  & 3213.1  & 4552.3\tabularnewline
red/D46/1200\_5000  & 1.1  & 5874.4  & 7194.9\tabularnewline
red/D46/300\_4230  & 4.6  & 4170.7  & 9052.5 \tabularnewline
 &  &  & \tabularnewline
\multicolumn{4}{c}{Jan 15 2002}\tabularnewline
blue/D46/830\_3460  & 1.1  & 3213.3  & 4552.7\tabularnewline
red/D46/300\_4230  & 4.6  & 4149.0  & 9032.9\tabularnewline
\hline
\end{tabular}
\par\end{centering}
\end{table}

In the following we will call blue, red, full, and high-resolution,
the spectra taken with the blue arm and grism $830/3460$, with the
red arm and grating $300/4230$ (plus dichroic D46), with the red
arm and no dichroic, and with the red arm and grating $1200/5000$,
respectively.

The frame pre-processing, wavelength calibration, and flux calibration
was done within \noun{eso-midas} (and its {}``context'' \emph{\noun{long}}),
with a modified version of the quick-look tool described above. The
procedure was similar to the EFOSC2 data reduction, except for the
bias subtraction, which is done directly by the \emph{Kast} acquisition
system. All the frames were trimmed to remove the overscan regions,
and then master flat-field (FF) and arc images were created. Three
to five dome flat-field frames per spectrograph configuration were
available, so a median frame was first created, and then the spectral
signature of the lamp was removed by fitting a polynomial surface
to the median FF, assuming that the polynomial does not vary along
the spatial direction.

Our main concern in the data reduction was to ensure that, in the
overlap region, the fluxes of blue and red spectra are calibrated
with the smallest systematic uncertainties. This was particularly
difficult for the shortest wavelengths of the red spectra (blueward
of $\sim5000$~\AA), where the response function drops steeply from
$50\%$ to zero in just $\sim300$~\AA\ (see below and Fig.~\ref{figcap:Response-functions-for}).
This gradient is extremely sensitive to the FF correction in this
spectral region, and it was found that a polynomial of degree $7$
gave the best response function for the red spectra. For all other
spectra a degree $6$ polynomial gave the best results.

The arcs were used to remap the bidimensional spectra from the pixel-pixel
space to the wavelength-pixel space, thereby also correcting the off-axis
distortions of the optics, such that the spectrum could be extracted
by just defining a window in the spatial direction. The two \emph{Kast}
lamps are filled with NeAr and HeHgCd, and since no cadmium or mercury
line lists are available in the \noun{midas} system, a suitable line
list was built from various sources. The computed dispersions and
wavelength coverages for each \emph{Kast} configuration are given
in Table~\ref{tab:Dispersions-and-wavelength}. The dispersion is
$\sim1$\AA~px$^{-1}$ for the blue and high resolution spectra,
and $4.6$\AA~px$^{-1}$ for the red and full range spectra. The
resolution, as measured on the arc lines, is $\sim3$\AA~FWHM and
$\sim10$\AA~FWHM for the high and low resolution spectra.

In the next step, all the spectra were divided by the master FF and
calibrated in wavelength, and then a preliminary extraction with a
large window was done, in order to create sky images for each spectrum,
that were used for the correction of the spectrograph flexures (see
below). The sky frames were constructed by first creating the sky
spectrum on the two sides of the central object, as an average within
two windows. The flanking sky spectra were used to fit the sky gradient
in the spatial direction, assumed to be linear and independent of
wavelength, and the bidimensional artificial sky frames were made
by simulating this gradient.

By comparing the position of sky lines in different spectra, it was
soon realized that the spectrograph is subject to large flexures,
with spectrum-to-spectrum differences of up to $10$~px, i.e. more
than $40$~\AA\ for the red spectra. This fact has several implications,
and at this point of the reduction it affects the definition of the
response function, in particular at the shortest wavelengths of the
red spectra. For this reason, the standard star spectra were registered
to rest-frame wavelengths before creating the response function. Due
to the short exposure times, the sky lines are almost absent from
the standard star spectra, so the shift in wavelength was computed
using the central wavelength of H$\alpha$ for the red spectra, and
H$\gamma$ for the blue spectra. This approach works if the effect
of radial velocities on the position of the stellar lines is negligible.
This assumption was checked by comparing the relative position of
H$\alpha$ and the telluric triplet at $\sim6870$\AA. The average
difference over all standard stars is $\Delta\lambda=309.1\pm0.7$
\AA, and since the scatter is much less than the spectral dispersion,
we conclude that indeed the effect of radial velocities (both intrinsic
and due to the solar motion) is negligible, and that the barycenter
of the telluric triplet is at $6871.9$ \AA.

The shift in wavelength was then subtracted from the starting wavelength
of the bidimensional spectra, the standard star spectra were extracted
again, and the response function was defined by comparison to the
published spectrum. Again, the most difficult task is following the
response function for the red spectra blueward of $\sim5000$\AA.  It was
found that a polynomial fit is not adequate even using the highest
degree implemented in \noun{midas} (30). We then used a spline of degree
1, which gave good results since the sample rate of the published fluxes
is a few \AA. The master response function was decided by
flux-calibrating all the standard stars with a given response function,
and then by taking the one that gave the best average calibration for
all standards and over a range of airmasses (see
Fig.~\ref{figcap:Response-functions-for}).  Due to the large wavelength
coverage, the most critical case is that of the red spectra, where we
expect a maximum difference of $\sim5\%$ from the red to the blue part
of the spectrum. For these spectra, the response function is able to
correct the instrumental flux down to $4300$\AA.

Finally the galaxy spectra were extracted and flux-calibrated with
a procedure resembling that adopted for the standard stars, but in
this case the flexure correction was done with the very strong sodium
sky line at $5892$\AA. In order to have a good subtraction of the
sky lines, the sky windows were chosen close to the object spectrum,
and the best windows were decided on an object-by-object basis with
several experiments. Also the windows for the object spectra needed
a careful selection, in order to enhance the faint lines over the
background, and to avoid as much as possible the places where the
sky line subtraction was less than optimal.

\subsection{Creation of merged spectra}

Using the continuum in the overlap regions, the final spectra have
been obtained by merging the blue, red (or full in the case of DDO~42),
and high resolution spectra. This implies that a reliable merging
could be produced only for those spectra where a meaningful continuum
could be defined (i.e. those with the higher $S/N$), namely for the
regions in DDO~42, DDO~53, and UGC~4483. First, the continua have
been equalized in the overlap regions, and then the spectra have been
rebinned to a common $1$\AA\ step, which is smaller than any of
the other dispersions. The equalization of the continua has been done
in the $\log(\lambda)-\log({\rm flux)}$ plane: a straight line has
been fitted to the continua, and then the difference $d$ of the two
fits at the midpoint of the overlap region has been taken. The normalization
factor is then $10^{d}$. In the merged spectrum H$\alpha$ belongs
to the high-resolution spectrum, H$\beta$ belongs to the red (or
full) spectrum, and H$\gamma$ belongs to the blue spectrum. These
hydrogen lines were then used to perform some health checks on the
merged spectra. Both H$\alpha$ and H$\beta$ are present in the red
(or full) spectrum, so we first checked that the ratio of these two
lines is the same in the merged spectrum as in the red (or full) spectrum.
If that was not the case, corrections to the equalization factors
were introduced. H$\gamma$ is present both in the blue and in the
red (or full) spectrum, however in that spectral region the response
function of the red (or full) spectrum reaches almost zero, so its
flux is rather uncertain. Comparing the flux of this line in the two
spectra is then not really meaningful, so we followed another approach.
For the Sculptor group dwarfs, Table~\ref{tabcap:Reddening-corrected-fluxes-for}
shows that the average flux of H$\gamma$ is $0.46\times{\rm H\beta}$,
with a dispersion of $0.04\times{\rm H\beta}$. Thus we made sure
that no strong deviations from this value were obtained: if the H$\gamma$
flux was more than $3\sigma$ discrepant from the quoted value, we
adjusted the equalization factor to bring the flux into agreement
with the average value while maintaining a good overlap of the continua.
This meant reducing the deviations to less than $2\sigma$.

\subsection{Flux measurements}

Line fluxes were measured in the way described in Sect.~\ref{sec:Flux-measurements},
regarding EFOSC2 measurements. The difference is that a different
set of lines had to be deblended. In particular, {[}\ion{O}{ii}]~$3727$
was deblended from two nearby unknown lines at $3722$\AA\ and $3736$\AA,
{[}\ion{O}{iii}]$4363$ was deblended from ${\rm H\gamma}$$4340$,
and finally the Gaussian deblending was employed to separate {[}\ion{O}{ii}]$7320$
from {[}\ion{O}{ii}]$7330$.

\section{Reduction of near-IR imaging \label{sec:Reduction-of-near-IR}}

The first step of the data reduction was to remove the sky background,
and simultaneously, the dark current and bias. We subtracted from
each object image the average of the preceding and the succeeding
sky images. This strategy accounts for sky background variations with
characteristic time comparable or a few times longer than the time
spent at each pointing. For the first and the last object, we could
only subtract the nearest sky image. Next, we carried out the flat
field correction. A common median zero point was imposed to all available
sky images, a median was obtained to produce a flat field, which was
normalized to unity. Then, all sky-subtracted object images were divided
by the flat-field. The bad pixels were masked out and all object images
were aligned and combined to produce a final image for each target.
In addition, for the SOFI data, we carried out an illumination correction
to account for the difference between the sky illumination and the
illumination of the dome screen used for flat-fielding, as described
on the instrument webpages. The entire data reduction was carried
out with standard \textit{\emph{IRAF}} tasks.

The photometric calibration was carried out by observations of standard
stars from Hunt et al. (\cite{hunt+98}) for the Northern objects,
and from Persson et al. (\cite{pers+98}) for the Southern ones. The
INGRID data were photometrically calibrated with measurements of 129
standard stars in 13 fields from the night of Jan 16, 2003, and the
SOFI data were calibrated with 13 standard stars observed during all
nights. Some galaxies were observed twice at the NTT to verify the
self consistency of the photometry and because the night of Oct. 15,
2002 was non-photometric, so all targets had to be re-observed during
the next night with shorter integration times. Finally, the calibration
of ESO\,473-024 is based on a single standard, observed at the same
airmass as the target. The transformation equations are:

\begin{equation}
H=h-(0.055\pm0.005)\times\sec(z)+(23.430\pm0.007)\end{equation}
 with r.m.s.=0.040 mag for INGRID,

\begin{equation}
H=h-(0.032\pm0.003)\times\sec(z)+(24.727\pm0.006)\end{equation}
 with r.m.s.=0.011 mag for the Oct 2002 SOFI run, and

\begin{equation}
H=h+(24.775\pm0.009)\end{equation}
 for the Aug 2003 SOFI run. Here the upper case indicates magnitudes
in the standard system, and the lower case indicates the instrumental
system. 
\end{document}